\documentclass[aps,prd,showpacs,notitlepage,nofootinbib,preprintnumbers,amsmath,amssymb]{revtex4-1}

\usepackage{graphics,graphicx}
\usepackage{dcolumn}
\usepackage{bm}
\usepackage{epsfig}
\usepackage[usenames]{color}
\usepackage{hyperref} 
\usepackage{mathbbol}
\usepackage{epstopdf}
\usepackage{simplewick}
\usepackage[utf8]{inputenc} 
\usepackage{slashed}

\def\eq#1{{Eq.~(\ref{#1})}}
\def\fig#1{{Fig.~\ref{#1}}}
\newcommand{\ben}{\begin{eqnarray*}}
\newcommand{\een}{\end{eqnarray*}}

\newcommand{\as}{\alpha_s}

\newcommand{\dhd}{{\textstyle d}
\lower.03ex\hbox{\kern-0.38em$^{\scriptstyle-}$}\kern-0.05em{}}
\newcommand{\dbar}{{\textstyle \delta}
\lower.03ex\hbox{\kern-0.38em$^{\scriptstyle-}$}\kern-0.05em{}}
\newcommand{\half}{{1\over 2}}

\newcommand{\ul}[1]{\underline{#1}}

\newcommand{\ubar}[1]{\overline{U}_{#1}}
\newcommand{\vbar}[1]{\overline{V}_{#1}}
\newcommand{\Tr}{\mathrm{Tr}}

\begin{document}

\title{Helicity Evolution at Small $x$}

\author{Yuri V. Kovchegov} 
         \email[Email: ]{kovchegov.1@osu.edu}
         \affiliation{Department of Physics, The Ohio State
           University, Columbus, OH 43210, USA}
\author{Daniel Pitonyak}
	\email[Email: ]{dpitonyak@quark.phy.bnl.gov}
	\affiliation{RIKEN BNL Research Center, Brookhaven National
          Laboratory, Upton, New York 11973, USA}
\author{Matthew D. Sievert}
	\email[Email: ]{msievert@bnl.gov}
	\affiliation{Physics Department, Brookhaven National
          Laboratory, Upton, NY 11973, USA \\}

\begin{abstract}
  We construct small-$x$ evolution equations which can be used to
  calculate quark and anti-quark helicity TMDs and PDFs, along with
  the $g_1$ structure function. These evolution equations resum powers
  of $\as \, \ln^2 (1/x)$ in the polarization-dependent evolution
  along with the powers of $\as \, \ln (1/x)$ in the unpolarized
  evolution which includes saturation effects. The equations are
  written in an operator form in terms of polarization-dependent
  Wilson line-like operators. While the equations do not close in
  general, they become closed and self-contained systems of non-linear
  equations in the large-$N_c$ and large-$N_c \, \& \, N_f$ limits. As
  a cross-check, in the ladder approximation, our equations map onto
  the same ladder limit of the infrared evolution equations for the $g_1$
  structure function derived previously by Bartels, Ermolaev and
  Ryskin \cite{Bartels:1996wc}.
\end{abstract}


\maketitle



\section{Introduction}

Our understanding of high-energy quantum chromodynamics (QCD) has
expanded greatly in the past two decades, in part due to the new and
exciting developments in saturation physics
\cite{Gribov:1984tu,Mueller:1986wy,McLerran:1993ni,McLerran:1993ka,McLerran:1994vd,Kovchegov:1996ty,Kovchegov:1997pc,Jalilian-Marian:1997xn}
and non-linear small-$x$ evolution
\cite{Mueller:1994rr,Mueller:1994jq,Mueller:1995gb,Balitsky:1996ub,Balitsky:1998ya,Kovchegov:1999yj,Kovchegov:1999ua,Jalilian-Marian:1997dw,Jalilian-Marian:1997gr,Iancu:2001ad,Iancu:2000hn}
(see
\cite{Gribov:1984tu,Iancu:2003xm,Weigert:2005us,Jalilian-Marian:2005jf,Gelis:2010nm,Albacete:2014fwa,KovchegovLevin}
for reviews and a book). During those developments, most of the
research effort was dedicated to calculating unpolarized observables,
such as total cross sections, unpolarized particle production and
correlations. In more recent years the interest in applying the
formalism of small-$x$ physics to calculating polarization-dependent
observables has been on the rise
\cite{Boer:2006rj,Boer:2008ze,Boer:2002ij,Dominguez:2011br,Metz:2011wb,Kovchegov:2012ga,Dominguez:2011br,Mueller:2012uf,Mueller:2013wwa,Kang:2011ni,Kang:2012vm,Schafer:2013mza,Zhou:2013gsa,Altinoluk:2014oxa,Kovchegov:2013cva,Kovchegov:2015zha,Balitsky:2014wna,Balitsky:2015qba,Tarasov:2015pxa,Altinoluk:2015gia,Boer:2015pni},
with particular attention being devoted to calculating the quark and
gluon transverse momentum-dependent parton distribution functions
(TMDs) of the proton \cite{Collins:1989gx,Collins:1981uk}. The goal of
the present work is to apply the small-$x$ saturation formalism to
helicity evolution in order to assess the amount of proton's spin
carried by the partons at small $x$.

One of the most profound mysteries in our understanding of the proton
structure is the so-called spin puzzle
\cite{Ashman:1987hv,Ashman:1989ig} (see
\cite{Accardi:2012qut,Aschenauer:2013woa,Aschenauer:2015eha} for a
comprehensive review of the experimental situation and additional
references): the presently measured spin and orbital angular momentum
(OAM) carried by the quarks and gluons inside the proton does not
appear to add up to $1/2$. The puzzle can be formalized in terms of
helicity sum rules \cite{Jaffe:1989jz,Ji:1996ek,Ji:2012sj} such as the
Jaffe-Manohar form~\cite{Jaffe:1989jz}
\begin{align}
  \label{eq:sum_rule}
  S_q + L_q + S_G + L_G = \frac{1}{2}. 
\end{align}
Here $L_q$ and $L_G$ are the orbital angular momenta of the quarks and
gluons respectively, while $S_q$ and $S_G$ denote the spin carried by
all the quarks and gluons and are defined as the following integrals
over Bjorken-$x$ at fixed momentum scale $Q^2$
\begin{align}
  \label{eq:net_spin}
  S_q (Q^2) = \frac{1}{2} \, \int\limits_0^1 dx \, \Delta \Sigma (x,
  Q^2), \ \ \ S_G (Q^2) = \int\limits_0^1 dx \, \Delta G (x, Q^2)
\end{align}
with 
\begin{align}
  \label{eq:Sigma}
  \Delta \Sigma (x, Q^2) = \left[ \Delta u + \Delta {\bar u} + \Delta
    d + \Delta {\bar d} + \ldots \right] \! (x, Q^2),
\end{align}
and the helicity parton distribution functions (hPDFs)
\begin{align}
  \label{eq:hPDFs}
  \Delta f (x, Q^2) \equiv f^+ (x, Q^2) - f^- (x, Q^2).
\end{align}
In \eqref{eq:hPDFs} $f^+$ ($f^-$) denote the number density of partons
with the same (opposite) helicity as the proton, and $f = u, {\bar u},
d, {\bar d}, \ldots , G$.

Indeed in the actual experiments due to finite center-of-mass energy
one cannot measure $\Delta \Sigma$ and $\Delta G$ down to $x=0$ as
required by \eq{eq:net_spin}: these quantities are measured down to
some minimal available value of $x$ instead. The current quark and
gluon spin values extracted from the experimental data are $S_q (Q^2 =
10\, \mbox{GeV}^2) \approx 0.15 \div 0.20$ (integrated over $0.001 < x
<1$) and $S_G (Q^2 = 10 \, \mbox{GeV}^2) \approx 0.13 \div 0.26$
(integrated over $0.05 < x <1$)
\cite{deFlorian:2014yva,Nocera:2014gqa,Aschenauer:2012ve}. Even the
largest values of $S_q$ and $S_G$ (in the above ranges) do not add up
to give $1/2$. The missing proton's spin is likely to be in the quark
and gluon OAM and/or at smaller values of $x$. It is, therefore,
important to gain good qualitative and quantitative understanding of
the behavior of $\Delta \Sigma$ and $\Delta G$ at small $x$. A strong
growth of these quantities at small $x$ may indicate that a large
amount of proton spin resides in that region of phase space, possibly
offering a solution to the proton spin puzzle.

The small-$x$ evolution of the quark polarization and flavor-dependent
observables was first considered by Kirschner and Lipatov in
\cite{Kirschner:1983di} (see also
\cite{Kirschner:1994rq,Kirschner:1994vc,Griffiths:1999dj}). The
calculation was similar to the well-known
Balitsky--Fadin--Kuraev--Lipatov (BFKL) equation
\cite{Kuraev:1977fs,Balitsky:1978ic}, using quarks instead of
gluons as the $t$-channel lines in the corresponding ladder. In
addition, non-ladder diagrams turned out to be important in the
calculation, making it very hard to resum all the relevant graphs
using the BFKL-like evolution equation in rapidity; the resummation in
\cite{Kirschner:1983di} was instead performed using a (non-linear) evolution
in the infrared cutoff. Perhaps most importantly it was demonstrated
that the small-$x$ limit of the quark-exchange amplitudes is dominated
by the resummation of the double-logarithmic (DLA) parameter $\as \,
\ln^2 s \sim \as \, \ln^2 (1/x)$ (with $s$ the center of mass energy
squared and $\as$ the strong coupling constant).\footnote{Henceforth
  we will refer to the resummation of powers of $\as \, \ln^2 s \sim
  \as \, \ln^2 (1/x)$ as the DLA resummation. This is not to be
  confused with the DLA limit of BFKL evolution, which resums powers
  of $\as \, \ln s \, \ln Q^2$.} This is in contrast to the standard
BFKL evolution, along with the nonlinear Balitsky--Kovchegov (BK)
\cite{Balitsky:1996ub,Balitsky:1998ya,Kovchegov:1999yj,Kovchegov:1999ua}
and Jalilian-Marian--Iancu--McLerran--Weigert--Leonidov--Kovner
(JIMWLK)
\cite{Jalilian-Marian:1997dw,Jalilian-Marian:1997gr,Iancu:2001ad,Iancu:2000hn}
evolution equations, all of which, at the leading order in $\as$ in
the kernel resum powers of single logarithm, $\as \, \ln (1/x)$. For
the $\as \, \ln^2 (1/x)$ parameter to become important one does not
need $x$ to be as small as would be needed for the $\as \, \ln (1/x)$
parameter to become important: hence the effects of small-$x$
evolution on polarization-dependent observables may be stronger and
easier to observe in experiments than the small-$x$ evolution for
unpolarized observables.

The small-$x$ limit of an observable relevant to the spin puzzle, the
structure function $g_1 (x, Q^2)$, was first considered by Bartels,
Ermolaev and Ryskin (from now on referred to as BER) in
\cite{Bartels:1995iu,Bartels:1996wc}. At the leading order in $\as$
and at leading twist this structure function is
\begin{align}
  \label{eq:g1}
  g_1 (x, Q^2) = \frac{1}{2} \sum_{f=\mbox{quarks}} Z_f^2 \, \left[
    \Delta f (x, Q^2) + \Delta {\bar f} (x, Q^2) \right]
\end{align}
with $Z_f$ the electric charge of a quark of flavor $f$ in units of the
electron's charge. Resumming double-logarithms $\as \, \ln^2 (1/x)$
using the infrared evolution equations, BER obtained a tantalizing
result: they argued that at small $x$ the $g_1 (x, Q^2)$
structure function, and, hence $\Delta \Sigma$, grow as
$(1/x)^{\omega_s}$ with a fairly large power $\omega_s = 1.01$ for
$\as = 0.18$ (with presumably larger $\omega_s$ for larger values of
the coupling $\as$). This result appears to indicate that small-$x$
partons may contribute a substantial amount to the net quark (and
gluon) spin in the proton $S_q$ ($S_G$).

The emerging physical picture of the proton would be quite
interesting: the majority of small-$x$ partons, generated through 
BFKL/BK/JIMWLK evolution are unpolarized. Nonetheless, there may exist
a minority of small-$x$ partons, generated though the non-eikonal
spin-dependent evolution, which carry a potentially large fraction of
the proton spin.

Our main aim in this work is to reproduce the results of BER using the
$s$-channel evolution language of
\cite{Mueller:1994rr,Mueller:1994jq,Mueller:1995gb,Balitsky:1996ub,Balitsky:1998ya,Kovchegov:1999yj,Kovchegov:1999ua,Jalilian-Marian:1997dw,Jalilian-Marian:1997gr,Iancu:2001ad,Iancu:2000hn},
which employs the light-cone Wilson line operators and, ultimately,
color dipoles. Moreover, we want to include saturation corrections
into the evolution equations, thus going beyond the evolution
considered by BER. Indeed saturation corrections resum
single logarithms and strictly speaking should come in only at the
level of the next-to-leading order (NLO) corrections to the DLA
helicity evolution. However, in the cases considered below it appears
possible to separate helicity evolution from the unpolarized BK/JIMWLK
evolution keeping the approximation and power counting under control,
akin to the evolution equation for the QCD Reggeon with saturation
corrections derived in \cite{Itakura:2003jp}.

The paper is structured as follows. In Sec.~\ref{sec:observables} we
define the observables we want to calculate, the quark and anti-quark
helicity TMDs and hPDFs. The hPDFs are related to the $g_1$ structure
function and are an essential ingredient in the spin sum rules. We
show that these observables depend on the ``polarized dipole''
operator \eqref{eq:correlators_sum}: this is a dipole operator in
which either the quark or the anti-quark interact with the target in a
polarization-dependent way. We construct the main ingredients of the
leading-order small-$x$ helicity evolution in Sec.~\ref{sec-ingr}. The
ingredients are the $q \to qG$, $G \to q\bar q$ and $G \to GG$
splitting kernels which include helicity dependence. We then study the
contributing diagrams in Sec.~\ref{sec:diag}. The main calculation is
carried out in Sec.~\ref{sec:operator}: there the evolution equations
for the fundamental and adjoint polarized dipoles are obtained after
one step of small-$x$ helicity evolution and are given in
Eqs.~\eqref{fund_evol} and \eqref{Gevol3} respectively. Just like the
equations in Balitsky hierarchy, equations \eqref{fund_evol} and
\eqref{Gevol3} are not closed, and involve higher-order Wilson-line
correlators on their right-hand sides. These are our main general
results.

At this point we cross-check our results against those of BER:
unfortunately the infrared evolution equations obtained by BER were
too complicated to allow for a complete analytic solution, and part of
their work including the calculation of the intercept (the power of
$1/x$ in the expression for $g_1$) was done numerically. However, BER
had an analytic result for the intercept of their evolution in the
case when one neglects the contributions of the non-ladder gluons
\cite{Bartels:1996wc}. Neglecting the non-ladder gluons is not
justified by the smallness of any parameter in the problem, and cannot
be used as an approximate solution of the problem in the sense of any
controlled approximation; nonetheless it is an interesting formal
limit where an analytic comparison is possible. In
Sec.~\ref{sec:operator} we neglect the non-ladder gluon contribution
and reproduce the analytic expression for the intercept obtained by
BER in the same limit (see \eq{ladder_intercept}). We thus accomplish
a successful cross-check of our technique.

Similar to the case of unpolarized evolution, to obtain a closed
equation out of, say, \eq{fund_evol} for the polarized dipole, one has
to take the large-$N_c$ limit \cite{'tHooft:1974hx} (with $N_c$ the
number of quark colors). Doing so in Sec.~\ref{sec:large_nc} we obtain
a closed system of equations \eqref{evol77} and \eqref{Gamma_evol}
(along with their linearized version \eqref{evol88}). The solution of
these equations (to be found in the future work) would yield the
energy dependence of the quark helicity TMD (along with the same
energy dependence of the $g_1$ structure function and $\Delta \Sigma$)
in the large-$N_c$ DLA limit.

Quarks are significantly more important for helicity small-$x$
evolution than for the unpolarized BFKL/BK/JIMWLK equations. The quark
contribution is neglected in the large-$N_c$ limit of
Sec.~\ref{sec:large_nc}. To preserve the contributions of the quarks
we take the large-$N_c \, \& \, N_f$ limit of Eqs.~\eqref{fund_evol}
and \eqref{Gevol3} in Sec.~\ref{sec:large_ncnf}. Here $N_f$ is the
number of quark flavors. More precisely, we assume that both $N_c$ and
$N_f$ are asymptotically large (in the evolution), while $\as \, N_c$
and $\as \, N_f$ are fixed and small. In the end we obtain a closed
system of equations \eqref{Qevol}, \eqref{Gevol}, \eqref{Aevol},
\eqref{Gamma_evol2} and \eqref{Gamma_evol3}. Solution of these
equations in their linearized form \eqref{Q_evol_lin},
\eqref{Gam_evol_lin} would yield an intercept governing the energy
dependence of helicity TMDs, the $g_1$ structure function and $\Delta
\Sigma$ which can be compared with the final (numerical) result of
BER. This is left for the future work.

We summarize our results in Sec.~\ref{sec:conc} and discuss potential
effects of saturation corrections on small-$x$ helicity evolution.


\section{The Observables}
\label{sec:observables}

To build helicity evolution we need to start by defining which
quantity we want to evolve. This is ultimately dictated by the
observable we wish to calculate. Let us start with the cross section
for semi-inclusive deep inelastic scattering (SIDIS) on a
longitudinally polarized target, $\gamma^* + {\vec p} \to {\vec q} +
X$ (with $p$ the target [e.g., proton] and $q$ the produced quark),
and the associated quark helicity TMD $g_{1L} (x, k_T)$, both
evaluated in small-$x$ kinematics, $s \gg Q^2 \gg k_T^2$. The
helicity TMD can be extracted from the SIDIS cross section.

\begin{figure}[htb]
\centering
\includegraphics[width= 0.9 \textwidth]{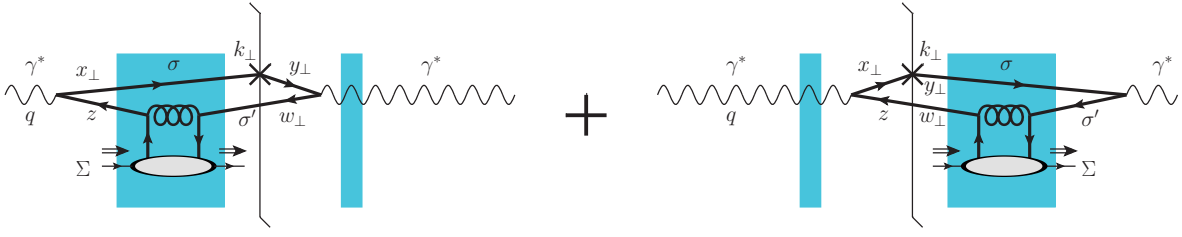}
\caption{Diagrams contributing to the quark helicity TMD $g^q_{1L} (x,
  k_T)$ at small-$x$.}
\label{fig:helicity_TMD}
\end{figure}

From the analysis of small-$x$ SIDIS and TMDs for polarized targets
carried out in \cite{Kovchegov:2015zha} we conclude that to calculate
$g_{1L} (x, k_T)$ one has to sum up the diagrams shown in
\fig{fig:helicity_TMD} (with all other potentially contributing
diagrams canceling out). We assume that the virtual photon is moving
along the light-cone ``+'' direction and work in the $A^+ =0$
light-cone gauge. As before, the shaded rectangles in
\fig{fig:helicity_TMD} denote the shock wave, though (an example of)
the polarization-dependent interaction with the target is shown
explicitly on top of the shock wave. If we model our proton as a large
nucleus \cite{McLerran:1993ni,McLerran:1993ka,McLerran:1994vd}, which
is the standard practice in saturation calculations, we would assume
that the polarization-dependent interaction happens with one of the
longitudinally polarized nucleons (with a sum over interactions with
all polarized nucleons implied). For the real proton one can think of
longitudinally polarized partons instead of nucleons, described by
some helicity TMD $g_{1L} (x_0, k_T)$ at the initial value $x_0$ of
Bjorken-$x$. Note that the virtual photon does not interact with the
shock wave when it goes through it: diagrams in \fig{fig:helicity_TMD}
are non-zero because we are using light-front perturbation theory
(LFPT) \cite{Lepage:1980fj,Brodsky:1997de}.

Just like in \cite{Kovchegov:2015zha} we begin by writing the quark
production cross section in SIDIS, this time based on the diagrams in
\fig{fig:helicity_TMD}.  The SIDIS cross section is weighted by the
polarization of the produced quark $\sigma$ and is summed over
$\sigma$ as well to give\footnote{Let us point out that the analysis
  in \cite{Kovchegov:2015zha} which led to the diagrams in
  \fig{fig:helicity_TMD} was performed for the quark TMD (in which the
  final-state interactions of the produced quark are defined to occur
  via an unpolarized gauge link), and not for the SIDIS cross section
  (where they may occur in a polarization-dependent way). Nonetheless,
  the diagrams in \fig{fig:helicity_TMD} give us the leading-energy
  contribution to the SIDIS cross section, which is extracted from
  \eq{eq:qprod_smallx} below and is given in
  \eq{eq:qprod_smallx2}. The terms in which the quark line scatters in
  a polarization-dependent way are not DLA:~note that
  \eq{eq:qprod_smallx} also contains some non-DLA terms that are
  neglected in obtaining \eq{eq:qprod_smallx2}.}
\begin{align}
  \label{eq:qprod_smallx}
  \sum_\sigma \sigma \, \frac{d \sigma^{SIDIS}_{T,L} (\sigma)}{d^2
    k_T} = - \int\limits_{z_i}^1 \frac{dz}{z \, (1-z)} & \int
  \frac{d^2 x_\perp \, d^2 y_\perp \, d^2 w_\perp}{2 (2 \pi)^3} \,
  e^{-i \ul{k} \cdot (\ul{x} - \ul{y})} \, \sum_{\sigma, \sigma', f}
  \sigma \, \Psi_{T,L}^{\gamma^* \to q \bar q} (\ul{x} - \ul{w}, z) \,
  \left[ \Psi_{T,L}^{\gamma^* \to q \bar q} (\ul{y} - \ul{w}, z) \right]^* \notag \\
  & \times \bigg\langle \mbox{tr} \left[ V_{\ul{x}} \,
    V^\dagger_{\ul{w}} (\sigma') \right] + \mbox{tr} \left[ V_{\ul{w}}
    (\sigma') \, V^\dagger_{\ul{y}} \right] \bigg\rangle_\Sigma \! \!
  (z),
\end{align}
where $\Psi^{\gamma^* \to q \bar q}$ is the light-cone wave function
for the $\gamma^* \to q \bar q$ splitting given by
\begin{subequations}
\begin{align}
\label{LCwfT}
\Psi_{T}^{\gamma^* \to q \bar q} (\ul{x} , z) = & \frac{e \, Z_f}{2
  \pi} \, \sqrt{z \, (1-z)} \, \left[ \delta_{\sigma, -\sigma'} \, (1
  - 2 z - \sigma \, \lambda ) \, i \, a_f \,
  \frac{\ul{\epsilon}_\lambda \cdot \ul{x}}{x_\perp} \, K_1 (x_\perp
  \, a_f) \right.  \notag \\ & \ \left. + \, \delta_{\sigma \sigma'}
  \, m_f \sqrt{2} \,
  \delta_{\sigma \lambda} \, K_0 (x_\perp \, a_f) \right], \\
\Psi_{L}^{\gamma^* \to q \bar q} (\ul{x} , z) = & \frac{e \, Z_f}{2
  \pi} \, [z \, (1-z)]^{3/2} \, 2 Q \, \delta_{\sigma, - \sigma'} \,
K_0 (x_\perp \, a_f) \label{LCwfL}
\end{align}
\end{subequations}
for transverse ($T$) and longitudinal ($L$) polarizations of the
virtual photon respectively. (Our normalization of light-cone wave
functions corresponds to that in \cite{KovchegovLevin}.) The notation
in the above formulas is largely explained in
\fig{fig:helicity_TMD}. Here $v^\pm \equiv (v^0 \pm v^3)/\sqrt{2}$ for
any four-vector $v^\mu = (v^+, v^-, \ul{v})$, where $\ul{v}$ is a
two-vector in the transverse plane with $v_\perp = v_T =
|\ul{v}|$. Note that $z = (q^+ - k^+)/k^+$ is the fraction of the
virtual photon's ($\gamma^*$) light-cone momentum carried by the
anti-quark: it is integrated over the range $z_i < z < 1$ with its
smallest value $z_i = \Lambda^2/s$, where $\Lambda$ is the infrared
(IR) cutoff and $s$ is the center-of-mass energy squared of the
$\gamma^* + p$ system. Above $\sigma, \sigma', \Sigma, \lambda$ are
the quark, anti-quark, target, and virtual photon polarizations
respectively, and $a_f^2 = z (1-z) \, Q^2 + m_f^2$ with $m_f$ the mass
of quarks of flavor $f$.

The interaction with the target is described by infinite light-cone
Wilson lines in the fundamental representation
\begin{align}
  \label{eq:fund_Wilson}
  V_{\ul{x}} = \mbox{P} \exp \left[ i \, g \,
    \int\limits_{-\infty}^\infty d x^+ \, A^- (x^+, x^- =0, \ul {x})
  \right].
\end{align}
The trace in \eq{eq:qprod_smallx} is over matrices in the fundamental
representation of SU($N_c$). Note that the anti-quark line interaction
with the shock wave is denoted by $V^\dagger_{\ul{w}} (\sigma')$: this
interaction is not completely eikonal, since we need to include a
non-eikonal interaction exchanging spin information with the target
(the interaction shown explicitly in
\fig{fig:helicity_TMD}). Therefore, $V^\dagger_{\ul{w}} (\sigma')$ is
not entirely a Wilson line of \eqref{eq:fund_Wilson}, but is a more
complicated operator including the spin information in it. The exact
formal definition of $V^\dagger_{\ul{w}} (\sigma')$ is not important
as long as we understand that it stands for an eikonal quark which
eventually undergoes a non-eikonal spin-dependent interaction that may
turn it into an eikonal gluon. The target averaging of the Wilson
line-like operators is denoted by $\langle \ldots \rangle_\Sigma$: in
this case it depends on the target helicity $\Sigma$. The argument $z$
of the angle brackets in \eq{eq:qprod_smallx} labels the longitudinal
momentum fraction carried by the polarized line, which in this case is
the anti-quark line.  (As discussed later, this simple picture is
modified by the effects of small-$x$ evolution, such that $z$ can
refer to softer partons in the dipole wave function).

Throughout this paper we are interested in extracting the leading-$\ln
s$ contributions to the observables we calculate, concentrating on the
DLA piece. In the unpolarized evolution case the quark loop depicted
in \fig{fig:helicity_TMD} does not give a logarithm of energy $s$. In
the polarized SIDIS case at hand it is possible to obtain one
logarithm of $s$ from the quark loop. To do so one first has to notice
that for $z \ll 1$ and $x_\perp^2 \ll 1/(z \, Q^2)$ the wave function
\eqref{LCwfT} simplifies to give (we put $m_f =0$ for simplicity)
\begin{align}
  \label{eq:qprod_smallx1}
  \sum_\sigma \sigma \, \frac{d \sigma^{SIDIS}_{T} (\sigma)}{d^2 k_T}
  \approx - & \int\limits_{z_i}^1 dz \int \frac{d^2 x_\perp \, d^2
    y_\perp \, d^2 w_\perp}{2 (2 \pi)^3} \, e^{-i \ul{k} \cdot (\ul{x}
    - \ul{y})} \, \sum_{\sigma', f} (-\sigma') \, \frac{\alpha_{EM} \,
    Z_f^2}{\pi} \, \frac{\ul{x} - \ul{w}}{|\ul{x} - \ul{w}|^2}
  \cdot \frac{\ul{y} - \ul{w}}{|\ul{y} - \ul{w}|^2} \notag \\
  & \times \bigg\langle \mbox{tr} \left[ V_{\ul{x}} \,
    V^\dagger_{\ul{w}} (\sigma') \right] + \mbox{tr} \left[ V_{\ul{w}}
    (\sigma') \, V^\dagger_{\ul{y}} \right] \bigg\rangle_\Sigma \! \!
  (z) \ \ \theta \left( \frac{1}{Q^2} - z \, |\ul{x} - \ul{w}|^2
  \right) \, \theta \left( \frac{1}{Q^2} - z \, |\ul{y} - \ul{w}|^2
  \right).
\end{align}
In arriving at \eq{eq:qprod_smallx1} we concentrated only on
transversely-polarized photons, averaging over the polarization
$\lambda$.  We have also dropped the term containing $(\ul{x} -
\ul{w}) \times (\ul{y} - \ul{w})$, since after integration over
$x_\perp$, $y_\perp$ and $w_\perp$ it should give something
proportional to $\ul{k} \times \ul{k} =0$ because there is no other
transverse vector in the problem apart from $\ul{k}$. The
$\theta$-functions in \eq{eq:qprod_smallx1} impose the approximation
we have employed ($x_\perp^2 \ll 1/(z \, Q^2)$ in \eq{LCwfT}). It is
also understood that $z \ll 1$, even though, in the
leading-logarithmic spirit we let the $z$ integration range go up to
$1$.

It is well-known that the leading power of energy in any perturbative
QCD interaction is given by the eikonal contribution, which is
polarization-independent. Therefore, as we also show explicitly in
Appendix~\ref{A}, the polarization-dependent contribution we are after
must be energy suppressed. It is convenient to show this dependence
explicitly by redefining the averaging in Eqs.~\eqref{eq:qprod_smallx}
and \eqref{eq:qprod_smallx1} using
\begin{align}\label{redef0}
  \left\langle \ldots \right\rangle_\Sigma (z) = \frac{1}{z \, s} \,
  \left\langle \! \left\langle \ldots \right\rangle \!
  \right\rangle_\Sigma (z),
\end{align}
where $z \, s$ is the center-of-mass energy squared for the
anti-quark--proton (or quark--proton) system.  (This identity can be
understood as the definition of double angle brackets.  Note that
after the effects of evolution, the factor of $1/(z s)$ always comes
in with the momentum fraction $z$ of the polarized line in the
dipole.) Employing the substitution \eqref{redef0} in
\eqref{eq:qprod_smallx1} yields
\begin{align}
  \label{eq:qprod_smallx2}
  \sum_\sigma \sigma \, \frac{d \sigma^{SIDIS}_{T} (\sigma)}{d^2 k_T}
  = - \frac{1}{s} & \, \int\limits_{z_i}^1 \frac{dz}{z} \int \frac{d^2
    x_\perp \, d^2 y_\perp \, d^2 w_\perp}{2 (2 \pi)^3} \, e^{-i
    \ul{k} \cdot (\ul{x} - \ul{y})} \, \sum_{\sigma', f} (-\sigma') \,
  \frac{\alpha_{EM} \, Z_f^2}{\pi} \, \frac{\ul{x} - \ul{w}}{|\ul{x} -
    \ul{w}|^2}
  \cdot \frac{\ul{y} - \ul{w}}{|\ul{y} - \ul{w}|^2} \notag \\
  & \times \left\langle \!\! \bigg\langle \mbox{tr} \left[ V_{\ul{x}}
      \, V^\dagger_{\ul{w}} (\sigma') \right] + \mbox{tr} \left[
      V_{\ul{w}} (\sigma') \, V^\dagger_{\ul{y}} \right] \bigg\rangle
    \!\!  \right\rangle_\Sigma \! \! (z) \ \ \theta \left(
    \frac{1}{Q^2} - z \, |\ul{x} - \ul{w}|^2 \right) \, \theta \left(
    \frac{1}{Q^2} - z \, |\ul{y} - \ul{w}|^2 \right).
\end{align}
Now we see explicitly that the $z$-integral is indeed logarithmic:
remembering that $z_i = \Lambda^2/s$ we conclude that it gives a
logarithm of energy. This is a peculiar feature of helicity-dependent
amplitudes: unlike the unpolarized case, quark loops here also
generate leading logarithms of energy
\cite{Kirschner:1983di,Kirschner:1994rq,Kirschner:1994vc,Griffiths:1999dj,Bartels:1995iu,Bartels:1996wc}.

\begin{figure}[htb]
\centering
\includegraphics[width= 0.4 \textwidth]{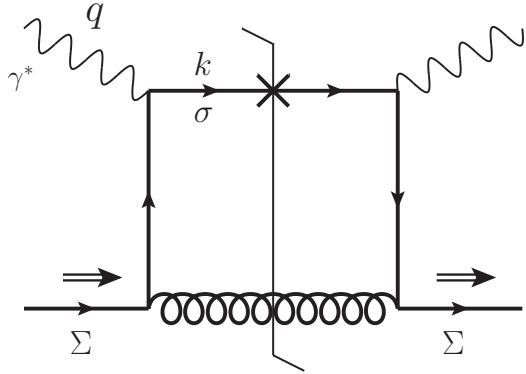}
\caption{Lowest-order diagram contributing to the polarized SIDIS
  cross section on a single quark.}
\label{SIDIS_LO}
\end{figure}

To extract the quark helicity TMD $g_{1L} (x, k_T)$ we consider the
lowest-order (LO) SIDIS process on a quark instead of the proton, as
pictured in \fig{SIDIS_LO}. A simple calculation yields
(cf. \cite{Kovchegov:2015zha} for the unpolarized quark TMD)
\begin{align}
\label{eq:SIDIS_LO}
\sum_\sigma \sigma \, \frac{d \sigma^{SIDIS}_{T} (\sigma)}{d^2 k_T}
\Bigg|_{LO} = \frac{4 \pi^2 \, \alpha_{EM}}{s} \, \sum_f Z_f^2 \,
g_{1L}^{quark} (x, k_T) \, \Sigma,
\end{align}
where $\Sigma$ is the polarization of the target quark and
\begin{align}
  \label{eq:quark_TMD}
  g_{1L}^{quark} (x, k_T) \bigg|_{x \ll 1, \, m=0} = \frac{\as \,
    C_F}{2 \, \pi^2} \, \frac{1}{k_T^2}
\end{align}
is the small-$x$ lowest-order quark helicity TMD of a target quark as
calculated in \cite{Meissner:2007rx}.

Comparing Eqs.~\eqref{eq:SIDIS_LO} and \eqref{eq:qprod_smallx2} we
read off the helicity TMD of a proton (or nuclear) longitudinally
polarized target in the leading logarithmic approximation in $s$:
\begin{align}
  \label{eq:g1L0}
  g_{1L} (x, k_T) = & \, - \frac{\Sigma}{4 \pi^3} \,
  \int\limits_{z_i}^1 \frac{dz}{z} \int \frac{d^2 x_\perp \, d^2
    y_\perp \, d^2 w_\perp}{2 (2 \pi)^3} \, e^{-i \ul{k} \cdot (\ul{x}
    - \ul{y})} \, \frac{\ul{x} - \ul{w}}{|\ul{x} - \ul{w}|^2}
  \cdot \frac{\ul{y} - \ul{w}}{|\ul{y} - \ul{w}|^2} \notag \\
  & \times \sum_{\sigma'} (-\sigma') \, \left\langle \!\! \bigg\langle
    \mbox{tr} \left[ V_{\ul{x}} \, V^\dagger_{\ul{w}} (\sigma')
    \right] + \mbox{tr} \left[ V_{\ul{w}} (\sigma') \,
      V^\dagger_{\ul{y}} \right] \bigg\rangle \!\!
  \right\rangle_\Sigma \! \! (z) \ \ \theta \left( \frac{1}{Q^2} - z
    \, |\ul{x} - \ul{w}|^2 \right) \, \theta \left( \frac{1}{Q^2} - z
    \, |\ul{y} - \ul{w}|^2 \right).
\end{align}

\eq{eq:g1L0} can be simplified further by noticing that outside the
double angle brackets the integrand is invariant under simultaneously
replacing $x \leftrightarrow y$ and $\ul{k} \to - \ul{k}$. At the same
time the integral is a function of $k_T$ instead of $\ul{k}$ because
there is no other transverse vector in the problem: hence the integral
is symmetric under $\ul{k} \to - \ul{k}$ transformation. Therefore we
can simply replace $y \to x$ in the second trace in the double angle
brackets without changing the equality.  We get
\begin{align}
  \label{eq:g1L}
  g_{1L} (x, k_T) = & \, - \frac{\Sigma}{4 \pi^3} \,
  \int\limits_{z_i}^1 \frac{dz}{z} \int \frac{d^2 x_\perp \, d^2
    y_\perp \, d^2 w_\perp}{2 (2 \pi)^3} \, e^{-i \ul{k} \cdot (\ul{x}
    - \ul{y})} \, \frac{\ul{x} - \ul{w}}{|\ul{x} - \ul{w}|^2}
  \cdot \frac{\ul{y} - \ul{w}}{|\ul{y} - \ul{w}|^2} \notag \\
  & \times \sum_{\sigma'} (-\sigma') \, \left\langle \!\! \bigg\langle
    \mbox{tr} \left[ V_{\ul{x}} \, V^\dagger_{\ul{w}} (\sigma')
    \right] + \mbox{tr} \left[ V_{\ul{w}} (\sigma') \,
      V^\dagger_{\ul{x}} \right] \bigg\rangle \!\!
  \right\rangle_\Sigma \! \! (z) \ \ \theta \left( \frac{1}{Q^2} - z
    \, |\ul{x} - \ul{w}|^2 \right) \, \theta \left( \frac{1}{Q^2} - z
    \, |\ul{y} - \ul{w}|^2 \right).
\end{align}

So far we have only calculated the energy dependence coming from a
single quark loop. The energy dependence of the longitudinally
polarized SIDIS cross section \eqref{eq:qprod_smallx} and the quark
helicity TMD \eqref{eq:g1L} mainly comes from the correlator of Wilson
lines.  We conclude that to find either of these quantities in the
small-$x$ regime we need to evolve the {\sl polarized dipole} operator
\begin{align}
  \label{eq:correlators_sum}
  \left\langle \!\! \bigg\langle \mbox{tr} \left[ V_{\ul{x}} \, V^\dagger_{\ul{y}}
      (\sigma) \right] + \mbox{tr} \left[ V_{\ul{y}} (\sigma) \, V^\dagger_{\ul{x}}
    \right] \bigg\rangle \!\!  \right\rangle_\Sigma \!  \!  (z)
\end{align}
in $z$ (the momentum fraction of the polarized line).

Other observables would benefit from constructing the small-$x$
evolution of the polarized dipole. Knowing the helicity TMD
\eqref{eq:g1L} would allow one to construct the quark hPDF,
\begin{align}
  \label{eq:Deltaq}
  \Delta q (x, Q^2) = \int d^2 k_T \, g_{1L} (x, k_T) = -
  \frac{\Sigma}{4 \pi^3} \, & \, \int\limits_{z_i}^1 \frac{dz}{z}
  \int\limits_{\rho^2} \frac{d^2 x_\perp \, d^2 w_\perp}{4 \pi} \,
  \frac{1}{|\ul{x} - \ul{w}|^2} \ \theta \left( \frac{1}{Q^2} - z
    \, |\ul{x} - \ul{w}|^2 \right) \notag \\
  & \times \sum_{\sigma'} (-\sigma') \, \left\langle \!\! \bigg\langle
    \mbox{tr} \left[ V_{\ul{x}} \, V^\dagger_{\ul{w}} (\sigma')
    \right] + \mbox{tr} \left[ V_{\ul{w}} (\sigma') \,
      V^\dagger_{\ul{x}} \right] \bigg\rangle \!\!
  \right\rangle_\Sigma \! \! (z) \ .
\end{align}
The integrals over $x_\perp$ and $w_\perp$ in \eq{eq:Deltaq} now have
a singularity at $\ul{x} = \ul{w}$: it is regulated by requiring that
$|\ul{x} - \ul{w}| > \rho$, where $\rho^2 = 1/(z \, s)$ is the
shortest allowed distance (squared) in the problem. We note in passing
that the transverse space integrals in \eq{eq:Deltaq} now bring in
another logarithm of energy, in part due to the ultra-violet (UV)
cutoff $\rho$, and in part due to the $z$-dependent IR cutoff
resulting from the $\theta$-function. Combining this with the energy
logarithm coming from the $z$ integration, we observe that the quark
loop in \fig{fig:helicity_TMD} leads to two logarithms of energy. This
is our first example of how the DLA appears in a calculation.

Knowing $\Delta q (x, Q^2)$ one can calculate the (leading-twist)
structure function $g_1 (x, Q^2)$ from \eq{eq:g1} above along with
$\Delta \Sigma$ and its contribution to $S_q$ in the spin sum rule
\eqref{eq:sum_rule}. Note that Eqs.~\eqref{eq:qprod_smallx},
\eqref{eq:SIDIS_LO} and \eqref{eq:g1} allow one to write down an
all-twist expression for $g_1 (x, Q^2)$, without extracting the
leading-energy (DLA) contribution from the quark loop:
\begin{align}
  \label{eq:g1_all_twists}
  g_1 (x, Q^2) & \, = \frac{s \, \Sigma}{2 \pi^2 \, \alpha_{EM}} \int
  d^2 k_\perp \, \sum_\sigma \sigma \, \left[ \frac{d
      \sigma^{SIDIS}_{T} (\sigma)}{d^2 k_T} + \frac{d
      \sigma^{SIDIS}_{L} (\sigma)}{d^2 k_T} \right] \notag \\ & = -
  \frac{\Sigma}{2 \pi^2 \, \alpha_{EM}} \int\limits_{z_i}^1
  \frac{dz}{z^2 \, (1-z)} \int \frac{d^2 x_\perp \, d^2 w_\perp}{4
    \pi} \, \sum_{\sigma, \sigma', f} \sigma \, \left[ \frac{1}{2}
    \sum_\lambda \left| \Psi_{T}^{\gamma^* \to q \bar q} (\ul{x} -
      \ul{w}, z) \right|^2 +
    \left| \Psi_{L}^{\gamma^* \to q \bar q} (\ul{x} - \ul{w}, z) \right|^2 \right] \notag \\
  & \times \left\langle \!\! \bigg\langle \mbox{tr} \left[ V_{\ul{x}}
      \, V^\dagger_{\ul{w}} (\sigma') \right] + \mbox{tr} \left[
      V_{\ul{w}} (\sigma') \, V^\dagger_{\ul{x}} \right] \bigg\rangle
    \!\!  \right\rangle_\Sigma \! \! (z).
\end{align}
We see that the $x$-dependence of $g_1 (x, Q^2)$ (either from
\eq{eq:g1} or its more precise version in \eq{eq:g1_all_twists}) and
$\Delta \Sigma (x, Q^2)$ is also determined by the energy dependence
of the correlator \eqref{eq:correlators_sum}. Hence the small-$x$
evolution of the operator \eqref{eq:correlators_sum} would allow us to
address many important observables, including the ones relevant for
the spin puzzle.


\section{The Ingredients}
\label{sec-ingr}

We want to derive $s$-channel helicity evolution equations, similar to
the BK/JIMWLK equations
\cite{Balitsky:1996ub,Balitsky:1998ya,Kovchegov:1999yj,Kovchegov:1999ua,Jalilian-Marian:1997dw,Jalilian-Marian:1997gr,Iancu:2001ad,Iancu:2000hn},
but now for the polarization-dependent dipole operator in
\eq{eq:correlators_sum}.  Following the standard technique we will
continue working in the light-cone gauge of the dipole, that is, as
our projectile is moving along the light-cone $x^+$ direction, we will
employ the $A^+ =0$ light-cone gauge.

Before we can even start discussing the diagrams relevant for the
helicity evolution, we observe that small-$x$ evolution in the $A^+
=0$ gauge takes place in the light-cone wave function
\cite{Lepage:1980fj,Brodsky:1997de} of the dipole projectile.  Hence,
to write down $s$-channel helicity evolution equations we will need
the basic helicity-dependent light-cone wave functions describing the
$q \to qG$, $G \to q\bar q$ and $G \to GG$ transitions, which are the
building blocks of any evolution equation. They can be obtained from
\cite{KovchegovLevin}, or they can be derived independently.  For
simplicity we assume that all the quarks are massless, $m=0$, since,
as one can show using the analysis below, mass corrections do not
contribute to the double-logarithmic (DLA) small-$x$ evolution.


\subsection{$q \to qG$ Splitting}

\begin{figure}[hbt]
 \centering
 \includegraphics[width=0.5\textwidth]{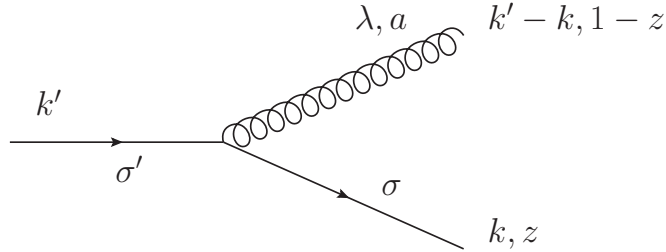}
 \caption{Leading-order diagram contributing to the light-cone wave
   function for $q \to qG$.}
\label{q_to_qG}
\end{figure}

Consider the $q \to qG$ splitting depicted in \fig{q_to_qG}. The exact
leading-order in $\as$ momentum-space light-cone wave function is
\begin{align}\label{eq:qtoqG}
  \psi^{q \to qG} = - g \, t^a \, \delta_{\sigma \sigma'} \, \sqrt{z}
  \ \frac{\ul{\epsilon}_\lambda^* \cdot (\ul {k} - z \ul{k}')}{|\ul
    {k} - z \ul{k}'|^2} \, \left[ 1+ z + \sigma \, \lambda \, (1-z)
  \right].
\end{align}
Here $0 \le z \le 1$, where $z = k^+/k'^+$, $\sigma$ and $\sigma'$ are
the quark polarizations, $\lambda$ and $a$ are the gluon polarization
and color, and $t^a$ are the fundamental SU($N_c$) generators. All the
labels are explained in \fig{q_to_qG}. We are using light-front
perturbation theory (LFPT) rules \cite{Lepage:1980fj,Brodsky:1997de}
with all the quark and gluon polarizations quantized along the same
$z$-axis (beam axis) direction.  The gluon polarization four-vector 
in the $A^+ =0$ light-cone gauge is
\begin{align}
  \epsilon^\mu_\lambda (k) = \left( 0, \frac{\ul{\epsilon}_\lambda
      \cdot \ul {k}}{k^+}, \ul{\epsilon}_\lambda \right).
\end{align} 

Since we are interested in small-$x$ evolution, we will only need the
cases when either the outgoing quark or the gluon are very soft, i.e.,
carry a small fraction of the ``$+$'' momentum of the incoming
quark. The relevant limits are
\begin{align}\label{softq1}
  \psi^{q \to qG} \big|_{z \to 0} \approx - g \, t^a \,
  \delta_{\sigma \sigma'} \, \sqrt{z} \ \frac{\ul{\epsilon}_\lambda^*
    \cdot \ul {k}}{\ul{k}^2} \, \left[ 1+ \sigma \, \lambda \right]
\end{align}
(soft outgoing quark) and
\begin{align}\label{softG1}
  \psi^{q \to qG}\big|_{z \to 1} \approx g \, t^a \, \delta_{\sigma
    \sigma'} \ \frac{\ul{\epsilon}_\lambda^* \cdot (\ul {k}' -
    \ul{k})}{|\ul {k}' - \ul{k}|^2} \, \left[ 2 + \sigma \, \lambda \,
    (1-z) + \ldots \right]
\end{align}
(soft gluon). In the latter we keep only the spin-dependent part of
the sub-eikonal corrections.

Fourier-transforming Eqs.~\eqref{softq1} and \eqref{softG1} into
transverse coordinate space we get
\begin{align}\label{softq1_coord}
  \psi^{q \to qG} \big|_{z \to 0} \approx - \frac{ i \, g \, t^a}{2
    \pi} \, \delta_{\sigma \sigma'} \, \sqrt{z} \
  \frac{\ul{\epsilon}_\lambda^* \cdot \ul {x}}{\ul{x}^2} \, \left[ 1+
    \sigma \, \lambda \right]
\end{align}
and
\begin{align}\label{softG1_coord}
  \psi^{q \to qG}\big|_{z \to 1} \approx \frac{ i \, g \, t^a}{2 \pi}
  \, \delta_{\sigma \sigma'} \ \frac{\ul{\epsilon}_\lambda^* \cdot
    \ul{x}}{\ul{x}^2} \, \left[ 2 + \sigma \, \lambda \, (1-z) +
    \ldots \right],
\end{align}
where $\ul{x}$ is the transverse vector separating the soft quark from
the position of the incoming quark in \eq{softq1_coord}, while in
\eq{softG1_coord} it denotes the transverse separation between the
soft gluon and the incoming quark.  Note that Eqs.~\eqref{eq:qtoqG},
\eqref{softq1}, \eqref{softG1}, \eqref{softq1_coord} and
\eqref{softG1_coord} are also valid for the anti-quark splitting,
${\bar q} \to {\bar q} G$.


\subsection{$G \to q\bar q$ Splitting}

\begin{figure}[hbt]
 \centering
 \includegraphics[width=0.5\textwidth]{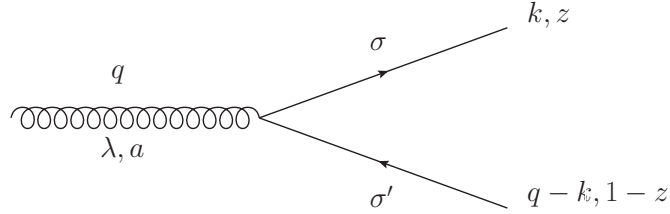}
 \caption{Leading-order diagram contributing to the light-cone wave
   function for $G \to q\bar q$.}
\label{G_to_qqbar}
\end{figure}

Now we consider the $G \to q\bar q$ splitting. The exact leading-$\as$
light-cone wave function is
\begin{align}
  \psi^{G \to q\bar q} = g \, t^a \, \sqrt{z \, (1-z)} \
  \frac{\ul{\epsilon}_\lambda \cdot (\ul {k} - z \ul{q})}{|\ul {k} -
    z \ul{q}|^2} \, \delta_{\sigma, - \sigma'} \, \left[ 1 - 2 \, z -
    \sigma \, \lambda \right].
\end{align}
The soft-quark limit is
\begin{align}\label{softq2}
  \psi^{G \to q\bar q} \big|_{z \to 0} \approx g \, t^a \, \sqrt{z} \
  \frac{\ul{\epsilon}_\lambda \cdot \ul{k}}{|\ul{k}|^2} \,
  \delta_{\sigma, - \sigma'} \, \left[ 1 - \sigma \, \lambda \right] ,
\end{align}
and the soft-antiquark limit is
\begin{align}
  \psi^{G \to q\bar q} \big|_{z \to 1} \approx g \, t^a \, \sqrt{1-z}
  \ \frac{\ul{\epsilon}_\lambda \cdot (\ul{q} - \ul{k})}{|\ul{q} -
    \ul{k}|^2} \, \delta_{\sigma, - \sigma'} \, \left[ 1 - \sigma' \,
    \lambda \right].
\end{align}
Here again we only keep the polarization-dependent sub-eikonal
corrections.
Fourier-transforming into transverse coordinate space gives
\begin{align}\label{softq2_coord}
  \psi^{G \to q\bar q} \big|_{z \to 0} \approx \frac{i \, g \, t^a}{2
    \pi} \, \sqrt{z} \ \frac{\ul{\epsilon}_\lambda \cdot
    \ul{x}}{|\ul{x}|^2} \, \delta_{\sigma, - \sigma'} \, \left[ 1 -
    \sigma \, \lambda \right]
\end{align}
and
\begin{align}
  \psi^{G \to q\bar q} \big|_{z \to 1} \approx \frac{i \, g \, t^a}{2
    \pi} \, \sqrt{1-z} \ \frac{\ul{\epsilon}_\lambda \cdot
    \ul{x}}{|\ul{x}|^2} \, \delta_{\sigma, - \sigma'} \, \left[ 1 -
    \sigma' \, \lambda \right],
\end{align}
where $\ul{x}$ is again the transverse vector connecting the position
of the incoming gluon with that of a soft (anti-)quark.


\subsection{$G \to GG$ Splitting}

\begin{figure}[hbt]
 \centering
 \includegraphics[width=0.5\textwidth]{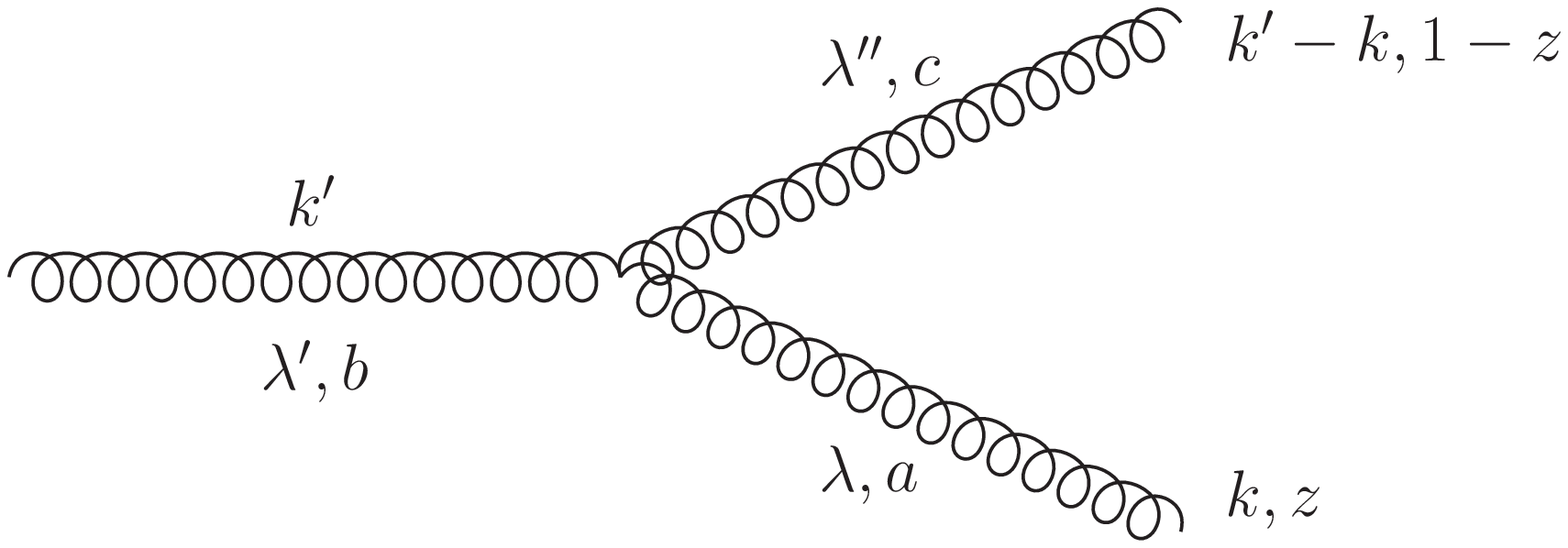}
 \caption{Leading-order diagram contributing to the light-cone wave
   function for $G \to GG$.}
 \label{G_to_GG}
\end{figure}

Finally, let us consider the $G \to GG$ splitting. The exact
leading-order light-cone wave function is
\begin{align}
  \psi^{G \to GG} = 2 \, i \, g \, f^{abc} \, \frac{z \, (1-z)}{|\ul
    {k} - z \ul{k}'|^2} \, \bigg[ \frac{1}{1-z} \,
  \ul{\epsilon}^*_{\lambda''} \cdot (\ul{k} - z \ul{k}') \,
  \ul{\epsilon}^*_{\lambda} \cdot \ul{\epsilon}_{\lambda'} & +
  \frac{1}{z} \, \ul{\epsilon}^*_{\lambda} \cdot (\ul{k} - z \ul{k}')
  \, \ul{\epsilon}^*_{\lambda''} \cdot \ul{\epsilon}_{\lambda'} \notag
  \\ & - \ul{\epsilon}_{\lambda'} \cdot (\ul{k} - z \ul{k}') \,
  \ul{\epsilon}^*_{\lambda''} \cdot \ul{\epsilon}^*_{\lambda} \bigg].
\end{align}

The soft-gluon limit is
\begin{align}\label{softG2}
  \psi^{G \to GG} \big|_{z \to 0} \approx 2 \, i \, g \, f^{abc} \,
  \frac{z}{\ul{k}^2} \, \bigg[ & \frac{1}{z} \,
  \ul{\epsilon}^*_{\lambda} \cdot \ul{k} \ \ul{\epsilon}^*_{\lambda''}
  \cdot \ul{\epsilon}_{\lambda'} - \ul{\epsilon}^*_{\lambda} \cdot
  \ul{k}' \, \ul{\epsilon}^*_{\lambda''} \cdot
  \ul{\epsilon}_{\lambda'} \notag \\ & + \,
  \ul{\epsilon}^*_{\lambda''} \cdot \ul{k} \ \ul{\epsilon}^*_{\lambda}
  \cdot \ul{\epsilon}_{\lambda'} - \ul{\epsilon}_{\lambda'} \cdot
  \ul{k} \ \ul{\epsilon}^*_{\lambda''} \cdot \ul{\epsilon}^*_{\lambda}
  + \ldots \bigg] \notag \\ = 2 \, i \, g \, f^{abc} \,
  \frac{z}{\ul{k}^2} \, \bigg[ & \frac{1}{z} \,
  \ul{\epsilon}^*_{\lambda} \cdot \ul{k} \ \delta_{\lambda'' \lambda'}
  - \ul{\epsilon}^*_{\lambda} \cdot \ul{k}' \ \delta_{\lambda''
    \lambda'} \notag \\ & + \, \ul{\epsilon}^*_{\lambda''} \cdot
  \ul{k} \ \delta_{\lambda \lambda'} + \ul{\epsilon}_{\lambda'} \cdot
  \ul{k} \ \delta_{\lambda'', -\lambda} + \ldots \bigg] ,
\end{align}
where again we only keep the sub-eikonal terms which transfer
polarization information to the softer gluon.
Fourier-transforming into transverse coordinate space gives
\begin{align}\label{softG2_coord}
  \psi^{G \to GG} \big|_{z \to 0} \approx \frac{- g \, f^{abc}}{\pi}
  \, \frac{z}{\ul{x}^2} \, \bigg[ & \frac{1}{z} \,
  \ul{\epsilon}^*_{\lambda} \cdot \ul{x} \ \delta_{\lambda'' \lambda'}
  + \ul{\epsilon}^*_{\lambda''} \cdot \ul{x} \ \delta_{\lambda
    \lambda'} + \ul{\epsilon}_{\lambda'} \cdot \ul{x} \
  \delta_{\lambda'', -\lambda} + \ldots \bigg] ,
\end{align}
where the transverse vector $\ul{x}$ connects the position of the
incoming gluon with that of the soft gluon.


\subsection{Evolution Kernels}

\begin{figure}[htb]
\centering
\includegraphics[width= 0.9  \textwidth]{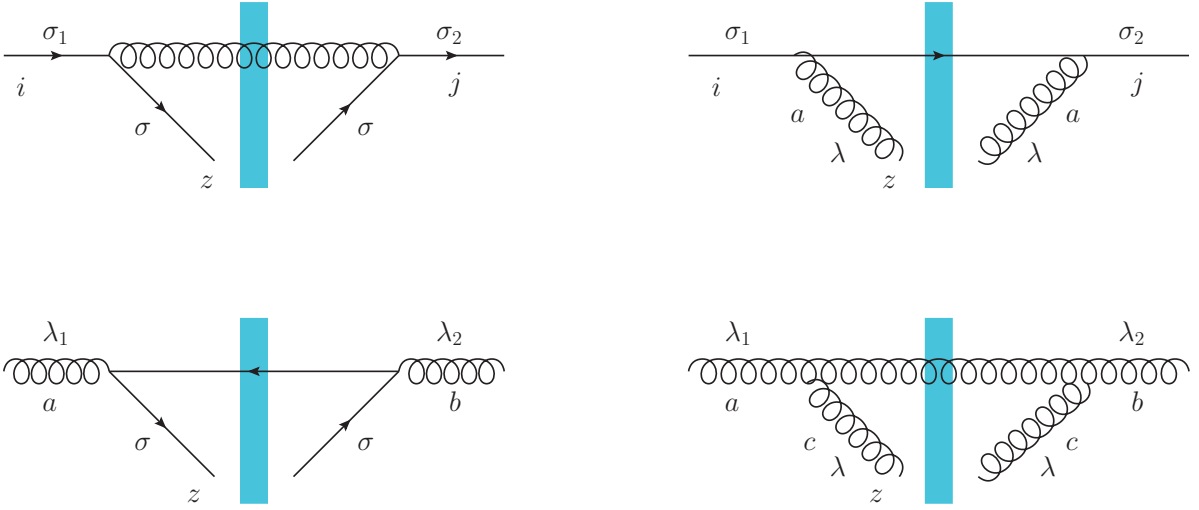}
\caption{Splitting kernels for the ladder part of helicity evolution.}
\label{splittings1}
\end{figure}

For future purposes it is instructive to square the wave functions
obtained above in order to understand the mechanism for generating the
double logarithms that we want to resum and to obtain evolution
kernels for the ladder part of the evolution. Squaring the
coordinate-space wave functions of Sections~\ref{sec-ingr} A, B and C
we can construct the splitting kernels illustrated in
\fig{splittings1} in a matrix form, by analogy to BER
\cite{Bartels:1996wc}. The shaded rectangles denote the shock wave,
which encodes both the subsequent evolution and the interaction with
the longitudinally polarized target. We assume that the amplitude
these wave functions squared connect to at the bottom is forward in
color and flavor (and of course in transverse positions); for a
longitudinally-polarized target, it is also automatically forward in
helicity. This assumption can be explicitly checked by the
lowest-order diagram calculations (akin to that done above for the
graph in \fig{SIDIS_LO}), which provide the initial conditions for
our evolution (see Appendix~\ref{A}). Since those amplitudes are
independent of color and flavor, we sum over the colors of soft quarks
and gluons, and also over flavors of the quarks in the lower left
panel of \fig{splittings1}. We also sum over polarizations and colors
of the hard quarks and gluons going through the shock wave, since in
the DLA the interactions of those particles with the shock wave can be
neglected. (Those interactions are at most single-logarithmic.)

The resulting kernels are (going left to right in each row of
\fig{splittings1} and keeping only the polarization-dependent
sub-eikonal part of the splitting kernel)
\begin{subequations}\label{kernels}
\begin{align}
  K^{q {\bar q} \to q {\bar q}} & = \frac{\as \, C_F}{2 \, \pi^2} \,
  \delta_{\sigma_1 \sigma} \, \delta_{\sigma_2 \sigma} \, \delta_{ij}
  \, \int dz \, \int \frac{d^2 x_\perp}{x_\perp^2}, \\
  K^{q {\bar q} \to GG} & = \frac{\as \, C_F}{2 \, \pi^2} \,
  \delta_{\sigma_1 \sigma_2} \, \sigma_1 \, \lambda \, \, \delta_{ij}
  \, \int dz \, \int \frac{d^2 x_\perp}{x_\perp^2}, \\
  K^{GG \to q {\bar q}} & = \frac{\as \, N_f}{4 \, \pi^2} \,
  \delta_{\lambda_1 \lambda_2} \, \delta_{\sigma, -\lambda_1} \, \,
  \delta^{ab} \, \int dz \, \int \frac{d^2 x_\perp}{x_\perp^2}, \\
  K^{GG \to GG} & = \frac{\as \, N_c}{\pi^2} \, \delta_{\lambda_1
    \lambda_2} \, \lambda \, \lambda_1 \, \, \delta^{ab} \, \int dz \,
  \int \frac{d^2 x_\perp}{x_\perp^2}.
\end{align}
\end{subequations}

At first glance the kernels in \eq{kernels} appear completely
irrelevant to the task at hand: the integrals over longitudinal
momentum fractions $z$ in those kernels are not logarithmic. Moreover,
the only integral in the kernels of \eq{kernels} that may generate a
logarithm is the integral over $x_\perp$. However, in the unpolarized
BFKL evolution
\cite{Mueller:1994rr,Mueller:1994jq,Mueller:1995gb,Balitsky:1996ub,Balitsky:1998ya,Kovchegov:1999yj,Kovchegov:1999ua,Jalilian-Marian:1997dw,Jalilian-Marian:1997gr,Iancu:2001ad,Iancu:2000hn}
(in coordinate space) such integrals are usually cut off by some
transverse momentum scale, e.g., by $Q^2$ in DIS, and do not become
logarithms of energy. Hence our usual unpolarized small-$x$ (gluon)
evolution intuition appears to tell us that the emission kernels in
\fig{splittings1} and \eq{kernels} cannot generate the DLA powers of
$\as \, \ln^2 s \sim \as \, \ln^2 (1/x)$.

Indeed the above concerns, while legitimate, are incorrect. An example
of the DLA evolution in the $s$-channel transverse coordinate space
formalism at hand is given in \cite{Itakura:2003jp} for the case of
the QCD Reggeon. Just like in \cite{Itakura:2003jp} we show in
Appendix~\ref{A} that the initial conditions for the
helicity-dependent evolution are given by an energy suppressed
cross-section ${\hat \sigma} \sim 1/(z \, s)$, where $s$ is the
center-of-mass energy squared of the system and $z$ is the
longitudinal momentum fraction of the softest parton in the
cascade. Hence all the kernels in \eq{kernels} would act on $1/z$ in
the initial condition or in the subsequent evolution. This would make
the $z$-integrals logarithmic, $\int dz/z$, yielding us one power of
$\ln s$ per each splitting. The other $\ln s$ arises from the
$x_\perp$-integral in \eq{kernels}. As will become apparent later, and
by analogy to what was already observed after \eq{eq:Deltaq} and in
\cite{Itakura:2003jp}, in the case of our helicity evolution the
$x_\perp$-integrals will be divergent in the ultra-violet (UV) and
will be regulated by the inverse center-of-mass energy of the
system. In addition, the infra-red (IR) cutoff on the
$x_\perp$-integral will be $z$-dependent, and would also generate a
logarithm of energy. We see that each integral in \eq{kernels}
contributes a logarithm of $s$, yielding $\as \, \ln^2 s$ per each
splitting, and hence contributing to the DLA approximation we are
constructing.


\section{DLA Diagrams: Ladders and non-ladders}
\label{sec:diag}

\begin{figure}[thb]
\centering
\includegraphics[width= 0.9 \textwidth]{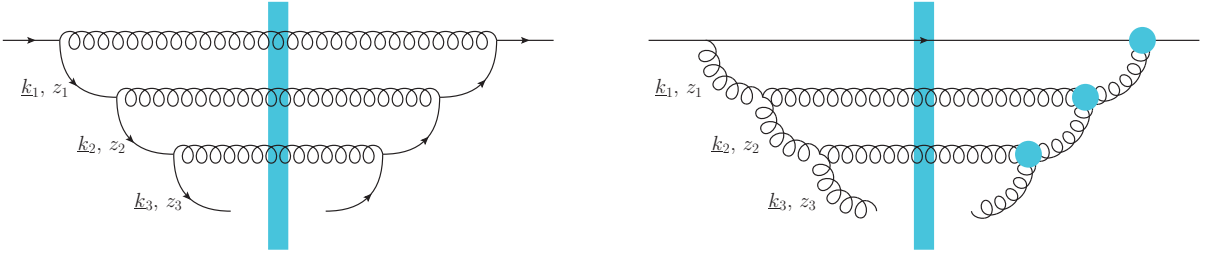}
\caption{Examples of the quark and gluon ladder diagrams. Non-eikonal
  gluon emission vertices are denoted by shaded circles.}
\label{ladders}
\end{figure}

Our goal is to construct the small-$x$ evolution of the polarized
dipole operator in \eq{eq:correlators_sum}. In this Section we will
explore the contributing diagrams using the evolution building blocks
developed in Sec.~\ref{sec-ingr}. The most natural guess for the types
of diagrams one has to resum to obtain DLA helicity evolution would be
quark and gluon ladders shown in \fig{ladders}. In
\cite{Itakura:2003jp} it was the quark ladder (i.e., a ladder with
quarks in the $t$-channel and with gluon rungs) that gave the
small-$x$ evolution of the QCD Reggeon.

Let us first discuss how the ladders give double logarithms. We begin
with the quark ladder, depicted also in the left panel of
\fig{ladders}. The longitudinal momentum fractions carried by the
quarks (and the gluons going through the shock wave) are ordered, $1
\gg z_1 \gg z_2 \gg z_3 \gg \ldots$. As one can see from \eq{softq1}
(and, in a general case, also from \eq{softq2}), each emission of a
soft quark generates a factor of $\sqrt{z}$. The quark ladder in
\fig{ladders} then yields
\begin{align}\label{zint}
  \int\limits^1_{z_i} \frac{d z_1}{z_1} \int\limits_{z_i}^{z_1}
  \frac{d z_2}{z_2} \int\limits_{z_i}^{z_2} \frac{d z_3}{z_3} \,
  (\sqrt{z_1})^2 \, \left( \sqrt{\frac{z_2}{z_1}} \right)^2 \, \left(
    \sqrt{\frac{z_3}{z_2}} \right)^2 = \int\limits^1_{z_i} \frac{d
    z_1}{z_1} \int\limits_{z_i}^{z_1} \frac{d z_2}{z_2}
  \int\limits_{z_i}^{z_2} \frac{d z_3}{z_3} \, z_3.
\end{align}
The factor of $z_3$ here is canceled by the energy-suppressed initial
conditions, which, just like in \eq{redef0} (or in \eq{eq:SIDIS_LO})
bring in a factor of $1/(z_3 \, s)$. We thus get a logarithmic
contribution from the integral in \eq{zint}. Hence one condition for
the DLA contribution is the longitudinal momentum ordering,
\begin{align}\label{cond1}
1 \gg z_1 \gg z_2 \gg z_3 \gg \ldots \ .
\end{align}

Gluon ladders also give a contribution similar to \eq{zint}. Consider
the ladder in the right panel of \fig{ladders}. To transfer
polarization information down the ladder, one of the quark-gluon
vertices (either to the left or right of the shock wave) has to be
sub-eikonal: according to \eq{softG1}, this means that the eikonal
vertex brings in a factor of $1$ (in term of counting the powers of
$z_1$), while the sub-eikonal vertex gives a power of $z_1$. We denote
sub-eikonal gluon emission vertices by the shaded circle in
\fig{ladders}: note that the circle could be on either side of the
shock wave, and the diagram on the right of \fig{ladders} represents
only one realization of the gluon ladder. The situation repeats itself
for the triple gluon vertices, as follows from \eq{softG2}: in the
emission of an $s$-channel gluon, one of the vertices (either to the
left or right of the shock wave, denoted by the shaded circle too) has
to be sub-eikonal. (We neglect the contribution of both vertices being
sub-eikonal, since this is even more energy-suppressed than helicity
evolution and is beyond our DLA here.)  For the gluon ladder on the
right of \fig{ladders} we get
 \begin{align}\label{zint2}
   \int\limits^1_{z_i} \frac{d z_1}{z_1} \int\limits_{z_i}^{z_1}
   \frac{d z_2}{z_2} \int\limits_{z_i}^{z_2} \frac{d z_3}{z_3} \, z_1
   \, \frac{z_2}{z_1} \, \frac{z_3}{z_2} = \int\limits^1_{z_i} \frac{d
     z_1}{z_1} \int\limits_{z_i}^{z_1} \frac{d z_2}{z_2}
   \int\limits_{z_i}^{z_2} \frac{d z_3}{z_3} \, z_3,
\end{align}
which again is a logarithmic contribution.

To obtain a DLA contribution we also need to get energy logarithms
coming from the integrals over transverse momenta. This is achievable
only if the softest gluon dominates in the energy denominator, just
like in small-$x$ evolution. After a little bit of work one can show
that this translates into the following condition (cf. Eq. (2.31) in
\cite{Bartels:1996wc})
\begin{align}\label{cond2}
  \frac{\ul{k}_1^2}{z_1} \ll \frac{\ul{k}_2^2}{z_2} \ll
  \frac{\ul{k}_3^2}{z_3} \ll \ldots \ .
\end{align} 
In transverse coordinate space this condition becomes
\begin{align}
  \label{eq:cond2_coord}
  z_1 \, \ul{x}_1^2 \gg z_2 \, \ul{x}_2^2 \gg z_3 \, \ul{x}_3^2 \gg
  \ldots \ ,
\end{align}
where the vectors $\ul{x}_n$ are Fourier conjugates of $\ul{k}_n$. The
ordering \eqref{eq:cond2_coord} leads to transverse coordinate
integrals like
\begin{align}\label{tr_int}
  \int\limits^{x_{n-1, \perp}^2 \, z_{n-1} / z_{n}}_{1/(z_n \, s)}
  \frac{d x_{n, \, \perp}^2}{x_{n, \, \perp}^2},
\end{align}
where the UV divergence gets regulated by the inverse of the largest
momentum scale squared associated with the emission of the $n$th
parton, $1/(z_n \, s)$. As discussed above (see discussions below
\eq{eq:Deltaq} and at the end of Sec.~\ref{sec-ingr}D), the transverse
integrals like \eqref{tr_int} in helicity evolution generate
logarithms of energy coming both from their UV and IR cutoffs.

We conclude that for ladder diagrams like those in \fig{ladders},
emission of each $s$-channel gluon generates a power of $\as$ along
with two logarithms of energy: one logarithm comes from the
integration over longitudinal momentum fractions, like that in
\eqref{zint}, while another logarithm of energy comes from the
transverse coordinate (or momentum) integrals \eqref{tr_int}. This
latter feature is the essential difference between the helicity
evolution and the unpolarized evolution: in the latter the transverse
integrals do not generate logarithms of energy and resummation is
single-logarithmic.

Let us point out that in helicity evolution quark and gluon ladders of
\fig{ladders} can mix, as follows from \fig{splittings1}
\cite{Bartels:1996wc}. This is in analogy to the $Q^2$ evolution of
the Dokshitzer--Gribov--Lipatov--Altarelli--Parisi (DGLAP) equation
\cite{Gribov:1972ri,Altarelli:1977zs,Dokshitzer:1977sg}.

\begin{figure}[htb]
\centering
\includegraphics[width= 0.45 \textwidth]{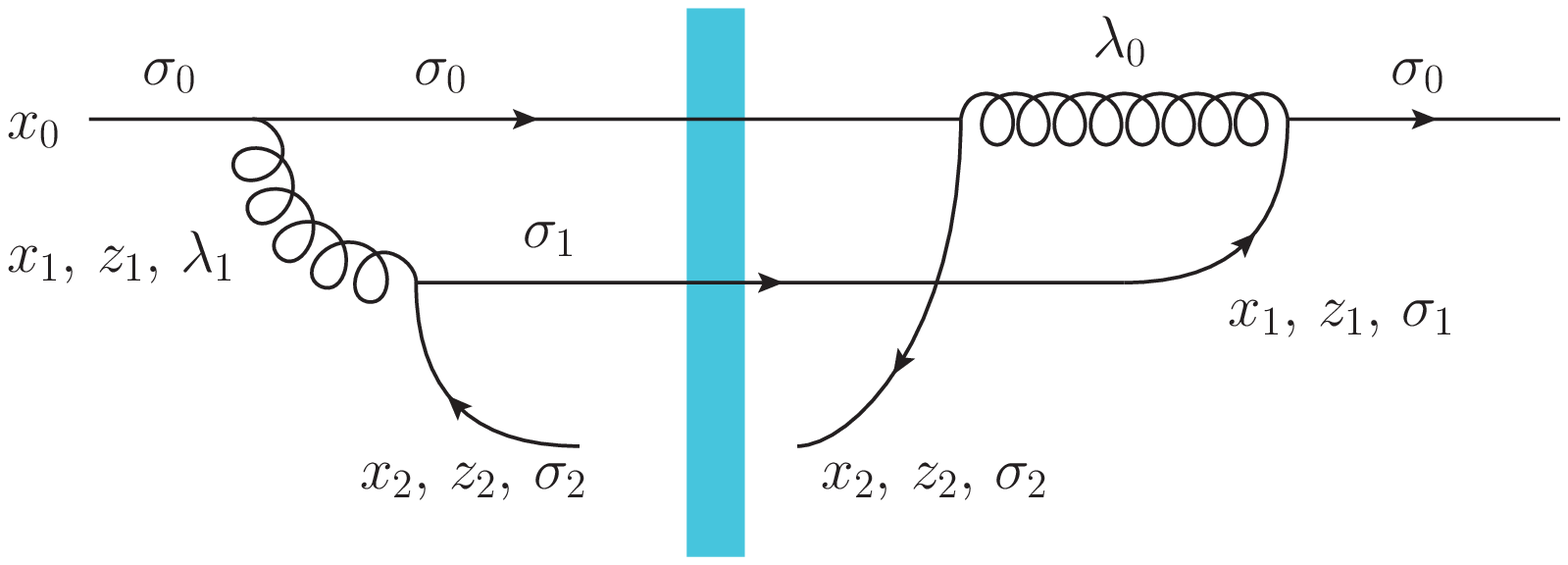}
\ \ \ \includegraphics[width= 0.45
\textwidth]{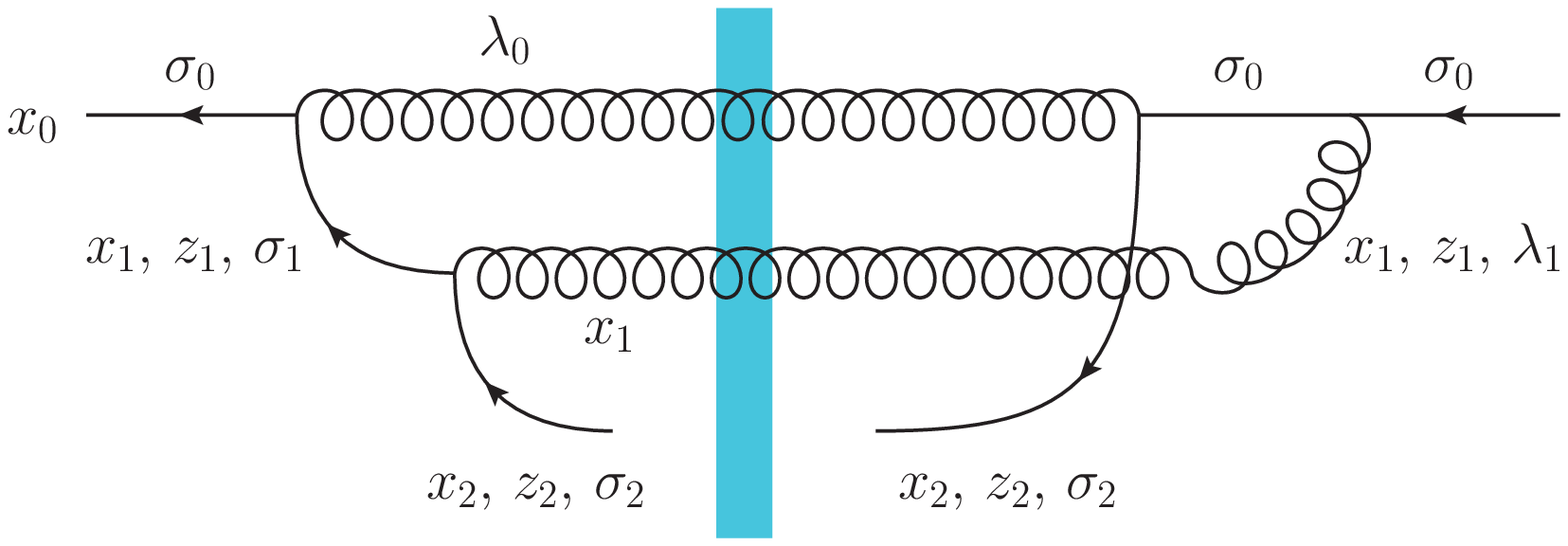}
\caption{Examples of non-ladder diagrams involving quarks.}
\label{non-ladders_quarks}
\end{figure}

However DLA evolution is not limited to ladder diagrams. There exist
non-ladder diagrams which also yield a DLA contribution.  We will
refer to diagrams as ``non-ladder'' if they contain an $s$-channel
parton with longitudinal momentum fraction $z_n$ which was not emitted
by the parton with the longitudinal momentum fraction $z_{n-1}$, but
instead was emitted by the parton with the fraction $z_m$ with $m <
n-1$. (The ordering \eqref{cond1} is implied in this definition.)
Examples of diagrams involving a non-ladder quark emission are shown
in \fig{non-ladders_quarks}. The diagram in the left panel depicts
helicity evolution of a polarized quark Wilson line, while the diagram
in the right panel depicts helicity evolution of a polarized
anti-quark Wilson line. Both graphs lead to a polarized anti-quark
Wilson line at the bottom of the evolution.

To analyze the diagrams in \fig{non-ladders_quarks} we first of all
notice that requiring that $1 \gg z_1 \gg z_2$ works on both sides of
the shock wave giving a ``logarithmic'' contribution
\begin{align}\label{zorder}
  \sim \int\limits^1_{z_i} \frac{d z_1}{z_1} \int\limits_{z_i}^{z_1}
  \frac{d z_2}{z_2} \, z_2
\end{align}
only if we assume that the vertex emitting gluon $z_1$ is eikonal in
both graphs of \fig{non-ladders_quarks}. The transverse position
integrals are more complicated. Noting that all other vertices are
non-eikonal and polarization-dependent, the transverse coordinate and
polarization dependence of the diagram in the left panel of
\fig{non-ladders_quarks} becomes (the factor of $\sigma_2$ comes from
the initial condition, or, equivalently, from the interaction with the
target)
\begin{align}\label{zero_nonladders}
  & - \! \! \sum_{\lambda_0, \lambda_1, \sigma_1, \sigma_2} \int d^2
  x_1 d^2 x_2 \, \frac{\ul{\epsilon}_{\lambda_1} \cdot
    \ul{x}_{21}}{x_{21}^2} \, \frac{\ul{\epsilon}^*_{\lambda_1} \cdot \ul{x}_{10}}{x_{10}^2} \, \frac{\ul{\epsilon}_{\lambda_0} \cdot \ul{x}_{10}}{x_{10}^2} \, \frac{\ul{\epsilon}^*_{\lambda_0} \cdot \ul{x}_{20}}{x_{20}^2} \, \delta_{\sigma_1, - \sigma_2} \, (1- \sigma_2 \, \lambda_1) \, \delta_{\sigma_0  \sigma_1} \, (1+\sigma_0 \, \lambda_0) \, \delta_{\sigma_0, - \sigma_2} \, (1- \sigma_2 \, \lambda_0)  \, \sigma_2 \notag \\
  & = - \sum_{\lambda_0, \lambda_1, \sigma_2} \int d^2 x_1 d^2 x_2 \,
  \frac{\ul{\epsilon}_{\lambda_1} \cdot \ul{x}_{21}}{x_{21}^2} \,
  \frac{\ul{\epsilon}^*_{\lambda_1} \cdot \ul{x}_{10}}{x_{10}^2} \,
  \frac{\ul{\epsilon}_{\lambda_0} \cdot \ul{x}_{10}}{x_{10}^2} \,
  \frac{\ul{\epsilon}^*_{\lambda_0} \cdot \ul{x}_{20}}{x_{20}^2} \,
  \delta_{\sigma_0, - \sigma_2} \, (1 - \sigma_2 \, \lambda_1) \,
  (1+\sigma_0 \, \lambda_0) \, (1- \sigma_2 \, \lambda_0) \, \sigma_2
  \notag \\ &
  = 
  4 \, \sigma_0 \, \sum_{\lambda_0, \lambda_1} \int d^2 x_1 d^2 x_2 \,
  \frac{\ul{\epsilon}_{\lambda_1} \cdot \ul{x}_{21}}{x_{21}^2} \,
  \frac{\ul{\epsilon}^*_{\lambda_1} \cdot \ul{x}_{10}}{x_{10}^2} \,
  \frac{\ul{\epsilon}_{\lambda_0} \cdot \ul{x}_{10}}{x_{10}^2} \,
  \frac{\ul{\epsilon}^*_{\lambda_0} \cdot \ul{x}_{20}}{x_{20}^2} \,
  \delta_{\lambda_0 \lambda_1} \, (1 + \lambda_1 \, \sigma_0) \notag
  \\ & = 2 \, \int d^2 x_1 d^2 x_2 \, \frac{1}{x_{10}^2 \, x_{21}^2 \,
    x_{20}^2} \, \left[ \sigma_0 \, \ul{x}_{21} \cdot \ul{x}_{20} - i
    \, \ul{x}_{21} \times \ul{x}_{20} \right],
\end{align} 
where the cross-product of two transverse vectors is defined as
$\ul{u} \times \ul{v} = u_x \, v_y - u_y \, v_x$, while $\ul{x}_{ij} =
\ul{x}_i - \ul{x}_j$ and $x_{ij} = |\ul{x}_{ij}|$. In arriving at
\eq{zero_nonladders} we have used the ingredients worked out in
Sec.~\ref{sec-ingr}. We note that the $x_2$-integral in the
$\sigma_0$-dependent term is logarithmic only in the IR when $x_{21}
\sim x_{20} \gg x_{10}$. The term with the cross product in
\eq{zero_nonladders} is independent of polarization and could be
neglected.

However there is no need to do that, because the whole contribution of
the diagram in the left panel of \fig{non-ladders_quarks} is canceled
by the diagram in the right panel of the same figure. Indeed a similar
calculation keeping only the transverse coordinate and polarization
dependence of the diagram in the right panel of
\fig{non-ladders_quarks} yields
\begin{align}\label{eq:non-ladder2}
  - & \sum_{\lambda_0, \lambda_1, \sigma_1, \sigma_2} \int d^2 x_1 d^2
  x_2 \, \frac{\ul{\epsilon}^*_{\lambda_0} \cdot
    \ul{x}_{10}}{x_{10}^2} \, \frac{\ul{\epsilon}^*_{\lambda_1} \cdot
    \ul{x}_{21}}{x_{21}^2} \, \frac{\ul{\epsilon}_{\lambda_0} \cdot
    \ul{x}_{20}}{x_{20}^2} \, \frac{\ul{\epsilon}_{\lambda_1} \cdot
    \ul{x}_{10}}{x_{10}^2} \, \delta_{\sigma_0 \sigma_1} \,
  (1+\sigma_0 \, \lambda_0) \, \delta_{\sigma_1 \sigma_2} \,
  (1+\sigma_1 \, \lambda_1) \, \delta_{\sigma_0 \sigma_2} \,
  (1+\sigma_0 \, \lambda_0) \, \sigma_2 \notag \\ & = -
  \sum_{\lambda_0, \lambda_1, \sigma_2} \int d^2 x_1 d^2 x_2 \,
  \frac{\ul{\epsilon}^*_{\lambda_0} \cdot \ul{x}_{10}}{x_{10}^2} \,
  \frac{\ul{\epsilon}^*_{\lambda_1} \cdot \ul{x}_{21}}{x_{21}^2} \,
  \frac{\ul{\epsilon}_{\lambda_0} \cdot \ul{x}_{20}}{x_{20}^2} \,
  \frac{\ul{\epsilon}_{\lambda_1} \cdot \ul{x}_{10}}{x_{10}^2} \,
  \delta_{\sigma_0 \sigma_2} \, (1+\sigma_0 \, \lambda_0)^2 \,
  (1+\sigma_2 \, \lambda_1) \, \sigma_2 \notag \\ & = -
  \sum_{\lambda_0, \lambda_1} \int d^2 x_1 d^2 x_2 \,
  \frac{\ul{\epsilon}^*_{\lambda_0} \cdot \ul{x}_{10}}{x_{10}^2} \,
  \frac{\ul{\epsilon}^*_{\lambda_1} \cdot \ul{x}_{21}}{x_{21}^2} \,
  \frac{\ul{\epsilon}_{\lambda_0} \cdot \ul{x}_{20}}{x_{20}^2} \,
  \frac{\ul{\epsilon}_{\lambda_1} \cdot \ul{x}_{10}}{x_{10}^2} \, 2
  \,(1+\sigma_0 \, \lambda_0) \, (1+\sigma_0 \, \lambda_1) \, \sigma_0
  \notag \\ & = - \sum_{\lambda_0, \lambda_1} \int d^2 x_1 d^2 x_2 \,
  \frac{\ul{\epsilon}^*_{\lambda_0} \cdot \ul{x}_{10}}{x_{10}^2} \,
  \frac{\ul{\epsilon}^*_{\lambda_1} \cdot \ul{x}_{21}}{x_{21}^2} \,
  \frac{\ul{\epsilon}_{\lambda_0} \cdot \ul{x}_{20}}{x_{20}^2} \,
  \frac{\ul{\epsilon}_{\lambda_1} \cdot \ul{x}_{10}}{x_{10}^2} \, 2 \,
  \left[(1+ \lambda_0 \, \lambda_1) \sigma_0 + \lambda_0 +
    \lambda_1\right] \notag \\ & = - 2 \int \frac{d^2 x_1 d^2
    x_2}{x_{10}^2 \, x_{21}^2 \, x_{20}^2} \, \left[ \sigma_0
    \,\ul{x}_{21} \cdot \ul{x}_{20} - i \, \ul{x}_{10} \times
    \ul{x}_{21} \right].
\end{align}
The remaining factors in both diagrams in \fig{non-ladders_quarks},
such as the color factors and $z$-integrals are identical in the
DLA. Hence the diagrams in the two panels of \fig{non-ladders_quarks}
cancel each other when added together, as can be seen by comparing
Eqs.~\eqref{zero_nonladders} and \eqref{eq:non-ladder2}. One may
rightfully worry that the left panel of \fig{non-ladders_quarks} gives
a contribution to the DLA evolution of a quark, while the right panel
contributes to the anti-quark evolution. However, our polarized dipole
operator \eqref{eq:correlators_sum} contains a sum of the polarized
quark and anti-quark line contributions taken at the same transverse
coordinates.  (The remaining anti-quark and quark lines are
unpolarized and cannot be involved in non-ladder quark
emission/absorption.) We conclude that the non-ladder quark diagrams
in \fig{non-ladders_quarks} do not contribute to the DLA evolution of
the polarized dipole operator \eqref{eq:correlators_sum}.

The argument can be generalized to include all non-ladder quark
emissions. For this one can imagine ``dressing'' the diagrams from
\fig{non-ladders_quarks} with higher-order DLA ladder corrections. One
can then argue that such corrections would not affect the
cancellation. For a complete analysis one has to include the
left-right mirror images of the diagrams in \fig{non-ladders_quarks}
and allow for non-ladder gluons too.  Hence non-ladder quarks in
general do not contribute to the DLA evolution of the polarized dipole
operator \eqref{eq:correlators_sum}.

\begin{figure}[htb]
\centering
\includegraphics[width= 0.9 \textwidth]{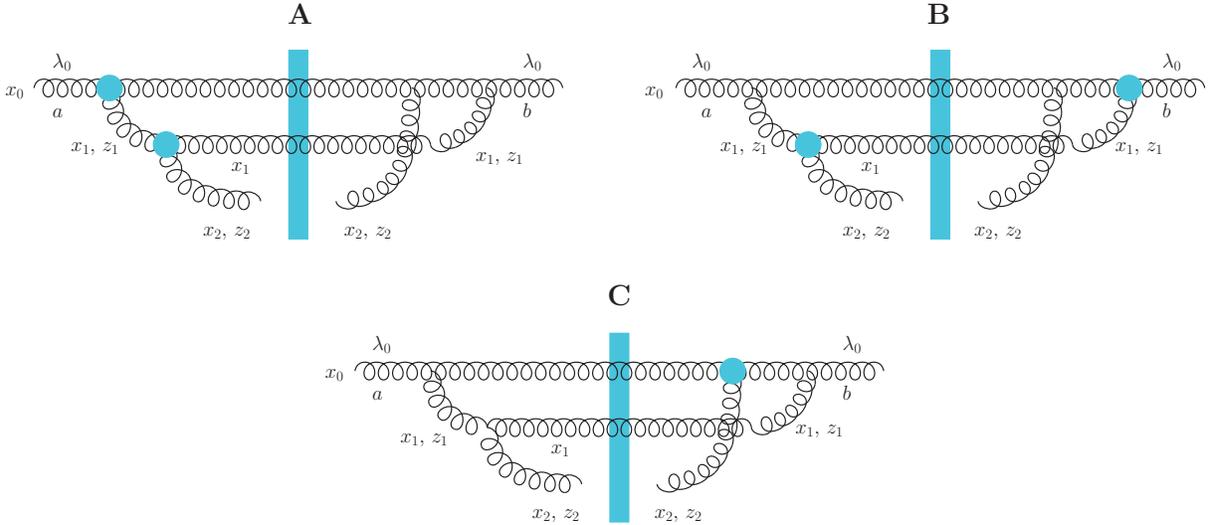}
\caption{Examples of non-ladder gluon diagrams. Non-eikonal gluon
  emission vertices are again denoted by shaded circles.}
\label{non-ladder_gluons}
\end{figure}

The situation with the non-ladder gluon diagrams is not so
straightforward. Examples of the non-ladder gluon graphs are shown in
\fig{non-ladder_gluons}. While the topology of all three diagrams in
\fig{non-ladder_gluons} is the same, the three graphs represent all
possible contributions coming from the non-eikonal spin-transferring
vertices. For $1 \gg z_1 \gg z_2$ the longitudinal integrals in
\fig{non-ladder_gluons} are also given by \eq{zorder} and are
logarithmic, generating two logarithms of energy when convoluted with
$\sim 1/(z_2 \, s)$ initial conditions. The transverse coordinate and
polarization dependence of the diagrams in \fig{non-ladder_gluons} can
be obtained similarly to the above calculation for the non-ladder
quark graphs. We get (after doing some algebra)
\begin{align}
  \label{eq:ABC}
  A = B = C \propto \lambda_0 \, \int \frac{d^2 x_1 d^2 x_2}{x_{10}^2
    \, x_{21}^2 \, x_{20}^2} \, \ul{x}_{21} \cdot \ul{x}_{20} \,
  \theta \! \left( x_{10}^2 \, z_1 - \mbox{max} \left\{ x_{21}^2,
      x_{20}^2 \right\} z_2 \right).
\end{align}
The origin of the $\theta$-function in \eq{eq:ABC} is as
follows. Imposing the ordering \eqref{eq:cond2_coord} onto the
diagrams in \fig{non-ladder_gluons} yields $x_{21}^2 \, z_2 \ll
x_{10}^2 \, z_1$ to the left of the shock wave and $x_{20}^2 \, z_2
\ll x_{10}^2 \, z_1$ to the right of the shock wave. To satisfy both
conditions we require
\begin{align}
x_{10}^2 \, z_1 \gg \mbox{max} \left\{ x_{21}^2, x_{20}^2 \right\}  z_2, 
\end{align}
which is imposed via the $\theta$-function in \eq{eq:ABC}.  The
logarithmic part of the $x_2$-integral in \eq{eq:ABC} comes from the
IR region where $x_{21} \sim x_{20} \gg x_{10}$, similar to
Eqs.~\eqref{zero_nonladders} and \eqref{eq:non-ladder2}. In that
region we get
\begin{align}
  \label{eq:ABC2}
  A = B = C \propto \lambda_0 \, \pi^2 \, \int\limits_{1/(z_1 \, s)}
  \frac{d x^2_{10}}{x_{10}^2} \int\limits^{x_{10}^2 \, z_1
    /z_2}_{x_{10}^2} \frac{d x_{21}^2} {x_{21}^2},
\end{align}
thus generating two logarithms of energy from the transverse
integration as well.

We see that the non-ladder gluon graphs in \fig{non-ladder_gluons} are
DLA diagrams. Moreover, unlike the non-ladder quark graphs in
\fig{non-ladders_quarks}, the non-ladder gluon graphs do not seem to
come with diagrams which cancel them. While some cancellations do
exist for certain classes of non-ladder gluon diagrams, they are far
insufficient to cancel all the non-ladder gluon contributions. We
conclude that to construct helicity evolution equations one has to
include non-ladder gluon diagrams as well.

\begin{figure}[htb]
\centering
\includegraphics[width= 0.9 \textwidth]{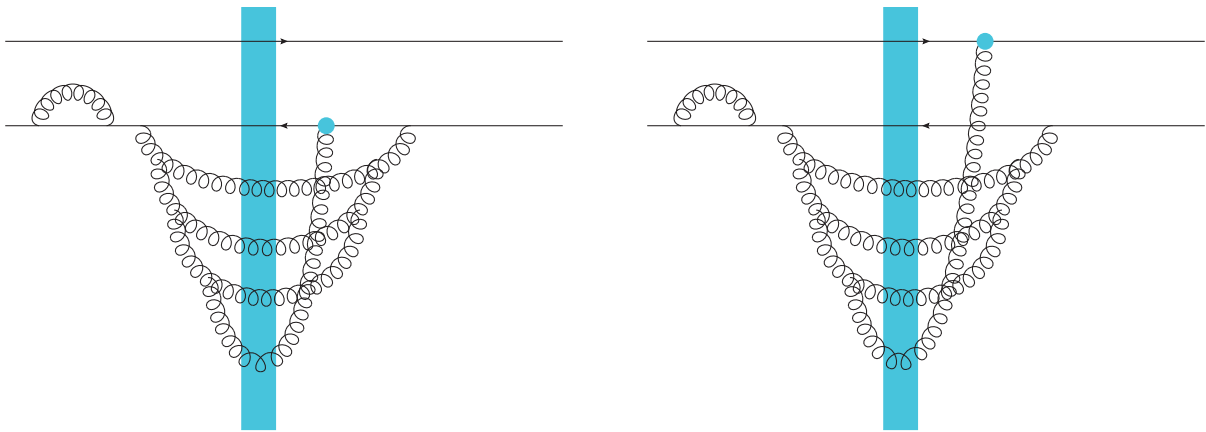}
\caption{Examples of non-ladder gluon diagrams with eikonal ``real''
  and ``virtual'' gluons.}
\label{new_nonladder}
\end{figure}

There is another subtlety here. Diagram C in \fig{non-ladder_gluons},
while DLA-type, contains the $s$-channel gluon with longitudinal
momentum fraction $z_1$: both the emission and absorption vertices for
that gluon are eikonal. The fact that the non-ladder gluon $z_2$
connects in the non-eikonal way to the initial (upper-most) gluon line
allowed gluon $z_1$ to be a ``regular'' eikonal gluon, just like in
the unpolarized LLA evolution. This observation opens up another range
of possibilities: it seems the gluons like $z_1$ can also be
completely virtual. In addition, instead of just one eikonal gluon, we
could emit (and absorb) several of them. Examples of such diagrams are
depicted in \fig{new_nonladder}, where we have both eikonal virtual
corrections and ``real'' gluons (in the dipole model terminology
\cite{Mueller:1994rr,Mueller:1994jq,Mueller:1995gb,Chen:1995pa}). All
such diagrams have to be included in the DLA helicity
evolution.\footnote{One may show that the non-ladder gluon projects
  the eikonal gluon evolution onto a color-octet (adjoint)
  channel. This appears to be similar to the unpolarized octet
  evolution which was employed by BER \cite{Bartels:1996wc}.}

Let us also point out that non-ladder gluons are not always non-planar
gluons, which can simply be eliminated by taking the large-$N_c$
limit. For instance the diagram in the left panel of
\fig{new_nonladder} is subleading in $N_c$, while diagram in the right
panel is leading-order in $N_c$. While taking the large-$N_c$ limit
would eliminate some non-ladder diagrams, it would also leave other
non-ladder diagrams almost unchanged.

To summarize this Section, let us reiterate that for DLA helicity
evolution one has to include non-eikonal soft quark and gluon
emissions. They may lead to a ladder (or mixing of quark and gluon
ladders), but soft gluons may also be emitted in a non-ladder way. In
such case the unpolarized eikonal-gluon evolution may contribute
``real'' and ``virtual'' corrections as well. We are now ready to
write down an equation for helicity evolution.


\section{Evolution Equation for the Polarized Dipole Operator}
\label{sec:operator}

Consider an eikonal quark or gluon, which may interact with the target
shock wave in a polarization-dependent way. The propagation of such a
particle is described by either one of the following operators:
\begin{align}\label{Wlines2}
  V_{\ul{x}} = V^{unp}_{\ul{x}} + \sigma \, V^{pol}_{\ul{x}}, \ \ \
  U_{\ul{x}} = U^{unp}_{\ul{x}} + \lambda \, U^{pol}_{\ul{x}}.
\end{align}
While $V^{unp}_{\ul{x}}$ and $U^{unp}_{\ul{x}}$ are the infinite
light-cone Wilson line operators \eqref{eq:fund_Wilson} (in the
fundamental and adjoint representations respectively),
$V^{pol}_{\ul{x}}$ and $U^{pol}_{\ul{x}}$ are more involved. These
latter objects are Wilson lines only in the sense of eikonal
propagation. They may change representation between fundamental and
adjoint as eikonal quarks become eikonal gluons (by emitting softer
quarks) and vice versa (but only one conversion may take place on each
side of the shock wave, since we only need the leading-order
sub-eikonal correction). Hence we will not write down an explicit
operatorial definition for them. It is likely that one can write down
the operator definitions of $V^{pol}_{\ul{x}}$ and $U^{pol}_{\ul{x}}$
along the lines of the calculation performed in
\cite{Balitsky:2015qba}. For the purposes of constructing the
evolution equations below we only need to know that these are eikonal
quarks or gluons which may emit softer particles in the
polarization-dependent way.

The polarized dipole operator is
\begin{align}
  \label{eq:pol_dip_fund}
  \left\langle \!\! \bigg\langle \mbox{tr} \left[ V_{\ul{x}} \,
      V^\dagger_{\ul{y}} (\sigma) \right] + \mbox{tr} \left[
      V_{\ul{y}} (\sigma) \, V^\dagger_{\ul{x}} \right] \bigg\rangle
    \!\!  \right\rangle_\Sigma \!  \!  (z) = & \, \sigma \,
  \left\langle \!\! \bigg\langle \mbox{tr} \left[ V^{unp}_{\ul{x}} \,
      V^{pol \, \dagger}_{\ul{y}} \right] + \mbox{tr} \left[
      V^{pol}_{\ul{y}} \, V^{unp \, \dagger}_{\ul{x}} \right]
    \bigg\rangle \!\!  \right\rangle \!  (z) \\ = & \, 2 \, \sigma \,
  \left\langle \!\! \bigg\langle \mbox{Re} \left( \mbox{tr} \left[
        V^{unp}_{\ul{x}} \, V^{pol \, \dagger}_{\ul{y}} \right]
    \right) \bigg\rangle \!\!  \right\rangle \!  (z) \notag
\end{align}
in the fundamental representation and
\begin{align}
  \label{eq:pol_dip_adj}
  \left\langle \!\! \bigg\langle \mbox{Tr} \left[ U_{\ul{x}} \,
      U^\dagger_{\ul{y}} (\lambda) \right] + \mbox{Tr} \left[
      U_{\ul{y}} (\lambda) \, U^\dagger_{\ul{x}} \right] \bigg\rangle
    \!\!  \right\rangle_\Sigma \!  \!  (z) = & \, \lambda \,
  \left\langle \!\! \bigg\langle \mbox{Tr} \left[ U^{unp}_{\ul{x}} \,
      U^{pol \, \dagger}_{\ul{y}} \right] + \mbox{Tr} \left[
      U^{pol}_{\ul{y}} \, U^{unp \, \dagger}_{\ul{x}} \right]
    \bigg\rangle \!\!  \right\rangle \!  (z) \\ = & \, 2 \, \lambda \,
  \left\langle \!\! \bigg\langle \mbox{Re} \left( \mbox{Tr} \left[
        U^{unp}_{\ul{x}} \, U^{pol \, \dagger}_{\ul{y}} \right]
    \right) \bigg\rangle \!\!  \right\rangle \!  (z) \notag
\end{align}
in the adjoint representation. (From now on we will suppress the
subscript $\Sigma$ in the angle brackets: without loss of generality
one may simply assume that the proton helicity is $\Sigma = +1$.) Note
that since $U^{pol}$ is not a standard adjoint Wilson line, it is not
clear whether it is purely real or whether it has an imaginary part:
this is why we left the Re sign in the last line of
\eq{eq:pol_dip_adj}.

Note that it is not very probable that the gluon helicity TMD is
related to the operator in \eq{eq:pol_dip_adj}.  As the unpolarized
gluon TMD in general is not related to the unpolarized adjoint dipole
\cite{Dominguez:2011gc,Dominguez:2011br,Balitsky:2014wna,Balitsky:2015qba,Tarasov:2015pxa},
it is likely that a proper operator governing the high-energy behavior
of the gluon helicity TMD is different from \eqref{eq:pol_dip_adj},
invoking higher-order correlators such as the quadrupole.  Here we
consider this operator simply because it will help us write a closed
system of evolution equations in the large-$N_c \, \& \, N_f$ case.

Below we will also sometimes employ the doublet
\begin{align}\label{Wdef2}
  W^{pol}_{\ul{x} \, \ul{y}} (z) \equiv \left(
\begin{array}{c}
  \frac{1}{N_c} \, \left\langle \!\! \bigg\langle \mbox{tr} \left[
      V^{unp}_{\ul{x}} \, V^{pol \, \dagger}_{\ul{y}} \right] +
    \mbox{tr} \left[ V^{pol}_{\ul{y}} \, V^{unp \, \dagger}_{\ul{x}}
    \right] \bigg\rangle \!\!  \right\rangle \! (z) \\
  \frac{1}{N_c^2 -1} \, \left\langle \!\! \bigg\langle \mbox{Tr} \left[
      U^{unp}_{\ul{x}} \, U^{pol \, \dagger}_{\ul{y}} \right] +
    \mbox{Tr} \left[ U^{pol}_{\ul{y}} \, U^{unp \, \dagger}_{\ul{x}}
    \right] \bigg\rangle \!\!  \right\rangle \! (z)
\end{array}
\right).
\end{align}

\subsection{Quark Dipole Evolution}

\begin{figure}[h!]
\centering
\includegraphics[width= 0.95 \textwidth]{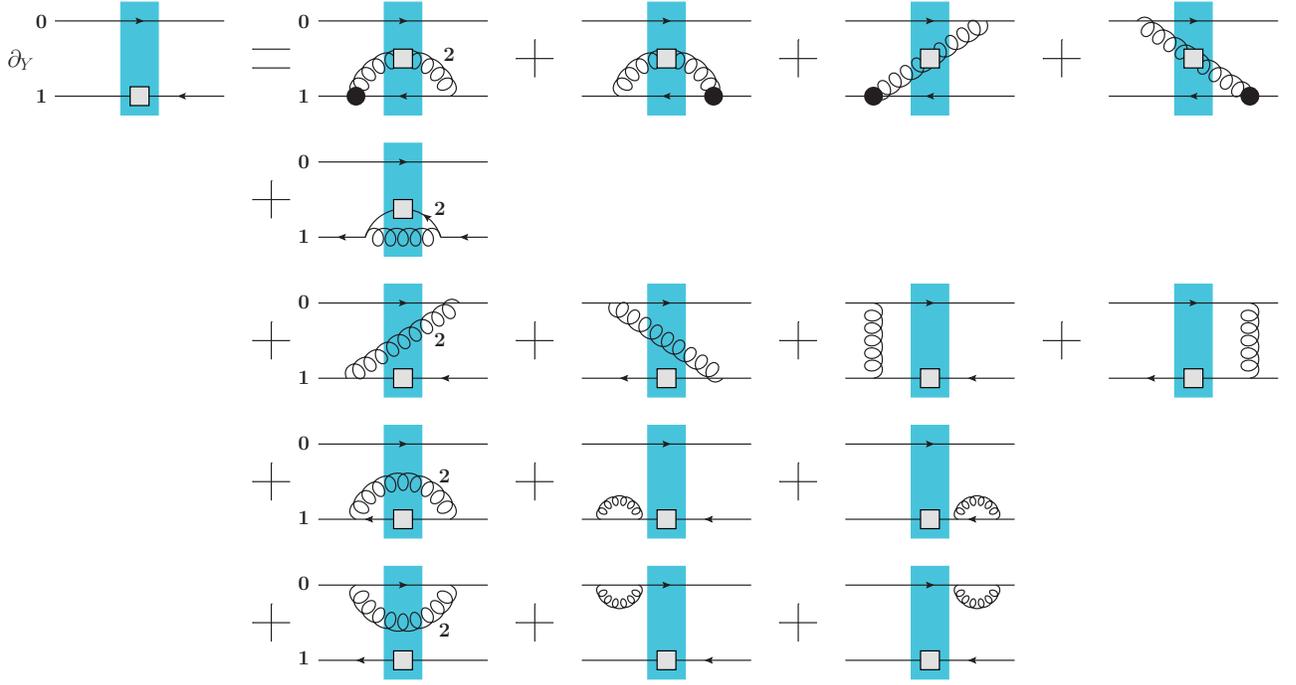}
\caption{One step of the polarized fundamental dipole
  evolution. Shaded rectangles denote the shock wave. Black circles
  represent spin-dependent (sub-eikonal) soft gluon emission
  vertices. Gray squares denote the lines carrying polarization
  information.}
\label{es}
\end{figure}

We begin with the fundamental dipole. To write an operator evolution
equation (similar to the first equation in the Balitsky hierarchy
\cite{Balitsky:1996ub,Balitsky:1998ya}), we have to perform one step
of evolution, which is shown in \fig{es}. Here we depict an equation
only for the first out of the two traces in \eq{eq:pol_dip_fund}:
however, it should be stressed that the resulting equations are only
valid when the trace is part of the whole polarized dipole operator
\eqref{eq:pol_dip_fund}, that is, when both traces are added together.

The first two diagrams on the right-hand side of the first row and the
diagram in the second row can be used to construct the ladder
approximation. The last two diagrams in the first row are the
non-ladder graphs. The last three rows consist of the usual
unpolarized small-$x$ evolution (of the BK/JIMWLK type) which also
contributes to the DLA as we have seen in Figs.~\ref{new_nonladder} and
\ref{non-ladder_gluons}.

Adding up the contributions of all the diagrams on the right hand side
of \eq{es} yields the following evolution equation for the fundamental
polarized dipole:
\begin{align}\label{evol1}
  & \frac{1}{N_c} \, \left\langle \mbox{tr} \left[ V^{unp}_{\ul{0}} \,
      V_{\ul{1}}^{pol \, \dagger} \right] \right\rangle (z) =
  \frac{1}{N_c} \, \left\langle \mbox{tr} \left[ V^{unp}_{\ul{0}} \,
      V_{\ul{1}}^{pol \, \dagger} \right] \right\rangle_0 (z) +
  \frac{\as}{2 \pi^2} \int\limits_{z_i}^z \frac{d z'}{z} \,
  \int\limits_{\rho'^2} d^2 x_{2} \\ & \times \left[ \left(
      \frac{1}{x_{21}^2} \, \theta (x_{10}^2 z - x_{21}^2 z') -
      \frac{\ul{x}_{21} \cdot \ul{x}_{20}}{x_{21}^2 \, x_{20}^2} \,
      \theta (x_{10}^2 z - \mbox{max} \{ x_{21}^2, x_{20}^2 \} z')
    \right) \right. \frac{2}{N_c} \left\langle \mbox{tr} \left[ t^b \,
      V^{unp}_{\ul{0}} \, t^a \, V_{\ul{1}}^{unp \, \dagger} \right]
    \, U^{pol \, ba}_{\ul{2}} \right\rangle (z') \notag \\ & \left. +
    \frac{1}{x_{21}^2} \, \theta (x_{10}^2 z - x_{21}^2 z') \,
    \frac{1}{N_c} \left\langle \mbox{tr} \left[ t^b \,
        V^{unp}_{\ul{0}} \, t^a \, V_{\ul{2}}^{pol \, \dagger} \right]
      \, U^{unp \, ba}_{\ul{1}} \right\rangle (z') \right] \notag \\ &
  + \frac{\as}{\pi^2} \int\limits_{z_i}^z \frac{d z'}{z'} \,
  \int\limits_{\rho'^2} d^2 x_{2} \, \frac{x_{10}^2}{x_{21}^2 \,
    x_{20}^2} \, \theta (x_{10}^2 z - x_{21}^2 z') \, \frac{1}{N_c}
  \left[ \left\langle \mbox{tr} \left[ t^b \, V^{unp}_{\ul{0}} \, t^a
        \, V_{\ul{1}}^{pol \, \dagger} \right] \, U^{unp \,
        ba}_{\ul{2}} \right\rangle (z') - C_F \left\langle \mbox{tr}
      \left[ V^{unp}_{\ul{0}} \, V_{\ul{1}}^{pol \, \dagger} \right]
    \right\rangle (z') \right]. \notag
\end{align}
The coordinates used above are explained in \fig{es}: we use an
abbreviated notation, e.g. $V_{\ul{0}}$ stands for $V_{\ul{x}_0}$. The
terms on the right of \eq{evol1} correspond to the diagrams on the
right of \fig{es}. Namely, the first two lines of \eq{evol1}
correspond to the first row of \fig{es}. The $\theta$-functions define
the range of $x_2$-integration giving the DLA contribution, while
$\rho'^{\, 2} = 1/(z' s)$ is the UV cutoff. The argument $z'$ denotes
the light-cone momentum fraction of the softest (polarized or
unpolarized) Wilson line operator in the correlator. Note that the UV
cutoff ``evolves" with $z'$. The third line of \eq{evol1} corresponds
to the second row of \fig{es}. Finally, the last line of \eq{evol1}
represents the last three rows of \fig{es}. The $\langle \ldots
\rangle_0$ term on the right of \eq{evol1} represents initial
conditions for the evolution, which, for simplicity, are not shown in
\fig{es}.

To further simplify \eq{evol1} we note that with the DLA accuracy
(remembering that $z \ge z'$, see \eqref{eq:ABC2})
\begin{align}\label{subst1}
  \frac{1}{x_{21}^2} \, \theta (x_{10}^2 z - x_{21}^2 z') -
  \frac{\ul{x}_{21} \cdot \ul{x}_{20}}{x_{21}^2 \, x_{20}^2} \, \theta
  (x_{10}^2 z - \mbox{max} \{ x_{21}^2, x_{20}^2 \} z') \approx
  \frac{1}{x_{21}^2} \, \theta (x_{10} - x_{21}).
\end{align}
Note that since
\begin{align}\label{adjoint}
  U^{unp \, ba}_{\ul{2}} = 2 \, \mbox{tr} \left[ t^a \,
    V^{unp}_{\ul{2}} \, t^b \, V^{unp \, \dagger}_{\ul{2}} \right]
\end{align}
one can use Fierz identity to show that
\begin{align}\label{Fierz}
  2 \, \mbox{tr} \left[ t^b \, V^{unp}_{\ul{0}} \, t^a \,
    V_{\ul{1}}^{pol \, \dagger} \right] \, U^{unp \ ba}_{\ul{2}} =
  \mbox{tr} \left[ V^{unp}_{\ul{0}} \, V^{unp \, \dagger}_{\ul{2}}
  \right] \, \mbox{tr} \left[ V^{unp}_{\ul{2}} \, V^{pol \,
      \dagger}_{\ul{1}} \right] - \frac{1}{N_c} \, \mbox{tr} \left[
    V^{unp}_{\ul{0}} \, V^{pol \, \dagger}_{\ul{1}} \right].
\end{align}
Using these results in \eq{evol1} yields
\begin{align}\label{evol2}
  \frac{1}{N_c} & \, \left\langle \mbox{tr} \left[ V^{unp}_{\ul{0}} \,
      V_{\ul{1}}^{pol \, \dagger} \right] \right\rangle (z) =
  \frac{1}{N_c} \, \left\langle \mbox{tr} \left[ V^{unp}_{\ul{0}} \,
      V_{\ul{1}}^{pol \, \dagger} \right] \right\rangle_0 (z) +
  \frac{\as}{2 \pi^2} \int\limits_{z_i}^z \frac{d z'}{z} \,
  \int\limits_{\rho'^2} d^2 x_{2} \notag \\ & \times \left[
    \frac{1}{x_{21}^2} \, \theta (x_{10} - x_{21}) \, \frac{2}{N_c}
    \left\langle \mbox{tr} \left[ t^b \, V^{unp}_{\ul{0}} \, t^a \,
        V_{\ul{1}}^{unp \, \dagger} \right] \, U^{pol \, ba}_{\ul{2}}
    \right\rangle (z') \right. \notag \\ & \left. + \frac{1}{x_{21}^2}
    \, \theta (x_{10}^2 z - x_{21}^2 z') \, \frac{1}{N_c} \left\langle
      \mbox{tr} \left[ t^b \, V^{unp}_{\ul{0}} \, t^a \,
        V_{\ul{2}}^{pol \, \dagger} \right] \, U^{unp \, ba}_{\ul{1}}
    \right\rangle (z') \right] \notag \\ & + \frac{\as}{2 \pi^2}
  \int\limits_{z_i}^z \frac{d z'}{z'} \, \int\limits_{\rho'^2} d^2
  x_{2} \, \frac{x_{10}^2}{x_{21}^2 \, x_{20}^2} \, \theta (x_{10}^2 z
  - x_{21}^2 z') \notag \\ & \times \frac{1}{N_c} \left[ \left\langle
      \mbox{tr} \left[ V^{unp}_{\ul{0}} \, V^{unp \, \dagger}_{\ul{2}}
      \right] \, \mbox{tr} \left[ V^{unp}_{\ul{2}} \, V^{pol \,
          \dagger}_{\ul{1}} \right] \right\rangle (z') - N_c
    \left\langle \mbox{tr} \left[ V^{unp}_{\ul{0}} \, V_{\ul{1}}^{pol
          \, \dagger} \right] \right\rangle (z') \right].
\end{align}
At this point one can notice that the last square brackets on the
right-hand side of \eq{evol2} go to zero as $\ul{2} \to \ul{0}$, hence
the $x_{20} \ll x_{10}$ region of integration is not logarithmic and
is beyond the DLA. The only DLA contribution in the last term on the
right of \eq{evol2} comes from the $x_{21} \ll x_{10}$ region, where
one can replace
\begin{align}\label{subst2}
  \frac{x_{10}^2}{x_{21}^2 \, x_{20}^2} \to \frac{1}{x_{21}^2} \,
  \theta (x_{10} - x_{21}) .
\end{align}
We obtain
\begin{align}\label{evol3}
  & \frac{1}{N_c} \, \left\langle \mbox{tr} \left[ V^{unp}_{\ul{0}} \,
      V_{\ul{1}}^{pol \, \dagger} \right] \right\rangle (z) =
  \frac{1}{N_c} \, \left\langle \mbox{tr} \left[ V^{unp}_{\ul{0}} \,
      V_{\ul{1}}^{pol \, \dagger} \right] \right\rangle_0 (z) +
  \frac{\as}{2 \pi^2} \int\limits_{z_i}^z \frac{d z'}{z} \,
  \int\limits_{\rho'^2} \frac{d^2 x_{2}}{x_{21}^2} \notag \\ & \times
  \left\{ \theta (x_{10} - x_{21}) \, \frac{2}{N_c} \left\langle
      \mbox{tr} \left[ t^b \, V^{unp}_{\ul{0}} \, t^a \,
        V_{\ul{1}}^{unp \, \dagger} \right] \, U^{pol \, ba}_{\ul{2}}
    \right\rangle (z') \right. + \theta (x_{10}^2 z -
  x_{21}^2 z') \, \frac{1}{N_c} \left\langle \mbox{tr} \left[ t^b \,
      V^{unp}_{\ul{0}} \, t^a \, V_{\ul{2}}^{pol \, \dagger} \right]
    \, U^{unp \, ba}_{\ul{1}} \right\rangle (z') \notag \\ & +
  \frac{z}{z'} \, \theta (x_{10} - x_{21}) \left. \frac{1}{N_c} \left[
      \left\langle \mbox{tr} \left[ V^{unp}_{\ul{0}} \, V^{unp \,
            \dagger}_{\ul{2}} \right] \, \mbox{tr} \left[
          V^{unp}_{\ul{2}} \, V^{pol \, \dagger}_{\ul{1}} \right]
      \right\rangle (z') - N_c \left\langle \mbox{tr} \left[
          V^{unp}_{\ul{0}} \, V_{\ul{1}}^{pol \, \dagger} \right]
      \right\rangle (z') \right] \right\}.
\end{align}

Finally, to cast \eq{evol3} in an explicitly DLA form we redefine
matrix elements in it using \eq{redef0} (cf. Eq.~(42) in
\cite{Itakura:2003jp}).  Note that $z$, used in the rescaling
\eq{redef0}, is not always the smallest longitudinal momentum fraction
in a diagram (and hence is not always in the argument of the matrix
elements). For instance, the operators in the last line of \eq{evol3}
have the polarized line carrying momentum fraction $z$, while the
softest unpolarized Wilson line carries the momentum fraction $z'$,
which enters in the argument of the matrix elements. Using the
redefinition \eqref{redef0} (and keeping in mind that the
$z$-rescaling is determined by the polarized line, not necessarily by
the softest parton which determines the argument of the Wilson line
matrix element) we obtain
\begin{align}\label{evol4}
  & \frac{1}{N_c} \, \left\langle \!\! \left\langle \mbox{tr} \left[
        V^{unp}_{\ul{0}} \, V_{\ul{1}}^{pol \, \dagger} \right]
    \right\rangle \!\! \right\rangle (z) = \frac{1}{N_c} \,
  \left\langle \!\! \left\langle \mbox{tr} \left[ V^{unp}_{\ul{0}} \,
        V_{\ul{1}}^{pol \, \dagger} \right] \right\rangle \!\!
  \right\rangle_0 (z) + \frac{\as}{2 \pi^2} \int\limits_{z_i}^z
  \frac{d z'}{z'} \, \int\limits_{\rho'^2} \frac{d^2 x_{2}}{x_{21}^2}
  \notag \\ & \times \left\{ \theta (x_{10} - x_{21}) \, \frac{2}{N_c}
    \left\langle \!\! \left\langle \mbox{tr} \left[ t^b \,
          V^{unp}_{\ul{0}} \, t^a \, V_{\ul{1}}^{unp \, \dagger}
        \right] \, U^{pol \, ba}_{\ul{2}} \right\rangle \!\!
    \right\rangle (z') \right. + \theta (x_{10}^2 z -
  x_{21}^2 z') \, \frac{1}{N_c} \left\langle \!\! \left\langle
      \mbox{tr} \left[ t^b \, V^{unp}_{\ul{0}} \, t^a \,
        V_{\ul{2}}^{pol \, \dagger} \right] \, U^{unp \, ba}_{\ul{1}}
    \right\rangle \!\! \right\rangle (z') \notag \\ & + \theta (x_{10}
  - x_{21}) \left. \frac{1}{N_c} \left[ \left\langle \!\! \left\langle
          \mbox{tr} \left[ V^{unp}_{\ul{0}} \, V^{unp \,
              \dagger}_{\ul{2}} \right] \, \mbox{tr} \left[
            V^{unp}_{\ul{2}} \, V^{pol \, \dagger}_{\ul{1}} \right]
        \right\rangle \!\! \right\rangle (z') - N_c \left\langle \!\!
        \left\langle \mbox{tr} \left[ V^{unp}_{\ul{0}} \,
            V_{\ul{1}}^{pol \, \dagger} \right] \right\rangle \!\!
      \right\rangle (z') \right] \right\}.
\end{align}
This is the simplest form of the evolution equation for the
fundamental dipole we could obtain without making any additional
assumptions. Note that \eq{adjoint}, and hence the identity
\eqref{Fierz}, would not be valid if we replace $U^{unp}_{\ul{2}}$ in
them with $U^{pol}_{\ul{2}}$, since it is not a standard Wilson line.

We would like to stress once again that \eq{evol4} is only valid if
the operator on its left hand side is considered as a part (a
``half'') of the polarized dipole operator
\eqref{eq:pol_dip_fund}. The corresponding equation for the full
polarized fundamental dipole operator \eqref{eq:pol_dip_fund} is
\begin{align}\label{fund_evol}
  & \frac{1}{N_c} \, \left\langle \!\! \left\langle \mbox{tr} \left[
        V^{unp}_{\ul{0}} \, V_{\ul{1}}^{pol \, \dagger} \right] +
      \mbox{tr} \left[ V_{\ul{1}}^{pol} \, V^{unp \, \dagger}_{\ul{0}}
      \right] \right\rangle \!\! \right\rangle (z) = \frac{1}{N_c} \,
  \left\langle \!\! \left\langle \mbox{tr} \left[ V^{unp}_{\ul{0}} \,
        V_{\ul{1}}^{pol \, \dagger} \right] + \mbox{tr} \left[
        V_{\ul{1}}^{pol} \, V^{unp \, \dagger}_{\ul{0}} \right]
    \right\rangle \!\!  \right\rangle_0 (z) + \frac{\as}{2 \pi^2}
  \int\limits_{z_i}^z \frac{d z'}{z'} \, \int\limits_{\rho'^2}
  \frac{d^2 x_{2}}{x_{21}^2} \notag \\ & \times \left\{ \theta (x_{10}
    - x_{21}) \, \frac{2}{N_c} \left\langle \!\! \left\langle
        \mbox{tr} \left[ t^b \, V^{unp}_{\ul{0}} \, t^a \,
          V_{\ul{1}}^{unp \, \dagger} \right] \, U^{pol \,
          ba}_{\ul{2}} + \mbox{tr} \left[ t^b \, V^{unp}_{\ul{1}} \,
          t^a \, V_{\ul{0}}^{unp \, \dagger} \right] \, U^{pol \,
          \dagger \, ab}_{\ul{2}} \right\rangle \!\!  \right\rangle
    (z') \right. \notag \\ & + \theta (x_{10}^2 z - x_{21}^2 z') \,
  \frac{1}{N_c} \left\langle \!\! \left\langle \mbox{tr} \left[ t^b \,
        V^{unp}_{\ul{0}} \, t^a \, V_{\ul{2}}^{pol \, \dagger} \right]
      \, U^{unp \, ba}_{\ul{1}} + \mbox{tr} \left[ t^b \,
        V^{pol}_{\ul{2}} \, t^a \, V_{\ul{0}}^{unp \, \dagger} \right]
      \, U^{unp \, \dagger \, ab}_{\ul{1}} \right\rangle \!\!
  \right\rangle (z') \notag \\ & + \theta (x_{10} - x_{21})
  \frac{1}{N_c} \left[ \left\langle \!\! \left\langle \mbox{tr} \left[
          V^{unp}_{\ul{0}} \, V^{unp \, \dagger}_{\ul{2}} \right] \,
        \mbox{tr} \left[ V^{unp}_{\ul{2}} \, V^{pol \,
            \dagger}_{\ul{1}} \right] + \mbox{tr} \left[
          V^{unp}_{\ul{2}} \, V^{unp \, \dagger}_{\ul{0}} \right] \,
        \mbox{tr} \left[ V^{pol}_{\ul{1}} \, V^{unp \,
            \dagger}_{\ul{2}} \right] \right\rangle \!\!
    \right\rangle (z') \right. \notag \\ & \left. \left. - N_c
      \left\langle \!\!  \left\langle \mbox{tr} \left[
            V^{unp}_{\ul{0}} \, V_{\ul{1}}^{pol \, \dagger} \right] +
          \mbox{tr} \left[ V^{pol}_{\ul{1}} \, V_{\ul{0}}^{unp \,
              \dagger} \right] \right\rangle \!\!  \right\rangle (z')
    \right] \right\}.
\end{align}


\subsection{Gluon Dipole}

\begin{figure}[h!]
\centering
\includegraphics[width= 0.95 \textwidth]{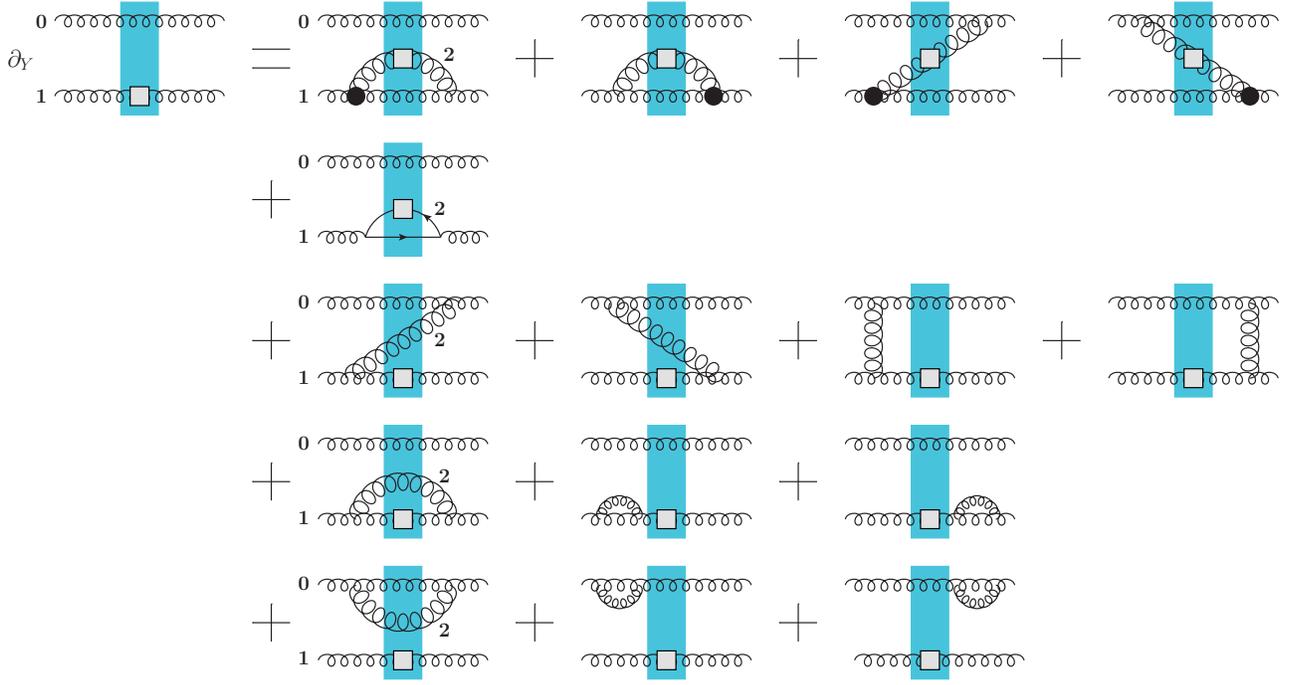}
\caption{One step of the polarized adjoint dipole evolution. Shaded
  rectangles denote the shock wave. Black circles represent
  spin-dependent (sub-eikonal) soft gluon emission vertices. Gray
  squares denote the lines carrying polarization information.}
\label{esG}
\end{figure}

Similar to the above, we can construct the evolution equation for the
polarized gluon dipole. Instead of \eq{evol1} we get
\begin{align}\label{Gevol1}
  & \frac{1}{N_c^2 -1} \, \left\langle \mbox{Tr} \left[
      U^{unp}_{\ul{0}} \, U_{\ul{1}}^{pol \, \dagger} \right]
  \right\rangle (z) = \frac{1}{N_c^2 -1} \, \left\langle \mbox{Tr}
    \left[ U^{unp}_{\ul{0}} \, U_{\ul{1}}^{pol \, \dagger} \right]
  \right\rangle_0 (z) + \frac{\as}{2 \pi^2} \int\limits_{z_i}^z
  \frac{d z'}{z} \, \int\limits_{\rho'^2} d^2 x_{2} \notag \\ & \times
  \left[ \left( \frac{1}{x_{21}^2} \, \theta (x_{10}^2 z - x_{21}^2
      z') - \frac{\ul{x}_{21} \cdot \ul{x}_{20}}{x_{21}^2 \, x_{20}^2}
      \, \theta (x_{10}^2 z - \mbox{max} \{ x_{21}^2, x_{20}^2 \} z')
    \right) \right. \notag \\ & \times \frac{4}{N_c^2 -1} \left\langle
    \mbox{Tr} \left[ T^b \, U^{unp}_{\ul{0}} \, T^a \, U_{\ul{1}}^{unp
        \, \dagger} \right] \, U^{pol \, ba}_{\ul{2}} \right\rangle
  (z') \notag \\ & \left. - \frac{1}{x_{21}^2} \, \theta (x_{10}^2 z -
    x_{21}^2 z') \, \frac{N_f}{N_c^2-1} \left\langle \mbox{tr} \left[
        t^b \, V^{unp}_{\ul{1}} \, t^a \, V_{\ul{2}}^{pol \, \dagger}
      \right] \, U^{unp \, ba}_{\ul{0}} + \mbox{tr} \left[ t^b \,
        V^{pol}_{\ul{2}} \, t^a \, V_{\ul{1}}^{unp \, \dagger} \right]
      \, U^{unp \, ba}_{\ul{0}} \right\rangle (z') \right] \notag \\ &
  + \frac{\as}{\pi^2} \int\limits_{z_i}^z \frac{d z'}{z'} \,
  \int\limits_{\rho'^2} d^2 x_{2} \, \frac{x_{10}^2}{x_{21}^2 \,
    x_{20}^2} \, \theta (x_{10}^2 z - x_{21}^2 z') \notag \\ & \times
  \frac{1}{N_c^2 -1} \left[ \left\langle \mbox{Tr} \left[ T^b \,
        U^{unp}_{\ul{0}} \, T^a \, U_{\ul{1}}^{pol \, \dagger} \right]
      \, U^{unp \, ba}_{\ul{2}} \right\rangle (z') - N_c \left\langle
      \mbox{Tr} \left[ U^{unp}_{\ul{0}} \, U_{\ul{1}}^{pol \, \dagger}
      \right] \right\rangle (z') \right].
\end{align}
Note that $t^a$ and $T^a$ are the fundamental and adjoint SU($N_c$)
generators correspondingly, with tr denoting the fundamental traces
and Tr denoting the adjoint ones. Equation \eqref{Gevol1} is
illustrated in \fig{esG}, where again we omit showing the initial
conditions term.

We can further simplify \eq{Gevol1} along the above lines. Instead of
\eq{Fierz} resulting from the Fierz identity, now we have
\begin{align}
  \mbox{Tr} \left[ T^b \, U^{unp}_{\ul{0}} \, T^a \, U_{\ul{1}}^{pol
      \, \dagger} \right] \, U^{unp \, ba}_{\ul{2}} = \mbox{Tr} \left[
    T^a \, U^{unp}_{\ul{2}} \, U^{unp \, \dagger}_{\ul{0}} \, T^a \,
    U^{unp}_{\ul{0}} \, U_{\ul{1}}^{pol \, \dagger} \right] ,
\end{align}
which can be used to justify the substitution \eqref{subst2} (that is,
only the region $x_{21} \ll x_{10}$ is logarithmic). Using
Eqs.~\eqref{subst1} and \eqref{subst2} in \eq{Gevol1} yields
\begin{align}\label{Gevol2}
  & \frac{1}{N_c^2 -1} \, \left\langle \mbox{Tr} \left[
      U^{unp}_{\ul{0}} \, U_{\ul{1}}^{pol \, \dagger} \right]
  \right\rangle (z) = \frac{1}{N_c^2 -1} \, \left\langle \mbox{Tr}
    \left[ U^{unp}_{\ul{0}} \, U_{\ul{1}}^{pol \, \dagger} \right]
  \right\rangle_0 (z) + \frac{\as}{2 \pi^2} \int\limits_{z_i}^z
  \frac{d z'}{z} \, \int\limits_{\rho'^2} \frac{d^2 x_{2}}{x_{21}^2}
  \notag \\ & \times \left\{ \theta (x_{10} - x_{21}) \,
    \frac{4}{N_c^2 -1} \left\langle \mbox{Tr} \left[ T^b \,
        U^{unp}_{\ul{0}} \, T^a \, U_{\ul{1}}^{unp \, \dagger} \right]
      \, U^{pol \, ba}_{\ul{2}} \right\rangle (z') \right. \notag \\ &
  - \theta (x_{10}^2 z - x_{21}^2 z') \, \frac{N_f}{N_c^2-1}
  \left\langle \mbox{tr} \left[ t^b \, V^{unp}_{\ul{1}} \, t^a \,
      V_{\ul{2}}^{pol \, \dagger} \right] \, U^{unp \, ba}_{\ul{0}} +
    \mbox{tr} \left[ t^b \, V^{pol}_{\ul{2}} \, t^a \, V_{\ul{1}}^{unp
        \, \dagger} \right] \, U^{unp \, ba}_{\ul{0}} \right\rangle
  (z') \notag \\ & \left. + \frac{z}{z'} \, \theta (x_{10} - x_{21})
    \frac{2}{N_c^2 -1} \left[ \left\langle \mbox{Tr} \left[ T^b \,
          U^{unp}_{\ul{0}} \, T^a \, U_{\ul{1}}^{pol \, \dagger}
        \right] \, U^{unp \, ba}_{\ul{2}} \right\rangle (z') - N_c
      \left\langle \mbox{Tr} \left[ U^{unp}_{\ul{0}} \,
          U_{\ul{1}}^{pol \, \dagger} \right] \right\rangle (z')
    \right] \right\}.
\end{align}

Finally, performing the redefinition \eqref{redef0} we arrive at
\begin{align}\label{Gevol3}
  & \frac{1}{N_c^2 -1} \, \left\langle \!\! \left\langle \mbox{Tr}
      \left[ U^{unp}_{\ul{0}} \, U_{\ul{1}}^{pol \, \dagger} \right]
    \right\rangle \!\! \right\rangle (z) = \frac{1}{N_c^2 -1} \,
  \left\langle \!\! \left\langle \mbox{Tr} \left[ U^{unp}_{\ul{0}} \,
        U_{\ul{1}}^{pol \, \dagger} \right] \right\rangle \!\!
  \right\rangle_0 (z) + \frac{\as}{2 \pi^2} \int\limits_{z_i}^z
  \frac{d z'}{z'} \, \int\limits_{\rho'^2} \frac{d^2 x_{2}}{x_{21}^2}
  \notag \\ & \times \left\{ \theta (x_{10} - x_{21}) \,
    \frac{4}{N_c^2 -1} \left\langle \!\! \left\langle \mbox{Tr} \left[
          T^b \, U^{unp}_{\ul{0}} \, T^a \, U_{\ul{1}}^{unp \,
            \dagger} \right] \, U^{pol \, ba}_{\ul{2}} \right\rangle
      \!\! \right\rangle (z') \right. \notag \\ & - \theta (x_{10}^2 z
  - x_{21}^2 z') \, \frac{N_f}{N_c^2-1} \left\langle \!\! \left\langle
      \mbox{tr} \left[ t^b \, V^{unp}_{\ul{1}} \, t^a \,
        V_{\ul{2}}^{pol \, \dagger} \right] \, U^{unp \, ba}_{\ul{0}}
      + \mbox{tr} \left[ t^b \, V^{pol}_{\ul{2}} \, t^a \,
        V_{\ul{1}}^{unp \, \dagger} \right] \, U^{unp \, ba}_{\ul{0}}
    \right\rangle \!\! \right\rangle (z') \notag \\ & \left. + \theta
    (x_{10} - x_{21}) \frac{2}{N_c^2 -1} \left[ \left\langle \!\!
        \left\langle \mbox{Tr} \left[ T^b \, U^{unp}_{\ul{0}} \, T^a
            \, U_{\ul{1}}^{pol \, \dagger} \right] \, U^{unp \,
            ba}_{\ul{2}} \right\rangle \!\! \right\rangle (z') - N_c
      \left\langle \!\! \left\langle \mbox{Tr} \left[ U^{unp}_{\ul{0}}
            \, U_{\ul{1}}^{pol \, \dagger} \right] \right\rangle \!\!
      \right\rangle (z') \right] \right\}.
\end{align}
The evolution equation for the full adjoint polarized dipole
\eqref{eq:pol_dip_adj} can be constructed by adding to \eq{Gevol3} its
hermitean conjugate, similar to the fundamental dipole case.

Equations \eqref{fund_evol} and \eqref{Gevol3} are our main
equations. Unfortunately they are not closed, like the equations in
the Balitsky hierarchy. That is, the objects (operator expectation
values) on their right-hand sides are not always the same as the
polarized dipole operators on their left-hand sides: hence, similar to
JIMWLK equation, they can only be solved by functional methods
\cite{Weigert:2000gi,Rummukainen:2003ns}.


\subsection{Ladder Approximation}
\label{sec:ladder}

We would now like to cross check our results against those of BER
\cite{Bartels:1996wc}. This is no easy task given that at the moment
we do not have a solution of Eqs.~\eqref{fund_evol} and \eqref{Gevol3}
and that our rapidity evolution method is very different from the
infrared cutoff evolution employed by BER. In addition, no analytic
solution for the DLA $g_1 (x, Q^2)$ evolution was derived in
\cite{Bartels:1996wc}, although an approximate expression was found in
\cite{Ermolaev:2003zx}.  However, it appears possible to identify ladder
diagrams both in the BER approach and in our calculation: comparing
the two would provide us at least with one cross check. Certainly, an
agreement between the two approaches in the ladder approximation does
not yet mean a complete agreement of the two calculations. Still, the
ladder cross check is a necessary condition which has to be satisfied
in order to achieve this full agreement.

As was already noted by BER \cite{Bartels:1995iu,Bartels:1996wc},
ladder diagrams are not the only ones contributing to helicity
evolution. Therefore, approximating the evolution by ladder graphs is
not justified. Still it may serve as a consistency check of our
calculations.

With that in mind, let us put all unpolarized Wilson lines to unity,
that is put all $U^{unp} = \mathbb{1}$ and $V^{unp} = \mathbb{1}$ in
Eqs.~\eqref{evol1} and \eqref{Gevol1}. In addition we need to discard
the contributions of the non-ladder gluons and quasi-LLA gluons: that
is, we only keep the contributions of the first two diagrams in the
first line and the diagram in the second line of Figs.~\ref{es} and
\ref{esG}. We then get
\begin{subequations}\label{eq:ladder1}
\begin{align}\label{evol5}
  \frac{1}{N_c} & \, \left\langle \mbox{tr} \left[ V_{\ul{1}}^{pol \, \dagger} \right] \right\rangle (z) = \frac{1}{N_c} \, \left\langle \mbox{tr} \left[ V_{\ul{1}}^{pol \, \dagger} \right] \right\rangle_0 (z) + \frac{\as}{2 \pi^2} \int\limits_{z_i}^z \frac{d z'}{z} \, \int\limits_{\rho'^2} \frac{d^2 x_{2}}{x_{21}^2} \, \theta (x_{10}^2 z - x_{21}^2 z')  \notag \\ & \times \left[ C_F \, \frac{1}{N_c} \left\langle \mbox{tr} \left[ V_{\ul{2}}^{pol \, \dagger} \right] \right\rangle (z') + 2 \, C_F \, \frac{1}{N_c^2 -1} \left\langle \mbox{Tr} \left[ U^{pol}_{\ul{2}} \right] \right\rangle (z')  \right], \\
  \frac{1}{N_c^2 -1} & \, \left\langle \mbox{Tr} \left[
      U_{\ul{1}}^{pol \, \dagger} \right] \right\rangle (z) = \frac{1}{N_c^2 -1} \, \left\langle \mbox{tr} \left[ U_{\ul{1}}^{pol \, \dagger} \right] \right\rangle_0 (z) +
  \frac{\as}{2 \pi^2} \int\limits_{z_i}^z \frac{d z'}{z} \,
  \int\limits_{\rho'^2} \frac{d^2 x_{2}}{x_{21}^2} \, \theta (x_{10}^2
  z - x_{21}^2 z') \notag \\ & \times \left[ 4 \, N_c \,
    \frac{1}{N_c^2 -1} \left\langle \mbox{Tr} \left[ U^{pol}_{\ul{2}}
      \right] \right\rangle (z') - \frac{N_f}{2 \, N_c} \left\langle
      \mbox{tr} \left[ V_{\ul{2}}^{pol \, \dagger} \right] + \mbox{tr}
      \left[ V_{\ul{2}}^{pol} \right] \right\rangle (z')
  \right]. \label{evol6}
\end{align}
\end{subequations}

In the same approximation the doublet \eqref{Wdef2} becomes
\begin{align}\label{Wdef}
W^{pol}_{\ul{x}\ul{y}} = \left( 
\begin{array}{c}
  \frac{1}{N_c} \, \left\langle \!\! \left\langle \mbox{tr} [V^{pol}_{\ul{x}}] + \mbox{tr} [V^{pol \, \dagger}_{\ul{x}}] \right\rangle\!\!
  \right\rangle \\
  \frac{1}{N_c^2 -1} \, \left\langle \!\! \left\langle \mbox{Tr} [U^{pol}_{\ul{x}}] + \mbox{Tr} [U^{pol \, \dagger}_{\ul{x}}] \right\rangle \!\!
  \right\rangle
\end{array}
\right).
\end{align}
In terms of the doublet Eqs.~\eqref{eq:ladder1} become
\begin{align}\label{evolution2}
  W^{pol}_{\ul{1} \, \ul{0}} (z) = W^{(0) \, pol}_{\ul{1} \, \ul{0}}
  (z) + \frac{\as}{2 \, \pi} \int\limits_{z_i}^z \frac{dz'}{z'} \,
  \int\limits_{\rho'^2}^{x_{01}^2 z/z'} \frac{d x_{21}^2}{x_{21}^2} \,
  M \, W^{pol}_{\ul{2} \, \ul{1}} (z'),
\end{align}
where we have defined 
\begin{align}\label{Matrix}
M \equiv \left(
\begin{array}{cc}
C_F & 2 \, C_F \\ - N_f & 4 \, N_c
\end{array}
\right)
\end{align}
in agreement with Eq.~(2.28) from \cite{Bartels:1996wc}. This
completes the cross-check between the ladder limits of BER and of our
calculations.  

One may worry that, because the calculation of BER was done in Feynman
gauge but our calculation has been done in the $A^+ = 0$ light-cone
gauge, the ``ladder graphs'' in the two cases may correspond to
different contributions.  However, a direct calculation of the ladder
diagrams in these two gauges shows that they coincide.  This occurs
because, in the light-cone gauge, all the evolution takes place within
the projectile wave function; for instance, the familiar DGLAP
evolution can also be re-derived within the same framework
\cite{KovchegovLevin}.

We can solve \eq{evolution2} to find the intercept in the ladder
case. We can look for the solution of \eq{evolution2} as a double
Mellin transform
\begin{align}
  \label{eq:Mellin}
  W^{pol}_{\ul{1} \, \ul{0}} (z) = \int \frac{d \omega}{2 \pi i} \,
  \frac{d \lambda}{2 \pi i} \, \left( \frac{z}{z_i} \right)^\omega \,
  \left( \frac{x_{01}^2}{\rho^2} \right)^\lambda \, W^{pol}_{\omega
    \lambda}
\end{align}
where $\omega$ and $\lambda$ integrals run along straight lines
parallel to the imaginary axis, $\rho^2 = 1/(z \, s)$ and $z_i =
\Lambda^2 /s$. Substituting this into \eq{evolution2} we obtain its
solution
\begin{align}\label{sol_ladder} 
  W^{pol}_{\ul{1} \, \ul{0}} (z) = \int \frac{d \omega}{2 \pi i} \,
  \frac{d \lambda}{2 \pi i} \, \left( \frac{z}{z_i} \right)^\omega \,
  \left( \frac{x_{01}^2}{\rho^2} \right)^\lambda \, \left[ 1 -
    \frac{\as}{2\pi} \, M \, \frac{1}{\omega \, \lambda} \right]^{-1}
  \, W^{pol \, (0)}_{\omega \lambda}.
\end{align}

The high-energy asymptotics of $W^{pol}_{\ul{1} \, \ul{0}} (z)$ are
dominated by the rightmost pole in $\omega$ of the integrand in
\eq{sol_ladder}. The pole position is given by the zeroes of the
determinant of the matrix in the square brackets. The poles are at
\begin{align}\label{omega_sp}
  \omega_\pm (\lambda) = \frac{\as}{8 \, \pi \, \lambda} \left[ 9 N_c
    - \frac{1}{N_c} \pm \frac{1}{N_c} \, \sqrt{(1+7 \, N_c^2)^2 +16 \,
      N_c \, N_f \, (1-N_c^2)} \right].
\end{align}
Clearly $\omega_+ (\lambda)$ is larger for positive $\lambda$ and
gives the high-energy asymptotics. Note that the energy dependence
comes into the integrand of \eqref{sol_ladder} as $(z \, s)^{\lambda +
  \omega_\pm (\lambda)}$. Picking up the $\omega_+ (\lambda)$ pole and
evaluating the remaining $\lambda$-integral using the saddle-point
method we find the (positive-$\lambda$) saddle point at
\begin{align}\label{lam_sp}
  \lambda_{s.p.} = \sqrt{\frac{\as}{8 \pi N_c}} \, \sqrt{9 \, N_c^2 -1
    + \sqrt{(1+7 \, N_c^2)^2 +16 \, N_c \, N_f \, (1-N_c^2)}}.
\end{align}
Substituting \eq{lam_sp} into \eq{omega_sp} yields
\begin{align}\label{ladder_intercept}
  {\tilde \omega}_+ (\lambda_{s.p.}) \equiv \lambda_{s.p.} + \omega_+
  (\lambda_{s.p.}) = \sqrt{\frac{\as}{2 \pi N_c}} \, \sqrt{9 \, N_c^2
    -1 + \sqrt{(1+7 \, N_c^2)^2 +16 \, N_c \, N_f \, (1-N_c^2)}}.
\end{align}

The high-energy asymptotics in the ladder limit is
\begin{align}\label{asymptotics}
  \Delta q (x, Q^2) \sim g_{1L} (x, k_T) \sim \left( \frac{1}{x}
  \right)^{{\tilde \omega}_+ (\lambda_{s.p.})}.
\end{align}
Following BER we use $\as = 0.18$ and $N_f =4$ along with $N_c=3$ to
obtain ${\tilde \omega}_+ (\lambda_{s.p.}) \approx 1.12$, in agreement
with the BER formula at the end of the paragraph following Eq.~(4.19)
in \cite{Bartels:1996wc}. Using ${\tilde \omega}_+ (\lambda_{s.p.})
\approx 1.12$ in \eq{asymptotics} yields
\begin{align}\label{asymptotics2}
  \Delta q (x, Q^2) \sim g_{1L} (x, k_T) \sim \left( \frac{1}{x}
  \right)^{1.12},
\end{align}
which is a slow but robust growth of the helicity TMDs and polarized
PDFs with with decreasing $x$, which may enable the low-$x$ region to
significantly contribute to $S_q (Q^2)$ from \eq{eq:net_spin} and,
therefore, to the helicity sum rule \eqref{eq:sum_rule}.

For a more realistic $\as= 0.3$ and $N_c = N_f =3$ (say, for RHIC
experimental kinematics) we get
\begin{align}\label{asymptotics3}
  \Delta q (x, Q^2) \sim g_{1L} (x, k_T) \sim \left( \frac{1}{x}
  \right)^{1.46},
\end{align}
which is a very fast growth with decreasing $x$. 

Here we have to remember that the ladder approximation used in this
Subsection is not a systematic physical approximation, but rather is
just a ``by hand'' truncation of the full calculation. Therefore, the
numbers obtained in Eqs.~\eqref{asymptotics2} and
\eqref{asymptotics3}, while encouraging, cannot be used to conclude
that there is a significant amount of spin at small $x$.


\subsection{Large-$N_c$ Limit}
\label{sec:large_nc}

By analogy with the unpolarized small-$x$ evolution, let us try to
produce a closed evolution equation out of \eq{fund_evol} by using the
large-$N_c$ approximation. Let us point out from the outset, that the
large-$N_c$ limit is far less precise in the helicity DLA evolution at
hand than it was in the unpolarized LLA small-$x$ evolution. As is
apparent from \eq{evol6}, gluon splitting into a $q\bar q$ pair is
$N_c$-suppressed, but is $N_f$-enhanced, such that it comes in with a
relative ``suppression factor'' of $N_f/N_c$. In phenomenological
applications with $N_f =3$ or $N_f =4$ this ``suppression factor'' is
$1$ or $4/3$ respectively, and leads to no suppression. This is in
contrast to the case of unpolarized BFKL/BK/JIMWLK evolution, where
quark bubbles enter only at NLO and are easily resummable
\cite{Gardi:2006rp,Balitsky:2006wa,Kovchegov:2006vj,Albacete:2007yr}. The
leading-order in unpolarized evolution involves gluons only, such that
the large-$N_c$ limit is accurate up to order-$1/N_c^2$ corrections
(see \cite{Rummukainen:2003ns,Kovchegov:2008mk} for more on this
issue). Hence the large-$N_c$ limit of the DLA helicity evolution we
will take below should be considered with caution and one has to be
very careful in interpreting the resulting numbers.

To take the large-$N_c$ limit of \eq{evol4} we first note that at
large $N_c$ only gluon emissions contribute. The soft quark emission
from the upper left corner of \fig{splittings1} is only allowed for
the parent quark. ($G \to q\bar q$ is $N_c$-suppressed.) If the parent
particle is a ``quark'' in a typical dipole, which is, in reality, is
simply a quark line in the large-$N_c$ representation of gluons as
$q\bar q$ pairs of different color, such splittings do not exist. This
means such splittings are limited to the case when quark or anti-quark
line of the dipole are from the original $q\bar q$ pair. Since, after
several steps of small-$x$ evolution at large-$N_c$ most dipoles will
be made out of gluons, it is probably safe to neglect the splitting in
the upper left corner of \fig{splittings1}. This means that we should
discard ``by hand'' the second term in the curly brackets of
\eq{fund_evol}.

The second observation we need to make is that $U^{pol \,
  ba}_{\ul{2}}$ is now (at large $N_c$) always an eikonal gluon line
(that is, it does not convert into an eikonal quark). We therefore
replace
\begin{align}\label{subst3}
  & \left\langle \!\! \left\langle \mbox{tr} \left[ t^b \,
        V^{unp}_{\ul{0}} \, t^a \, V_{\ul{1}}^{unp \, \dagger} \right]
      \, U^{pol \, ba}_{\ul{2}} \right\rangle \!\! \right\rangle \to
  \notag \\ & \frac{1}{2} \, \left\langle \!\! \left\langle \mbox{tr}
      \left[ V^{unp}_{\ul{0}} \, V^{pol \, \dagger}_{\ul{2}} \right]
    \right\rangle \!\! \right\rangle \, \left\langle \!\! \left\langle
      \mbox{tr} \left[ V^{unp}_{\ul{2}} \, V^{unp \, \dagger}_{\ul{1}}
      \right] \right\rangle \!\! \right\rangle + \frac{1}{2} \,
  \left\langle \!\! \left\langle \mbox{tr} \left[ V^{unp}_{\ul{0}} \,
        V^{unp \, \dagger}_{\ul{2}} \right] \right\rangle \!\!
  \right\rangle \, \left\langle \!\! \left\langle \mbox{tr} \left[
        V^{pol}_{\ul{2}} \, V^{unp \, \dagger}_{\ul{1}} \right]
    \right\rangle \!\! \right\rangle
\end{align}
while identifying the gluon helicity $\lambda_2$ with the polarized
quark and anti-quark line helicities $\sigma_2$ in the large-$N_c$
limit. Equation \eqref{subst3} can be obtained by substituting
\eq{Wlines2} into \eq{Fierz} and by expanding the latter to the linear
order in polarized Wilson lines.

Using \eq{subst3} and discarding the the second term in the curly
brackets of \eq{evol4} yields in the large-$N_c$ limit
\begin{align}\label{evol7}
  & \frac{1}{N_c} \, \left\langle \!\! \left\langle \mbox{tr} \left[
        V^{unp}_{\ul{0}} \, V_{\ul{1}}^{pol \, \dagger} \right]
    \right\rangle \!\! \right\rangle (z) = \frac{1}{N_c} \,
  \left\langle \!\! \left\langle \mbox{tr} \left[ V^{unp}_{\ul{0}} \,
        V_{\ul{1}}^{pol \, \dagger} \right] \right\rangle \!\!
  \right\rangle_0 (z) + \frac{\as}{2 \pi^2} \int\limits_{z_i}^z
  \frac{d z'}{z'} \, \int\limits_{\rho'^2} \frac{d^2 x_{2}}{x_{21}^2}
  \, \theta (x_{10} - x_{21}) \notag \\ & \times \left\{ \frac{1}{N_c}
    \, \left\langle \!\! \left\langle \mbox{tr} \left[
          V^{unp}_{\ul{0}} \, V^{pol \, \dagger}_{\ul{2}} \right]
      \right\rangle \!\! \right\rangle \, \left\langle \!\!
      \left\langle \mbox{tr} \left[ V^{unp}_{\ul{2}} \, V^{unp \,
            \dagger}_{\ul{1}} \right] \right\rangle \!\! \right\rangle
  \right.  + \frac{1}{N_c} \, \left\langle \!\!
    \left\langle \mbox{tr} \left[ V^{unp}_{\ul{0}} \, V^{unp \,
          \dagger}_{\ul{2}} \right] \right\rangle \!\! \right\rangle
  \, \left\langle \!\! \left\langle \mbox{tr} \left[ V^{pol}_{\ul{2}}
        \, V^{unp \, \dagger}_{\ul{1}} \right] \right\rangle \!\!
  \right\rangle \notag \\ & + \left. \frac{1}{N_c} \left\langle \!\!
      \left\langle \mbox{tr} \left[ V^{unp}_{\ul{0}} \, V^{unp \,
            \dagger}_{\ul{2}} \right] \right\rangle \!\! \right\rangle
    \, \left\langle \!\! \left\langle \mbox{tr} \left[
          V^{unp}_{\ul{2}} \, V^{pol \, \dagger}_{\ul{1}} \right]
      \right\rangle \!\! \right\rangle - \left\langle \!\!
      \left\langle \mbox{tr} \left[ V^{unp}_{\ul{0}} \,
          V_{\ul{1}}^{pol \, \dagger} \right] \right\rangle \!\!
    \right\rangle \right\} (z').
\end{align}

Let us rewrite this result in terms of the polarized dipole amplitude
\begin{align}\label{Gdef0}
  G_{10} (z) \equiv \frac{1}{2 N_c} \, \Big\langle \!\! \Big\langle
  \mbox{tr} \left[ V^{unp}_{\ul{0}} \, V_{\ul{1}}^{pol \, \dagger}
  \right] \Big\rangle \!\! \Big\rangle (z) + \frac{1}{2 N_c} \,
  \Big\langle \!\! \Big\langle \mbox{tr} \left[ V^{pol}_{\ul{1}} \,
    V_{\ul{0}}^{unp \, \dagger} \right] \Big\rangle \!\! \Big\rangle
  (z)
\end{align}
and the standard (albeit symmetrized) unpolarized dipole $S$-matrix
\begin{align}\label{Sdef0}
  S_{01} (z) = \frac{1}{2 N_c} \, \Big\langle \!\! \Big\langle
  \mbox{tr} \left[ V^{unp}_{\ul{0}} \, V_{\ul{1}}^{unp \, \dagger}
  \right] \Big\rangle \!\! \Big\rangle (z) + \frac{1}{2 N_c} \,
  \Big\langle \!\! \Big\langle \mbox{tr} \left[ V^{unp}_{\ul{1}} \,
    V_{\ul{0}}^{unp \, \dagger} \right] \Big\rangle \!\! \Big\rangle
  (z) = \frac{1}{N_c} \, \Big\langle \!\! \Big\langle \mbox{tr} \left[
    V^{unp}_{\ul{0}} \, V_{\ul{1}}^{unp \, \dagger} \right]
  \Big\rangle \!\! \Big\rangle (z),
\end{align}
where we assumed that 
\begin{align}\label{unp_tr}
  \mbox{tr} \left[ V^{unp}_{\ul{0}} \, V^{unp \, \dagger}_{\ul{1}}
  \right] = \mbox{tr} \left[ V^{unp}_{\ul{1}} \, V^{unp \,
      \dagger}_{\ul{0}} \right],
\end{align}
which is indeed true for the unpolarized LLA evolution with
leading-order (in powers of the color charge density of the target)
$C$-even initial conditions
\cite{Mueller:1994rr,Mueller:1994jq,Mueller:1995gb,Balitsky:1996ub,Balitsky:1998ya,Kovchegov:1999yj,Kovchegov:1999ua,Jalilian-Marian:1997dw,Jalilian-Marian:1997gr,Iancu:2001ad,Iancu:2000hn}.

Adding to \eq{evol7} its hermitean conjugate with $0 \leftrightarrow
1$ we get
\begin{align}\label{evol77}
  G_{10} (z) = G_{10}^{(0)} (z)+ \frac{\as \, N_c}{2 \pi^2}
  \int\limits_{z_i}^z \frac{d z'}{z'} \, \int\limits_{\rho'^2}
  \frac{d^2 x_{2}}{x_{21}^2} \, \theta (x_{10} - x_{21}) & \left[
    2 \, \Gamma_{02, \, 21} (z') \, S_{21} (z') + 2\, G_{21} (z') \, S_{02}
    (z') \right. \notag \\ & \left. + \, G_{12} (z') \, S_{02} (z') -
    \Gamma_{01, \, 21} (z') \right],
\end{align}
where we employed the fact that \eq{unp_tr} is valid in LLA and for
standard quasi-classical initial conditions (that is, we neglected the
odderon contributions).

\begin{figure}[h!]
\centering
\includegraphics[width= \textwidth]{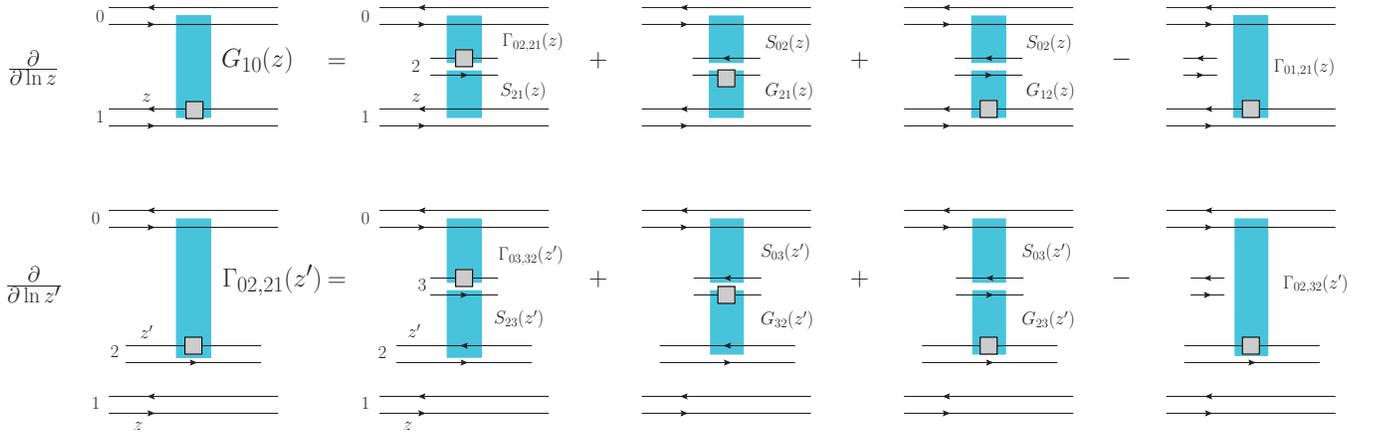}
\caption{Large-$N_c$ helicity evolution for the polarized dipole
  amplitude $G$ and the neighbor dipole amplitude $\Gamma$. As before,
  for pictorial simplicity we do not show the contributions of the
  initial condition terms. Double lines denote gluons at large
  $N_c$. Only one of the virtual diagrams is shown (last diagram in
  each line): virtual corrections to the right of the shock wave are
  implied, but not shown explicitly. }
\label{Large-N_Evol}
\end{figure}

We had to double the coefficients of the first two terms on the
right-hand side of \eqref{evol77} to account for the fact that, in the
large-$N_c$ limit, the factor associated with the polarized gluon
emission by the gluon is four times larger than the factor associated
with the polarized gluon emission by a quark (see e.g. the right
column of \eq{Matrix}): only half of that difference is due to the
color factors.

The evolution equation \eqref{evol77} is illustrated in the upper
panel of \fig{Large-N_Evol}, where the gluon lines in the large-$N_c$
limit are represented by the double quark lines, and the lack of
quark-gluon vertices denotes summation over all gluon connections from
Figs.~\ref{es} and \ref{esG} \cite{Mueller:1994rr}.

\begin{figure}[h!]
\centering
\includegraphics[width= 0.65 \textwidth]{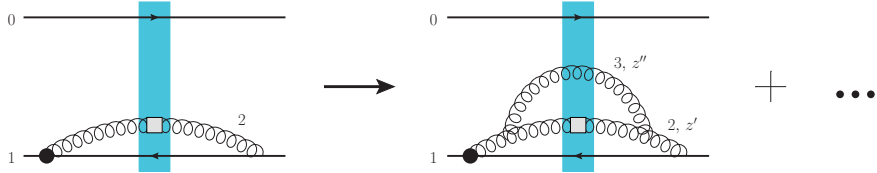}
\caption{An illustration of evolution of the neighbor dipole.}
\label{gamma_fig}
\end{figure}

Equation \eqref{evol77} contains a new object - the ``neighbor''
dipole $\Gamma_{02, \, 21} (z')$. Note the following peculiarity of
the dipole $02$ formed in the first two diagrams on the right hand
side of the diagrammatic evolution equation in \fig{es}: to be DLA,
the subsequent evolution in that dipole has to ``know'' about the
dipole $21$. The reason for that is illustrated in \fig{gamma_fig},
where, in the right panel, we show a sample diagram contributing to
one step of the evolution of the dipole $02$. Note that ordering
\eqref{eq:cond2_coord} implies that we have to have
\begin{align}\label{cond22}
x_{21}^2 \, z' \gg x_{32}^2 \, z''. 
\end{align}
Just like in \eq{evol77}, the dipole $02$ evolution is also cut off by
$x_{20}$ in the IR, $x_{32} \ll x_{20}$. We see that for $x_{20}^2 >
x_{21}^2 \, z'/z''$ the condition \eqref{cond22} becomes more
constraining than $x_{32} \ll x_{20}$. We conclude that the evolution
of the dipole $02$ ``knows'' about the dipole $21$. This violates the
naive dipole independence at large $N_c$. The issue is that while
dipole $02$ contains a polarized gluon at $x_2$, it does not carry the
large transverse momentum needed for generating dipole $21$. It is
clear from \fig{gamma_fig} that the origin of dipoles $02$ and $21$ is
different topologically too. Note that the virtual (last) term in
\eq{evol77} can be shown to bring in a neighbor dipole amplitude
$\Gamma_{01, \, 21} (z')$.

To remedy the problem of such neighbor dipole amplitudes, let us write
down an evolution equation for $\Gamma_{02, \, 21} (z')$:
\begin{align}
  \Gamma_{02, \, 21} (z') = \Gamma^{(0)}_{02, \, 21} (z') + \frac{\as
    \, N_c}{2 \pi} \int\limits_{z_i}^{z'} \frac{d z''}{z''} \,
  \int\limits_{{\rho''}^2}^{\mbox{min} \left\{ x_{02}^2 , x_{21}^2 \,
      z'/z'' \right\}} \frac{d x_{32}^2}{x_{32}^2} \, & \left[ 2 \,
    \Gamma_{03, \, 32} (z'') \, S_{23} (z'') + 2 \, G_{32} (z'') \,
    S_{03} (z'') \right. \notag \\ & \left. + \, G_{23} (z'') \,
    S_{03} (z'') - \Gamma_{02, \, 32} (z'')
  \right]. \label{Gamma_evol}
\end{align}
Here ${\rho''}^2 = 1/(z'' \, s)$. Equation \eqref{Gamma_evol} is
illustrated diagrammatically in the lower panel of \fig{Large-N_Evol}.

Equations \eqref{evol77} and \eqref{Gamma_evol} give a coupled closed
system of equations which need to be solved to find the
energy-dependence of $G$. We leave the solution of these equations for
the future work: it may be that, just like with the evolution
considered by BER, only numerical solution of these equations is
possible.

Equations~\eqref{evol77} and \eqref{Gamma_evol} combine DLA evolution
of helicity distributions with the LLA evolution of $S$, akin to
Eq.~(43) of \cite{Itakura:2003jp}. Note that in the strict DLA
approximation $S=1$. By keeping $S$ in Eqs.~\eqref{evol77} and
\eqref{Gamma_evol} we are combining the DLA evolution for the
polarized dipole operator with the LLA saturation corrections in a
single equation. Again strictly-speaking such an approximation is not
justified, and only a complete LLA calculation (that is,
next-to-leading order (NLO) helicity evolution calculation) would show
whether we have the right to keep LLA saturation corrections
here. Still we argue that such an NLO calculation is not going to
change the operator structure of \eq{fund_evol}, and, hence
\eq{fund_evol} would probably remain valid, while augmented by the LLA
helicity evolution term. Therefore, as was also argued in
\cite{Itakura:2003jp}, it is probably justified to mix DLA helicity
evolution with the LLA unpolarized evolution in Eqs.~\eqref{evol77}
and \eqref{Gamma_evol}. Moreover, if one models the target as a
nucleus with atomic number $A \gg 1$, then even in the strict DLA
limit one can justify keeping the unpolarized Wilson lines $S$ as a
resummation of the multiple rescattering parameter $\as^2 A^{1/3} \sim
1$ \cite{KovchegovLevin}.

In the strict DLA limit we can simplify Eqs.~\eqref{evol77} and
\eqref{Gamma_evol} then as follows. The unpolarized dipoles do not
have any DLA evolution (in our $\as \, \ln^2 s$ resummation
parameter). Hence, with the DLA accuracy, we put $S =1$ obtaining
\begin{subequations}\label{evol88}
\begin{align}
  G_{01} (z) & = G_{01}^{(0)} (z)+ \frac{\as \, N_c}{2 \pi}
  \int\limits_{z_i}^z \frac{d z'}{z'} \, \int\limits_{\rho'^2}^{x_{10}^2}
  \frac{d x_{21}^2}{x_{21}^2} \, 
  \left[  \Gamma_{02, \, 21} (z')  + 3 \, G_{21} (z')  \right], \\
  \Gamma_{02, \, 21} (z') & = \Gamma^{(0)}_{02, \, 21} (z') +
  \frac{\as \, N_c}{2 \pi} \int\limits_{z_i}^{z'} \frac{d z''}{z''}
  \, \int\limits_{{\rho''}^2}^{\mbox{min} \left\{ x_{02}^2 , x_{21}^2
      \, z'/z'' \right\}} \frac{d x_{32}^2}{x_{32}^2} \, \left[
    \Gamma_{03, \, 32} (z'') + 3 \, G_{23} (z'') \right]. \label{Glin}
\end{align}
\end{subequations}
Here we have employed the assumption $G_{21} = G_{12}$ which is valid
for a longitudinally polarized target (which possesses no preferred
transverse direction, if one neglects transverse gradients of the
target density). The solution of the coupled equations \eqref{evol88}
would give one the intercept of small-$x$ helicity evolution in the
large-$N_c$ limit.


\subsection{Large-$N_c \, \&  \, N_f$ Limit}
\label{sec:large_ncnf}

Consider now the case when both $N_c$ and $N_f$ are comparably large,
such that we are keeping powers of $\as \, N_c =$~const$\ll 1$ and
$\as \, N_f =$~const$' \ll 1$, while neglecting the subleading powers
of $1/N_c$ and $1/N_f$.

\begin{figure}[h!]
\centering
\includegraphics[width= 0.35 \textwidth]{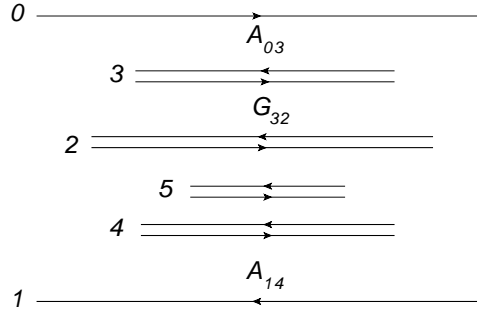}
\caption{Different types of dipoles in our large-$N_c \, \& \, N_f$
  resummation. For illustration purposes, any line can be polarized.}
\label{large_Nc}
\end{figure}

In addition to $G_{10} (z)$ defined in \eq{Gdef0} above, which is made
out of quark and anti-quark lines of gluons (with $x_1$ line
polarized), let us define
\begin{align}\label{QGdef}
  A_{10} (z) = \frac{1}{2 N_c} \, \Big\langle \!\! \Big\langle
  \mbox{tr} \left[ V^{unp}_{\ul{0}} \, V_{\ul{1}}^{pol \, \dagger}
  \right] \Big\rangle \!\! \Big\rangle (z) + \frac{1}{2 N_c} \,
  \Big\langle \!\! \Big\langle \mbox{tr} \left[ V^{pol}_{\ul{1}} \,
    V_{\ul{0}}^{unp \, \dagger} \right] \Big\rangle \!\! \Big\rangle
  (z)
\end{align}
with $x_1$ being a true quark or anti-quark polarized line and $x_0$
being the (anti-)quark line of the gluon, and
\begin{align}\label{Qdef}
  Q_{10} (z) = \frac{1}{2 N_c} \, \Big\langle \!\! \Big\langle
  \mbox{tr} \left[ V^{unp}_{\ul{0}} \, V_{\ul{1}}^{pol \, \dagger}
  \right] \Big\rangle \!\! \Big\rangle (z) + \frac{1}{2 N_c} \,
  \Big\langle \!\! \Big\langle \mbox{tr} \left[ V^{pol}_{\ul{1}} \,
    V_{\ul{0}}^{unp \, \dagger} \right] \Big\rangle \!\! \Big\rangle
  (z)
\end{align}
with both $x_0$ and $x_1$ being true quark and anti-quark lines and
$x_1$ polarized. The definition of $A$ is illustrated in
\fig{large_Nc}.

An analogue of \eq{evol7}, but now without neglecting the second term
in the curly brackets of \eq{evol4}, in the large-$N_c \, \& \, N_f$
limit is
\begin{align}\label{evol12}
  & \frac{1}{N_c} \, \left\langle \!\! \left\langle \mbox{tr} \left[ V^{unp}_{\ul{0}} \, V_{\ul{1}}^{pol \, \dagger} \right] \right\rangle \!\! \right\rangle (z) = \frac{1}{N_c} \, \left\langle \!\! \left\langle \mbox{tr} \left[ V^{unp}_{\ul{0}} \, V_{\ul{1}}^{pol \, \dagger} \right] \right\rangle \!\! \right\rangle_0 (z) + \frac{\as}{2 \pi^2} \int\limits_{z_i}^z \frac{d z'}{z'} \, \int\limits_{\rho'^2} \frac{d^2 x_{2}}{x_{21}^2} \, \theta (x_{10} - x_{21})  \notag \\ & \times \left\{  \frac{1}{N_c} \, \left\langle \!\! \left\langle \mbox{tr} \left[ V^{unp}_{\ul{0}} \, V^{pol \, \dagger}_{\ul{2}} \right] \right\rangle \!\! \right\rangle \, \left\langle \!\! \left\langle \mbox{tr} \left[ V^{unp}_{\ul{2}} \, V^{unp \, \dagger}_{\ul{1}} \right] \right\rangle \!\! \right\rangle \right.  + \frac{1}{N_c} \, \left\langle \!\! \left\langle \mbox{tr} \left[ V^{unp}_{\ul{0}} \, V^{unp \, \dagger}_{\ul{2}} \right] \right\rangle \!\! \right\rangle \, \left\langle \!\! \left\langle \mbox{tr} \left[ V^{pol}_{\ul{2}} \, V^{unp \, \dagger}_{\ul{1}} \right] \right\rangle \!\! \right\rangle \notag \\ & + \left. \frac{1}{N_c}  \left\langle \!\! \left\langle \mbox{tr} \left[ V^{unp}_{\ul{0}} \, V^{unp \, \dagger}_{\ul{2}} \right] \right\rangle \!\! \right\rangle \, \left\langle \!\! \left\langle \mbox{tr} \left[ V^{unp}_{\ul{2}} \, V^{pol \, \dagger}_{\ul{1}} \right]  \right\rangle \!\! \right\rangle  - \left\langle \!\! \left\langle \mbox{tr} \left[ V^{unp}_{\ul{0}} \, V_{\ul{1}}^{pol \, \dagger} \right] \right\rangle \!\! \right\rangle  \right\}  (z') \notag \\
  & + \frac{\as}{4 \pi^2} \int\limits_{z_i}^z \frac{d z'}{z'} \,
  \int\limits_{\rho'^2} \frac{d^2 x_{2}}{x_{21}^2} \, \theta (x_{10}^2
  z - x_{21}^2 z') \, \frac{1}{N_c} \, \left\langle \!\! \left\langle
      \mbox{tr} \left[ V^{unp}_{\ul{0}} \, V^{unp \, \dagger}_{\ul{1}}
      \right] \right\rangle \!\! \right\rangle \, \left\langle \!\!
    \left\langle \mbox{tr} \left[ V^{unp}_{\ul{1}} \, V^{pol \,
          \dagger}_{\ul{2}} \right] \right\rangle \!\! \right\rangle
  (z').
\end{align}

\begin{figure}[htb]
\centering
\includegraphics[width= \textwidth]{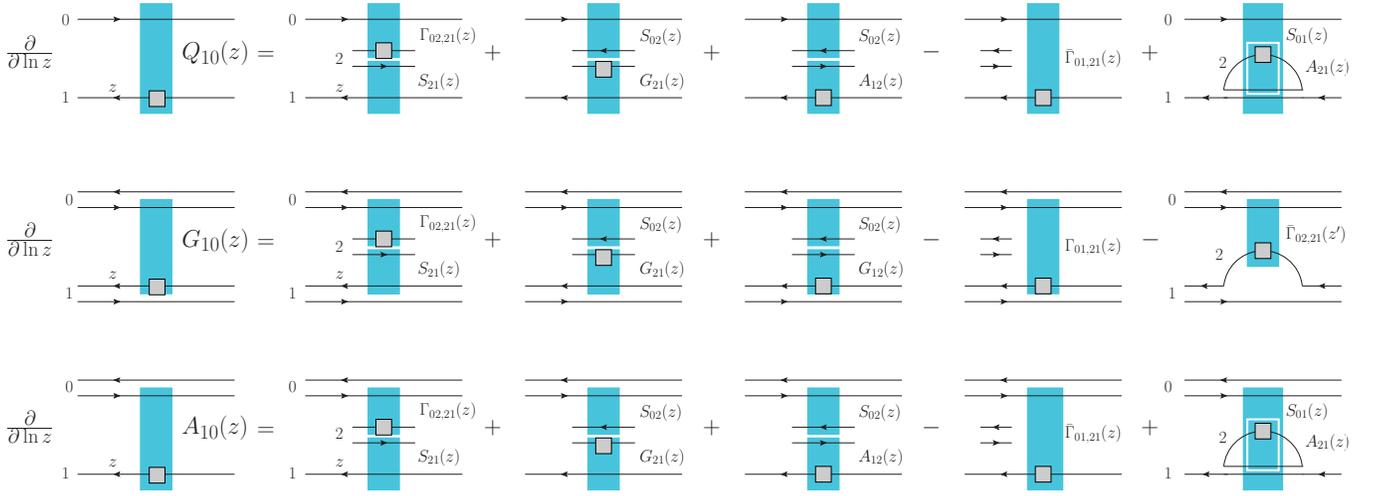}
\caption{Large-$N_c \, \& \, N_f$ helicity evolution for the polarized
  dipole amplitudes $Q$, $G$ and $A$.}
\label{Large-NcNf_Evol1}
\end{figure}

Adding a complex conjugate to \eq{evol12} and dividing by 2 to get
$Q_{10} (z)$ on the left we obtain
\begin{align}\label{Qevol}
  & Q_{10} (z) = Q_{10}^{(0)} (z) + \frac{\as \, N_c}{2 \pi^2}
  \int\limits_{z_i}^z \frac{d z'}{z'} \, \int\limits_{\rho'^2} \frac{d^2 x_{2}}{x_{21}^2} \, \theta (x_{10} - x_{21}) \notag \\ & \times \left[ S_{21} (z') \, \Gamma_{02, \, 21} (z') + S_{02} (z') \, G_{21} (z')  + S_{02} (z') \, A_{12} (z') - {\bar \Gamma}_{01, \, 21} (z') \right] \notag \\
  & + \frac{\as \, N_c}{4 \pi^2} \int\limits_{z_i}^z \frac{d z'}{z'}
  \, \int\limits_{\rho'^2} \frac{d^2 x_{2}}{x_{21}^2} \, \theta
  (x_{10}^2 z - x_{21}^2 z') \, S_{01} (z') \, A_{21} (z').
\end{align}
Equation \eqref{Qevol} is illustrated diagrammatically in the first
line of \fig{Large-NcNf_Evol1}, where again we do not show the initial
condition term for simplicity.

\begin{figure}[h!]
\centering
\includegraphics[width= 0.5 \textwidth]{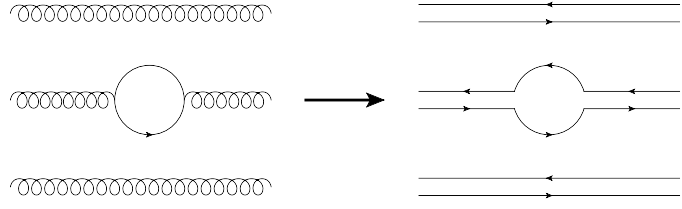}
\caption{Diagram illustrating the origin of the last term in
  \eq{Gevol}.}
\label{large_Nf}
\end{figure}

Now, we turn to the evolution for $G_{10} (z)$. Adding to \eq{evol12}
its complex conjugate and dividing by two gives
\begin{align}\label{Gevol}
  & G_{10} (z) = G_{10}^{(0)} (z) + \frac{\as \, N_c}{2 \pi^2}
  \int\limits_{z_i}^z \frac{d z'}{z'} \, \int\limits_{\rho'^2}
  \frac{d^2 x_{2}}{x_{21}^2} \, \theta (x_{10} - x_{21}) \notag \\ &
  \times \left[ 2\, S_{21} (z') \, \Gamma_{02, \, 21} (z') + 2\,
    S_{02} (z') \, G_{21} (z') + S_{02} (z') \, G_{12} (z') -
    \Gamma_{01, \, 21} (z') \right] \notag \\ & - \frac{\as \, N_f}{4
    \pi^2} \int\limits_{z_i}^z \frac{d z'}{z'} \,
  \int\limits_{\rho'^2} \frac{d^2 x_{2}}{x_{21}^2} \, \theta (x_{10}^2
  z - x_{21}^2 z') \, {\bar \Gamma}_{02, \, 21} (z'),
\end{align}
where we have ``by hand'' removed the last term in \eq{evol12} as an
impossible one in the evolution of a fundamental dipole made out of
quark and anti-quark lines of two gluons. We have also added ``by
hand'' the $N_f$-term arising due to $G \to q \bar q$ splitting. The
term comes from the gluon dipole evolution \eqref{Gevol3} and is
illustrated in \fig{large_Nf}: one has to take the large-$N_c \, \& \,
N_f$ limit of \eqref{Gevol3} to obtain this term in \eq{Gevol}. Note a
new object, ${\bar \Gamma}_{02, \, 21}$, which is the neighbor dipole
amplitude with line $2$ being an actual polarized quark (or
anti-quark), and, unlike in $\Gamma_{02, \, 21}$, not a quark (or
anti-quark) line of a large-$N_c$ gluon. Equation \eqref{Gevol} is
illustrated diagrammatically in the second line of
\fig{Large-NcNf_Evol1}.

Finally, the evolution for $A_{01} (z)$ reads
\begin{align}\label{Aevol}
  & A_{10} (z) = A_{10}^{(0)} (z) + \frac{\as \, N_c}{2 \pi^2}
  \int\limits_{z_i}^z \frac{d z'}{z'} \, \int\limits_{\rho'^2}
  \frac{d^2 x_{2}}{x_{21}^2} \, \theta (x_{10} - x_{21}) \notag \\ &
  \times \left[ S_{21} (z') \, \Gamma_{02, \, 21} (z') + S_{02} (z')
    \, G_{21} (z') + S_{02} (z') \, A_{12} (z') - {\bar \Gamma}_{01,
      \, 21} (z') \right] \notag \\ & +\frac{\as \, N_c}{4 \pi^2}
  \int\limits_{z_i}^z \frac{d z'}{z'} \, \int\limits_{\rho'^2}
  \frac{d^2 x_{2}}{x_{21}^2} \, \theta (x_{10}^2 z - x_{21}^2 z') \,
  S_{01} (z') \, A_{21} (z'),
\end{align}
as can also be obtained from \eq{evol12}. It is depicted in the last
line of \fig{Large-NcNf_Evol1}.

\begin{figure}[htb]
\centering
\includegraphics[width= \textwidth]{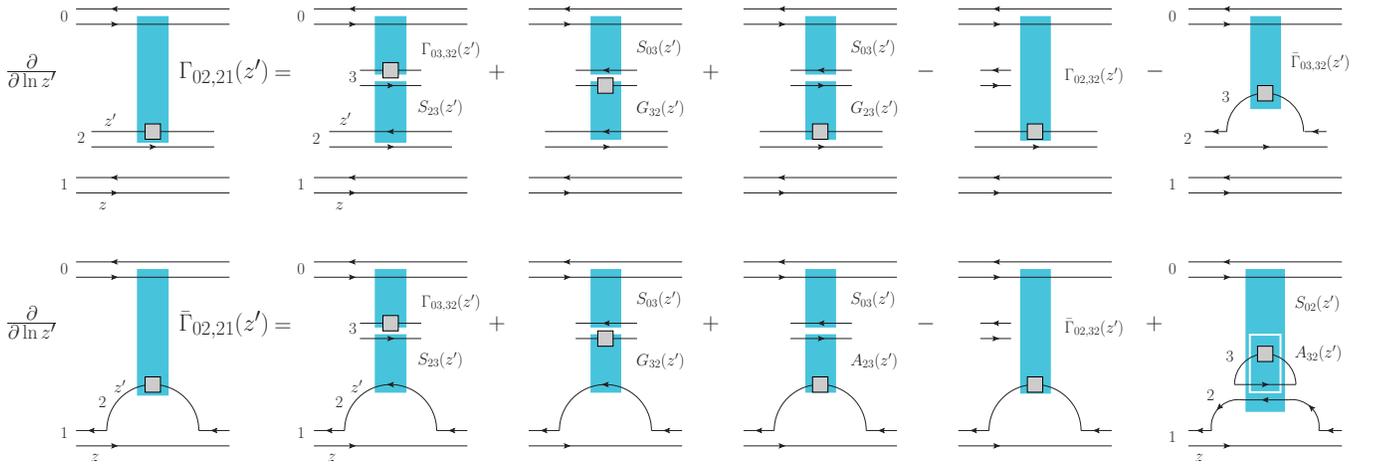}
\caption{Large-$N_c \, \& \, N_f$ helicity evolution for the polarized
  neighbor dipole amplitudes $\Gamma$ and $\bar \Gamma$.}
\label{Large-NcNf_Evol2}
\end{figure}

Note that \eq{Gamma_evol} also has to be modified yielding
\begin{align}
  & \, \Gamma_{02, \, 21} (z') = \Gamma^{(0)}_{02, \, 21} (z') +
  \frac{\as \, N_c}{2 \pi} \int\limits_{z_i}^{z'} \frac{d z''}{z''} \,
  \int\limits_{{\rho''}^2}^{\mbox{min} \left\{ x_{02}^2 , x_{21}^2 \,
      z'/z'' \right\}} \frac{d x_{32}^2}{x_{32}^2} \, \notag \\ &
  \times \left[ 2\, \Gamma_{03, \, 32} (z'') \, S_{23} (z'') + 2 \,
    G_{32} (z'') \, S_{03} (z'') + G_{23}
    (z'') \, S_{03} (z'') - \Gamma_{02, \, 32} (z'') \right] \notag \\
  & - \frac{\as \, N_f}{4 \pi} \int\limits_{z_i}^{z'} \frac{d
    z''}{z''} \, \int\limits_{\rho''^2}^{x_{21}^2 \, z'/z''} \frac{d
    x_{32}^2}{x_{32}^2} \, {\bar \Gamma}_{03, \, 32}
  (z'). \label{Gamma_evol2}
\end{align}
We also need an equation for $\bar \Gamma$:
\begin{align}
  & \, {\bar \Gamma}_{02, \, 21} (z') = {\bar \Gamma}^{(0)}_{02, \,
    21} (z') + \frac{\as \, N_c}{2 \pi} \int\limits_{z_i}^{z'} \frac{d
    z''}{z''} \, \int\limits_{{\rho''}^2}^{\mbox{min} \left\{ x_{02}^2
      , x_{21}^2 \, z'/z'' \right\}} \frac{d x_{32}^2}{x_{32}^2} \,
  \notag \\ & \times \left[ \Gamma_{03, \, 32} (z'') \, S_{23} (z'') +
    G_{32} (z'') \, S_{03} (z'') + A_{23} (z'') \, S_{03} (z'') -
    {\bar \Gamma}_{02, \, 32} (z'') \right] \notag \\ & + \frac{\as \,
    N_c}{4 \pi} \int\limits_{z_i}^{z'} \frac{d z''}{z''} \,
  \int\limits_{\rho''^2}^{x_{21}^2 \, z'/z'' } \frac{d
    x_{32}^2}{x_{32}^2} \, S_{02} (z') \, A_{32}
  (z'). \label{Gamma_evol3}
\end{align}
Both of these equations follow from Eqs.~\eqref{fund_evol} and
\eqref{Gevol3}. They are diagrammatically illustrated in
\fig{Large-NcNf_Evol2}.

Equations~\eqref{Qevol}, \eqref{Gevol}, \eqref{Aevol},
\eqref{Gamma_evol2}, and \eqref{Gamma_evol3} are the large-$N_c \, \&
\, N_f$ evolution equations which are DLA in polarization-dependent
terms, but also include LLA saturation corrections through the
$S$-matrices. They are the main result of this Subsection.

In the pure DLA limit we linearize all these equations by putting
$S=1$ in them (we again assume that $G_{01} = G_{10}$, which is true
for a large, longitudinally polarized target):
\begin{subequations}\label{Q_evol_lin}
\begin{align}\label{Qevol2}
  Q_{01} (z) = & \, Q_{01}^{(0)} (z) + \frac{\as \, N_c}{2 \pi^2}
  \int\limits_{z_i}^z \frac{d z'}{z'} \, \int\limits_{\rho'^2} \frac{d^2 x_{2}}{x_{21}^2} \, \theta (x_{10} - x_{21}) \, \left[ G_{12} (z') +  \Gamma_{02, \, 21} (z') + A_{21} (z') - {\bar \Gamma}_{01, \, 21} (z') \right] \notag \\
  & + \frac{\as \, N_c}{4 \pi^2} \int\limits_{z_i}^z \frac{d z'}{z'}
  \, \int\limits_{\rho'^2} \frac{d^2 x_{2}}{x_{21}^2} \, \theta
  (x_{10}^2 z - x_{21}^2 z') \, A_{21} (z'), \\
  G_{10} (z) = & \, G_{10}^{(0)} (z) + \frac{\as \, N_c}{2 \pi^2} \int\limits_{z_i}^z \frac{d z'}{z'} \, \int\limits_{\rho'^2} \frac{d^2 x_{2}}{x_{21}^2} \, \theta (x_{10} - x_{21})  \, \left[ \Gamma_{02, \, 21} (z') + 3 \, G_{12} (z') \right] \notag \\ & - \frac{\as \, N_f}{4 \pi^2} \int\limits_{z_i}^z \frac{d z'}{z'} \, \int\limits_{\rho'^2} \frac{d^2 x_{2}}{x_{21}^2} \,  \theta (x_{10}^2 z - x_{21}^2 z') \, {\bar \Gamma}_{02, \, 21} (z'), \label{GGevol2} \\
  A_{01} (z) = & \, A_{01}^{(0)} (z) + \frac{\as \, N_c}{2 \pi^2}
  \int\limits_{z_i}^z \frac{d z'}{z'} \, \int\limits_{\rho'^2}
  \frac{d^2 x_{2}}{x_{21}^2} \, \theta (x_{10} - x_{21}) \left[ G_{12}
    (z') + \Gamma_{02, \, 21} (z') + A_{21} (z') - {\bar \Gamma}_{01,
      \, 21} (z') \right] \notag \\ & +\frac{\as \, N_c}{4 \pi^2}
  \int\limits_{z_i}^z \frac{d z'}{z'} \, \int\limits_{\rho'^2}
  \frac{d^2 x_{2}}{x_{21}^2} \, \theta (x_{10}^2 z - x_{21}^2 z') \,
  A_{12} (z').
\end{align}
\end{subequations}
The linearized equations for $\Gamma$ and $\bar \Gamma$ in the
large-$N_c \, \& \, N_f$ limit become
\begin{subequations}\label{Gam_evol_lin}
\begin{align}\label{Gamma_evol4}
  \Gamma_{02, \, 21} (z') = & \, \Gamma^{(0)}_{02, \, 21} (z') +
  \frac{\as \, N_c}{2 \pi} \int\limits_{z_i}^{z'} \frac{d z''}{z''} \, \int\limits_{{\rho''}^2}^{\mbox{min} \left\{ x_{02}^2 , x_{21}^2 \, z'/z'' \right\}} \frac{d x_{32}^2}{x_{32}^2} \, \left[ \Gamma_{03, \, 32} (z'')  + 3 \, G_{23} (z'') \right] \notag \\ & - \frac{\as \, N_f}{4 \pi} \int\limits_{z_i}^{z'} \frac{d z''}{z''} \, \int\limits_{\rho''^2}^{x_{21}^2 \, z'/z''} \frac{d x_{32}^2}{x_{32}^2} \,  {\bar \Gamma}_{03, \, 32} (z'), \\
  {\bar \Gamma}_{02, \, 21} (z') = & \, {\bar \Gamma}^{(0)}_{02, \,
    21} (z') + \frac{\as \, N_c}{2 \pi} \int\limits_{z_i}^{z'} \frac{d
    z''}{z''} \, \int\limits_{{\rho''}^2}^{\mbox{min} \left\{ x_{02}^2
      , x_{21}^2 \, z'/z'' \right\}} \frac{d x_{32}^2}{x_{32}^2} \,
  \left[ \Gamma_{03, \, 32} (z'') + G_{23} (z'') + A_{23} (z'') -
    {\bar \Gamma}_{02, \, 32} (z'') \right] \notag \\ & + \frac{\as \,
    N_c}{4 \pi} \int\limits_{z_i}^{z'} \frac{d z''}{z''} \,
  \int\limits_{\rho''^2}^{x_{21}^2 \, z'/z'' } \frac{d
    x_{32}^2}{x_{32}^2} \, A_{32} (z').
\end{align}
\end{subequations}

Clearly in the large-$N_c$ / fixed-$N_f$ limit the linearized
equations for $G_{01} (z)$ and $\Gamma_{02, \, 21} (z')$ become a
closed system of equations \eqref{evol88} again, as employed in the
previous Subsection. Since our final observable, quark helicity TMD or
hPDF, is related to $Q$, for the large-$N_c$ limit to be relevant, $G$
should dominate (or at least be comparable to) $A$.

The linearized equations \eqref{Q_evol_lin} and \eqref{Gam_evol_lin},
when solved, should yield the helicity evolution intercept in the
large-$N_c \, \& \, N_f$ limit. This number should be compared to the
all-orders in $N_c \, \& \, N_f$ result of BER
\cite{Bartels:1996wc}. Solution of Eqs.~\eqref{Q_evol_lin} and
\eqref{Gam_evol_lin} is left for the future (probably numerical) work.


\section{Conclusions and Outlook}
\label{sec:conc}

In this paper we have constructed small-$x$ evolution equations
governing the leading $x$ dependence of the fundamental and adjoint
polarized dipole operators \eqref{eq:pol_dip_fund} and
\eqref{eq:pol_dip_adj} given by Eqs.~\eqref{fund_evol} and
\eqref{Gevol3} respectively. The equations resum double logarithms of
$x$ (powers of $\as \, \ln^2 (1/x)$) in helicity evolution, and also
include saturation effects through unpolarized dipoles, to be evolved
with the LLA BK/JIMWLK evolution (resumming powers of $\as \, \ln
(1/x)$).

The equations are not closed, but they do become closed in the two
limits considered in this work: the large-$N_c$ and the large-$N_c \,
\& \, N_f$ limits. In the large-$N_c$ limit the helicity evolution is
given by a closed system of two equations, \eqref{evol77} and
\eqref{Gamma_evol}. In the large-$N_c \, \& \, N_f$ limit the closed
system consists of five equations, \eqref{Qevol}, \eqref{Gevol},
\eqref{Aevol}, \eqref{Gamma_evol2}, and \eqref{Gamma_evol3}.

Solution of these equations, while not straightforward, should be
possible in principle, and is left for future work. The polarized
quark dipole operator gives us energy dependence of the quark helicity
TMD $g_{1L} (x, k_T)$ via \eq{eq:g1L} and of the quark hPDF $\Delta q
(x, Q^2)$ via \eq{eq:Deltaq}. Therefore, it is very important to solve
the helicity evolution equations obtained here, in order to evaluate
the small-$x$ quark contribution to the helicity sum rule and to
address a possible resolution of the spin puzzle.

Indeed higher-order (LLA) corrections to our DLA results need to be
calculated as well. While the ladder approximation is only a rough
estimate of the result, it is perhaps troubling that for realistic
$\as$ the quark hPDF grows with $x$ faster than the unpolarized
structure functions, as one can see from \eq{asymptotics3}. If this
steep growth is also the case with the full DLA solution, it is clear
that higher-order LLA corrections must be numerically large for $\as
\approx 0.3$ to reduce the growth of the
intercept. It is, therefore, possible that LLA corrections are very
important for the present phenomenology.

Another important possible source of the intercept reduction is
saturation effects. Indeed saturation effects are defined as a
slowdown of an observable's growth with increasing energy or
decreasing $x$. The impact of saturation on the DLA Reggeon evolution,
which is similar to our helicity evolution considered here, was
studied in \cite{Itakura:2003jp}. It was shown that indeed saturation
effects reduce the Reggeon intercept.

There is another consideration which makes it important to include
saturation effects in the problem at hand. In all of the above DLA
evolution equations we integrate over $x_{21}^2$ up to $x_{10}^2 \,
z/z'$. (For definitiveness let us consider \eq{fund_evol}.) Since
$\Lambda^2/s = z_i < z' < z<1$ we get the absolute upper bound on the
$x_{21}$ integral,
\begin{align}
  x_{21}^2 < x_{10}^2 \, \frac{z}{z'} < x_{10}^2 \,
  \frac{s}{\Lambda^2}.
\end{align} 
Since $x_{10} \sim 1/k_T$ with $k_T$ the typical transverse momentum
in the problem, and $s \, x_{10}^2 \sim s/k_T^2 \gg 1$, we have
\begin{align}\label{IRconcern}
  x_{10}^2 \, \frac{s}{\Lambda^2} \gg \frac{1}{\Lambda^2}
\end{align} 
and the upper bound of the $x_{21}^2$-integration may end up in the
non-perturbative IR region. Throughout the calculations in this paper
we simply assumed that everything is perturbative, and never worried
about the transverse coordinate (or momentum) integration
regions. However, this is indeed potentially dangerous. While the
equations \eqref{evol77} and \eqref{Gamma_evol} appear to be
insensitive to the deep IR region, Eqs.~\eqref{Qevol}, \eqref{Gevol},
\eqref{Aevol}, \eqref{Gamma_evol2}, and \eqref{Gamma_evol3} need
saturation effects to stay IR safe. As usual, saturation effects come
in through the unpolarized dipole $S$-matrices, effectively cutting
off the IR regions of integration, and justifying our perturbative
approach to the problem. Their impact on the intercept of helicity
evolution is likely to be important and needs to be studied in detail
in the future.


\section*{Acknowledgments}

The authors are grateful to Raju Venugopalan for his support and
encouragement throughout the project and a critical reading of the
manuscript.  The authors would also like to thank Ian Balitsky,
Joachim Bartels, Giovanni Chirilli, and Al Mueller for informative
discussions. YK would like to thank Andreas Metz and Kirill Tuchin for
encouraging him to think about the subject in the course of the past
several years. This material is based upon work supported by the
U.S. Department of Energy, Office of Science, Office of Nuclear
Physics under Award Number DE-SC0004286 (YK), the RIKEN BNL Research
Center (DP), and DOE Contract No. DE-SC0012704 (MS).  MS receives
additional support from an EIC program development fund from BNL.\\


\appendix

\section{}
\renewcommand{\theequation}{A\arabic{equation}}
  \setcounter{equation}{0}
\label{A}

\begin{figure}[htb]
\centering
\includegraphics[width=0.4\textwidth]{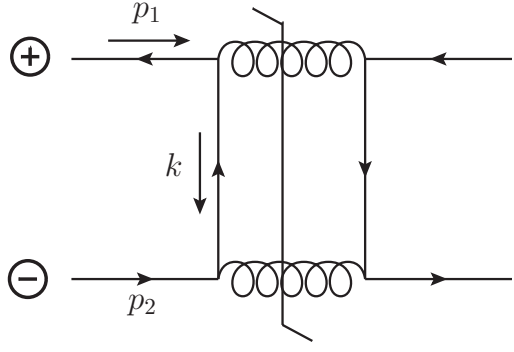}
\caption{The tree-level forward Reggeon exchange amplitude.}
\label{Fig-reg_tree}
\end{figure}
Let us calculate the tree-level (Born) quark exchange amplitude
contributing to the initial conditions of our evolution equations; for
a similar discussion, see Appendix B of \cite{Itakura:2003jp}.  We
consider the annihilation amplitude of an antiquark with momentum
$p_1$ and spin $\sigma_1$ on a quark with momentum $p_2$ and spin
$\sigma_2$ at high energy (Fig.~\ref{Fig-reg_tree}). We will treat all
particles as massless, and we take $p_1$ to be moving along the
light-cone $x^+$ axis and $p_2$ to be moving along the light-cone
$x^-$ axis.  Let us fix the external momenta to be
\begin{align} \label{eq-regkin1}
   p_1^\mu &= (p_1^+, 0^-, \ul{0}) = (zP^+, 0^-, \ul{0})\,, \notag \\
   p_2^\mu &= (0^+, p_2^-, \ul{0})\,,
\end{align}
where $z$ is the longitudinal momentum fraction of the antiquark, with
$zs = (p_1+p_2)^2 = 2 zP^+ p_2^-$.  We also define Sudakov momentum
fractions $\alpha \equiv k^+/p_1^+$ and $\beta \equiv |k^-|/p_2^-$ and
employ the on-shell conditions for the produced gluons to write
\begin{align} \label{eq-regkin2}
   k^- &= p_1^- - (p_1 - k)^- = - \half\frac{k_T^2}{(1-\alpha)p_1^+} \equiv - \beta p_2^-\,, \notag \\
   k^+ &= (p_2 + k)^+ - p_2^+ = \half\frac{k_T^2}{(1-\beta)p_2^-} \equiv \alpha p_1^+.
\end{align}
Regge kinematics corresponds to the limit $k_T^2 \ll zs$ and,
therefore, $\alpha \approx \beta \approx k_T^2 /(zs) \ll 1$.  In this
limit one then has $k^2 = - \alpha \beta zs - k_T^2 \approx - k_T^2$.

The $q \bar{q} \rightarrow GG$ squared amplitude (averaged over colors
of the quark and antiquark and summed over the outgoing gluon spins
$\lambda_1,\lambda_2$) is
\begin{align} \label{eq-regtree2}
    \langle M^2 \rangle &\equiv \frac{1}{N_c^2} \sum_{colors} \sum_{\lambda_1 , \lambda_2} | M |^2 
    \notag\\
      &=
    \frac{C_F^2}{N_c} \frac{g^4}{k_T^4} 
    \left[ \vbar{\sigma_1}(p_1) \gamma_\mu \slashed{k} \gamma_\nu U_{\sigma_2}(p_2) \right]
    \left[ \ubar{\sigma_2}(p_2) \gamma^\nu \slashed{k} \gamma^\mu V_{\sigma_1}(p_1) \right],
\end{align}
where $U_\sigma$ ($V_\sigma$) is a quark (antiquark) spinor with spin
$\sigma$.  Rather than computing the spinor products independently, it
is convenient to isolate the desired polarizations $\sigma_1 ,
\sigma_2$ using the chirality projectors:
\begin{align}
\left[ U_\sigma (p) \right]_\alpha &\left[ \ubar{\sigma} (p) \right]_\beta \rightarrow \left[ \mathcal{P}_\sigma \slashed{p} \right]_{\alpha \beta}\,,
\end{align}
where
\begin{align} \label{eq-chiproj}
   \mathcal{P}_\sigma &\equiv \half \left( 1 + \sigma \gamma^5 \right),
\end{align}
which satisfies
\begin{align}
   \mathcal{P}_\sigma \mathcal{P}_{\sigma'} &= \delta_{\sigma \sigma'} \mathcal{P}_\sigma\,, \notag \\
   \mathcal{P}_\sigma \gamma^\mu &= \gamma^\mu \mathcal{P}_{-\sigma}\,.
\end{align}
For $V_\sigma$ we similarly make the replacement $V_\sigma
\vbar{\sigma} \rightarrow \mathcal{P}_{-\sigma} \slashed{p}$.  Thus
$\mathcal{P}_+ = \mathcal{P}_R$ is the right-handed projector and
$\mathcal{P}_- = \mathcal{P}_L$ is the left-handed projector.  Using
this in \eqref{eq-regtree2} gives
\begin{align} 
   \langle{M^2}\rangle &= \frac{C_F^2}{N_c} \frac{g^4}{k_T^4}
   \Tr [ \mathcal{P}_{-\sigma_1} \slashed{p_1} \gamma_\mu \slashed{k} \gamma_\nu \mathcal{P}_{\sigma_2} \slashed{p_2} \gamma^\nu \slashed{k} \gamma^\mu ]
      \notag \\ &=
   4 \frac{C_F^2}{N_c} \frac{g^4}{k_T^4} \delta_{-\sigma_1 \sigma_2} \Tr[\mathcal{P}_{-\sigma_1} \, \slashed{k} \, \slashed{p_2} \slashed{k} \slashed{p_1}]   
 \notag \\
  &= 
   4 (4\pi)^2 \frac{\alpha_s^2 C_F^2}{N_c} \, \frac{zs}{k_T^2} \, \delta_{-\sigma_1 \sigma_2} \label{eq-regtree5} .
\end{align}
The differential cross-section for Reggeon exchange at this order is
then
\begin{align} \label{eq-regtree88}
   \frac{d\sigma^{Born}_{Reg}} {d^2k} &= \frac{1}{16\pi^2 z^2s^2} \langle M^2 \rangle
  = 4 \frac{\alpha_s^2 C_F^2}{N_c} \, \frac{1}{zs} \, \frac{1} {k_T^2} \delta_{-\sigma_1 \sigma_2}\,,
\end{align}
or written as (the numerator of) a spin asymmetry is
\begin{align} \label{eq-regtree8}
    \frac{d\Delta \sigma^{Born}_{Reg}} {d^2k} &\equiv \half \left[ \frac{d\sigma^{Born}_{Reg}(+,+)} {d^2 k} - \frac{d\sigma^{Born}_{Reg}(+,-)} {d^2 k} \right] = -2 \frac{\alpha_s^2 C_F^2}{N_c} \, \frac{1}{zs} \, \frac{1} {k_T^2} \,.
\end{align}
We see explicitly that this polarization-dependent interaction is
suppressed by one power of $zs$.

\begin{figure}[htb]
\centering
\includegraphics[width=0.4\textwidth]{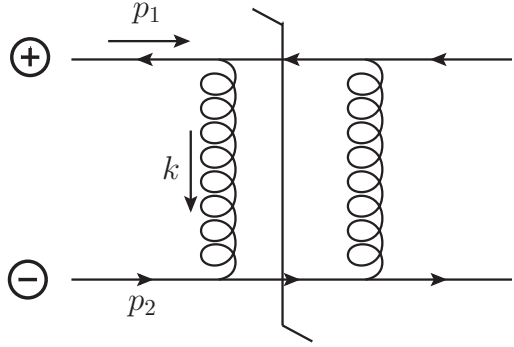}
\caption{The tree-level Pomeron exchange.}
\label{Fig-pom_tree}
\end{figure}

Next, let us calculate the tree-level gluon exchange in
Fig.~\ref{Fig-pom_tree}. This diagram can also lead to a spin
asymmetry if one of the gluons in the squared amplitude is sub-eikonal
and spin dependent, as will be seen clearly in what follows.  At the
level of the $q\bar{q}\to q\bar{q}$ squared amplitude, we find
\begin{align} \label{e:M2pom}
\langle M^2 \rangle &=  g^4\frac{C_F} {2N_c}\frac{1} {(k^2)^2}\,
{\rm Tr}\!\left[ \mathcal{P}_{-\sigma_1} \slashed{p}_1 \gamma^\mu(\slashed{p}_1-\slashed{k})\gamma^\rho\right]
{\rm Tr}\!\left[ \mathcal{P}_{\sigma_2} \slashed{p}_2\gamma_\rho(\slashed{p}_2+\slashed{k})\gamma_\mu\right].
\end{align}
%
The usual Pomeron exchange keeps the contribution from \eqref{e:M2pom}
that is dominant at high energies, i.e., eikonal gluon vertices.  This
leads to the dominant high-energy scattering cross-section, but is
insensitive to the spins $\sigma_1 , \sigma_2$ and, therefore, does
not contribute to the initial conditions of the helicity evolution.
On the other hand, the spin-dependent part of the first trace factor
in Eq.~(\ref{e:M2pom}) gives
\begin{equation}
-\frac{1} {2}\sigma_1{\rm Tr}\!\left[\gamma_5\slashed{p}_1\gamma^\mu(\slashed{p}_1-\slashed{k})\gamma^\rho\right] = 
 \half \sigma_1 \Tr[ \gamma^5 \slashed{p_1} \gamma^\mu \slashed{k} \gamma^\rho] =
-2 i \sigma_1 \: p_1^+ k_\alpha \: \epsilon^{- \mu \alpha \rho} ,
\end{equation}
which tells us, because of the antisymmetry of the epsilon tensor,
that either $\mu$ or $\rho$ must be a transverse index, i.e., one must
have a sub-eikonal vertex.  In the end we obtain
\begin{equation} \label{e:tree_Pom}
\frac{d\Delta \sigma^{Born}_{Pom}} {d^2 k} = \frac{2 \, C_F} {N_c}\alpha_s^2\,\frac{1} {zs}\frac{1} {k_T^2}\,,
\end{equation}
which is of the same order in the energy $zs$ as the Born diagram for
Reggeon exchange given in Eq.~(\ref{eq-regtree8}).

The initial conditions for our helicity evolution equations can be
constructed out of the cross sections Eqs.~(\ref{eq-regtree8}) and
(\ref{e:tree_Pom}), ``dressed" by unpolarized multiple
rescatterings~\cite{Itakura:2003jp}.



\begin{thebibliography}{10}

\bibitem{Bartels:1996wc}
J.~Bartels, B.~Ermolaev, and M.~Ryskin, {\it {Flavor singlet contribution to
  the structure function G(1) at small x}},  {\em Z.Phys.} {\bf C72} (1996)
  627--635, [\href{http://xxx.lanl.gov/abs/hep-ph/9603204}{{\tt
  hep-ph/9603204}}].

\bibitem{Gribov:1984tu}
L.~V. Gribov, E.~M. Levin, and M.~G. Ryskin, {\it {Semihard Processes in QCD}},
   {\em Phys. Rept.} {\bf 100} (1983) 1--150.

\bibitem{Mueller:1986wy}
A.~H. Mueller and J.-w. Qiu, {\it Gluon recombination and shadowing at small
  values of x},  {\em Nucl. Phys.} {\bf B268} (1986) 427.

\bibitem{McLerran:1993ni}
L.~D. McLerran and R.~Venugopalan, {\it Computing quark and gluon distribution
  functions for very large nuclei},  {\em Phys. Rev.} {\bf D49} (1994)
  2233--2241, [\href{http://xxx.lanl.gov/abs/hep-ph/9309289}{{\tt
  hep-ph/9309289}}].

\bibitem{McLerran:1993ka}
L.~D. McLerran and R.~Venugopalan, {\it Gluon distribution functions for very
  large nuclei at small transverse momentum},  {\em Phys. Rev.} {\bf D49}
  (1994) 3352--3355, [\href{http://xxx.lanl.gov/abs/hep-ph/9311205}{{\tt
  hep-ph/9311205}}].

\bibitem{McLerran:1994vd}
L.~D. McLerran and R.~Venugopalan, {\it Green's functions in the color field of
  a large nucleus},  {\em Phys. Rev.} {\bf D50} (1994) 2225--2233,
  [\href{http://xxx.lanl.gov/abs/hep-ph/9402335}{{\tt hep-ph/9402335}}].

\bibitem{Kovchegov:1996ty}
Y.~V. Kovchegov, {\it Non-Abelian {Weizs\"{a}cker-Williams} field and a two-
  dimensional effective color charge density for a very large nucleus},  {\em
  Phys. Rev.} {\bf D54} (1996) 5463--5469,
  [\href{http://xxx.lanl.gov/abs/hep-ph/9605446}{{\tt hep-ph/9605446}}].

\bibitem{Kovchegov:1997pc}
Y.~V. Kovchegov, {\it {Quantum structure of the non-Abelian
  Weizs\"{a}cker-Williams field for a very large nucleus}},  {\em Phys. Rev.}
  {\bf D55} (1997) 5445--5455,
  [\href{http://xxx.lanl.gov/abs/hep-ph/9701229}{{\tt hep-ph/9701229}}].

\bibitem{Jalilian-Marian:1997xn}
J.~Jalilian-Marian, A.~Kovner, L.~D. McLerran, and H.~Weigert, {\it The
  intrinsic glue distribution at very small x},  {\em Phys. Rev.} {\bf D55}
  (1997) 5414--5428, [\href{http://xxx.lanl.gov/abs/hep-ph/9606337}{{\tt
  hep-ph/9606337}}].

\bibitem{Mueller:1994rr}
A.~H. Mueller, {\it Soft gluons in the infinite momentum wave function and the
  {BFKL} pomeron},  {\em Nucl. Phys.} {\bf B415} (1994) 373--385.

\bibitem{Mueller:1994jq}
A.~H. Mueller and B.~Patel, {\it Single and double {BFKL} pomeron exchange and
  a dipole picture of high-energy hard processes},  {\em Nucl. Phys.} {\bf
  B425} (1994) 471--488, [\href{http://xxx.lanl.gov/abs/hep-ph/9403256}{{\tt
  hep-ph/9403256}}].

\bibitem{Mueller:1995gb}
A.~H. Mueller, {\it Unitarity and the {BFKL} pomeron},  {\em Nucl. Phys.} {\bf
  B437} (1995) 107--126, [\href{http://xxx.lanl.gov/abs/hep-ph/9408245}{{\tt
  hep-ph/9408245}}].

\bibitem{Balitsky:1996ub}
I.~Balitsky, {\it Operator expansion for high-energy scattering},  {\em Nucl.
  Phys.} {\bf B463} (1996) 99--160,
  [\href{http://xxx.lanl.gov/abs/hep-ph/9509348}{{\tt hep-ph/9509348}}].

\bibitem{Balitsky:1998ya}
I.~Balitsky, {\it Factorization and high-energy effective action},  {\em Phys.
  Rev.} {\bf D60} (1999) 014020,
  [\href{http://xxx.lanl.gov/abs/hep-ph/9812311}{{\tt hep-ph/9812311}}].

\bibitem{Kovchegov:1999yj}
Y.~V. Kovchegov, {\it Small-x {$F_2$} structure function of a nucleus including
  multiple pomeron exchanges},  {\em Phys. Rev.} {\bf D60} (1999) 034008,
  [\href{http://xxx.lanl.gov/abs/hep-ph/9901281}{{\tt hep-ph/9901281}}].

\bibitem{Kovchegov:1999ua}
Y.~V. Kovchegov, {\it Unitarization of the {BFKL} pomeron on a nucleus},  {\em
  Phys. Rev.} {\bf D61} (2000) 074018,
  [\href{http://xxx.lanl.gov/abs/hep-ph/9905214}{{\tt hep-ph/9905214}}].

\bibitem{Jalilian-Marian:1997dw}
J.~Jalilian-Marian, A.~Kovner, and H.~Weigert, {\it The {Wilson}
  renormalization group for low x physics: Gluon evolution at finite parton
  density},  {\em Phys. Rev.} {\bf D59} (1998) 014015,
  [\href{http://xxx.lanl.gov/abs/hep-ph/9709432}{{\tt hep-ph/9709432}}].

\bibitem{Jalilian-Marian:1997gr}
J.~Jalilian-Marian, A.~Kovner, A.~Leonidov, and H.~Weigert, {\it The {Wilson}
  renormalization group for low x physics: Towards the high density regime},
  {\em Phys. Rev.} {\bf D59} (1998) 014014,
  [\href{http://xxx.lanl.gov/abs/hep-ph/9706377}{{\tt hep-ph/9706377}}].

\bibitem{Iancu:2001ad}
E.~Iancu, A.~Leonidov, and L.~D. McLerran, {\it {The renormalization group
  equation for the color glass condensate}},  {\em Phys. Lett.} {\bf B510}
  (2001) 133--144.

\bibitem{Iancu:2000hn}
E.~Iancu, A.~Leonidov, and L.~D. McLerran, {\it Nonlinear gluon evolution in
  the color glass condensate. {I}},  {\em Nucl. Phys.} {\bf A692} (2001)
  583--645, [\href{http://xxx.lanl.gov/abs/hep-ph/0011241}{{\tt
  hep-ph/0011241}}].

\bibitem{Iancu:2003xm}
E.~Iancu and R.~Venugopalan, {\it The color glass condensate and high energy
  scattering in {QCD}},  \href{http://xxx.lanl.gov/abs/hep-ph/0303204}{{\tt
  hep-ph/0303204}}.

\bibitem{Weigert:2005us}
H.~Weigert, {\it Evolution at small {$x_{bj}$: The Color Glass Condensate}},
  {\em Prog. Part. Nucl. Phys.} {\bf 55} (2005) 461--565,
  [\href{http://xxx.lanl.gov/abs/hep-ph/0501087}{{\tt hep-ph/0501087}}].

\bibitem{Jalilian-Marian:2005jf}
J.~Jalilian-Marian and Y.~V. Kovchegov, {\it Saturation physics and deuteron
  gold collisions at {RHIC}},  {\em Prog. Part. Nucl. Phys.} {\bf 56} (2006)
  104--231, [\href{http://xxx.lanl.gov/abs/hep-ph/0505052}{{\tt
  hep-ph/0505052}}].

\bibitem{Gelis:2010nm}
F.~Gelis, E.~Iancu, J.~Jalilian-Marian, and R.~Venugopalan, {\it {The Color
  Glass Condensate}},  {\em Ann.Rev.Nucl.Part.Sci.} {\bf 60} (2010) 463--489,
  [\href{http://xxx.lanl.gov/abs/1002.0333}{{\tt arXiv:1002.0333}}].

\bibitem{Albacete:2014fwa}
J.~L. Albacete and C.~Marquet, {\it {Gluon saturation and initial conditions
  for relativistic heavy ion collisions}},  {\em Prog.Part.Nucl.Phys.} {\bf 76}
  (2014) 1--42, [\href{http://xxx.lanl.gov/abs/1401.4866}{{\tt
  arXiv:1401.4866}}].

\bibitem{KovchegovLevin}
Y.~V. Kovchegov and E.~Levin, {\em Quantum Chromodynamics at High Energy}.
\newblock Cambridge University Press, 2012.

\bibitem{Boer:2006rj}
D.~Boer, A.~Dumitru, and A.~Hayashigaki, {\it {Single transverse-spin
  asymmetries in forward pion production at high energy: Incorporating small-x
  effects in the target}},  {\em Phys.Rev.} {\bf D74} (2006) 074018,
  [\href{http://xxx.lanl.gov/abs/hep-ph/0609083}{{\tt hep-ph/0609083}}].

\bibitem{Boer:2008ze}
D.~Boer, A.~Utermann, and E.~Wessels, {\it {The Saturation scale and its
  x-dependence from Lambda polarization studies}},  {\em Phys.Lett.} {\bf B671}
  (2009) 91--98, [\href{http://xxx.lanl.gov/abs/0811.0998}{{\tt
  arXiv:0811.0998}}].

\bibitem{Boer:2002ij}
D.~Boer and A.~Dumitru, {\it {Polarized hyperons from pA scattering in the
  gluon saturation regime}},  {\em Phys.Lett.} {\bf B556} (2003) 33--40,
  [\href{http://xxx.lanl.gov/abs/hep-ph/0212260}{{\tt hep-ph/0212260}}].

\bibitem{Dominguez:2011br}
F.~Dominguez, J.-W. Qiu, B.-W. Xiao, and F.~Yuan, {\it {On the linearly
  polarized gluon distributions in the color dipole model}},  {\em Phys.Rev.}
  {\bf D85} (2012) 045003, [\href{http://xxx.lanl.gov/abs/1109.6293}{{\tt
  arXiv:1109.6293}}].

\bibitem{Metz:2011wb}
A.~Metz and J.~Zhou, {\it {Distribution of linearly polarized gluons inside a
  large nucleus}},  {\em Phys.Rev.} {\bf D84} (2011) 051503,
  [\href{http://xxx.lanl.gov/abs/1105.1991}{{\tt arXiv:1105.1991}}].

\bibitem{Kovchegov:2012ga}
Y.~V. Kovchegov and M.~D. Sievert, {\it {A New Mechanism for Generating a
  Single Transverse Spin Asymmetry}},  {\em Phys.Rev.} {\bf D86} (2012) 034028,
  [\href{http://xxx.lanl.gov/abs/1201.5890}{{\tt arXiv:1201.5890}}].

\bibitem{Mueller:2012uf}
A.~Mueller, B.-W. Xiao, and F.~Yuan, {\it {Sudakov Resummation in Small-$x$
  Saturation Formalism}},  {\em Phys.Rev.Lett.} {\bf 110} (2013), no.~8 082301,
  [\href{http://xxx.lanl.gov/abs/1210.5792}{{\tt arXiv:1210.5792}}].

\bibitem{Mueller:2013wwa}
A.~Mueller, B.-W. Xiao, and F.~Yuan, {\it {Sudakov double logarithms
  resummation in hard processes in the small-x saturation formalism}},  {\em
  Phys.Rev.} {\bf D88} (2013), no.~11 114010,
  [\href{http://xxx.lanl.gov/abs/1308.2993}{{\tt arXiv:1308.2993}}].

\bibitem{Kang:2011ni}
Z.-B. Kang and F.~Yuan, {\it {Single Spin Asymmetry Scaling in the Forward
  Rapidity Region at RHIC}},  {\em Phys.Rev.} {\bf D84} (2011) 034019,
  [\href{http://xxx.lanl.gov/abs/1106.1375}{{\tt arXiv:1106.1375}}].

\bibitem{Kang:2012vm}
Z.-B. Kang and B.-W. Xiao, {\it {Sivers asymmetry of Drell-Yan production in
  small-$x$ regime}},  {\em Phys.Rev.} {\bf D87} (2013) 034038,
  [\href{http://xxx.lanl.gov/abs/1212.4809}{{\tt arXiv:1212.4809}}].

\bibitem{Schafer:2013mza}
A.~Schäfer and J.~Zhou, {\it {Process dependent nuclear $k_\perp$ broadening
  effect}},  {\em Phys.Rev.} {\bf D88} (2013), no.~7 074012,
  [\href{http://xxx.lanl.gov/abs/1305.5042}{{\tt arXiv:1305.5042}}].

\bibitem{Zhou:2013gsa}
J.~Zhou, {\it {Transverse single spin asymmetries at small x and the anomalous
  magnetic moment}},  {\em Phys.Rev.} {\bf D89} (2014), no.~7 074050,
  [\href{http://xxx.lanl.gov/abs/1308.5912}{{\tt arXiv:1308.5912}}].

\bibitem{Altinoluk:2014oxa}
T.~Altinoluk, N.~Armesto, G.~Beuf, M.~Martinez, and C.~A. Salgado, {\it
  {Next-to-eikonal corrections in the CGC: gluon production and spin
  asymmetries in pA collisions}},  {\em JHEP} {\bf 07} (2014) 068,
  [\href{http://xxx.lanl.gov/abs/1404.2219}{{\tt arXiv:1404.2219}}].

\bibitem{Kovchegov:2013cva}
Y.~V. Kovchegov and M.~D. Sievert, {\it {Sivers function in the quasiclassical
  approximation}},  {\em Phys. Rev.} {\bf D89} (2014), no.~5 054035,
  [\href{http://xxx.lanl.gov/abs/1310.5028}{{\tt arXiv:1310.5028}}].

\bibitem{Kovchegov:2015zha}
Y.~V. Kovchegov and M.~D. Sievert, {\it {Calculating TMDs of an Unpolarized
  Target: Quasi-Classical Approximation and Quantum Evolution}},
  \href{http://xxx.lanl.gov/abs/1505.0117}{{\tt arXiv:1505.0117}}.

\bibitem{Balitsky:2014wna}
I.~Balitsky and A.~Tarasov, {\it {Evolution of gluon TMD at low and moderate
  x}},  {\em Int.J.Mod.Phys.Conf.Ser.} {\bf 37} (2015) 0058,
  [\href{http://xxx.lanl.gov/abs/1411.0714}{{\tt arXiv:1411.0714}}].

\bibitem{Balitsky:2015qba}
I.~Balitsky and A.~Tarasov, {\it {Rapidity evolution of gluon TMD from low to
  moderate x}},  {\em JHEP} {\bf 10} (2015) 017,
  [\href{http://xxx.lanl.gov/abs/1505.0215}{{\tt arXiv:1505.0215}}].

\bibitem{Tarasov:2015pxa}
A.~Tarasov, {\it {Evolution of gluon TMDs from small to moderate x}},  in {\em
  {Proceedings, QCD Evolution Workshop (QCD 2015)}}, 2015.
\newblock \href{http://xxx.lanl.gov/abs/1510.0693}{{\tt arXiv:1510.0693}}.

\bibitem{Altinoluk:2015gia}
T.~Altinoluk, N.~Armesto, G.~Beuf, and A.~Moscoso, {\it
  {Next-to-next-to-eikonal corrections in the CGC}},
  \href{http://xxx.lanl.gov/abs/1505.0140}{{\tt arXiv:1505.0140}}.

\bibitem{Boer:2015pni}
D.~Boer, M.~G. Echevarria, P.~Mulders, and J.~Zhou, {\it {Single spin
  asymmetries from a single Wilson loop}},
  \href{http://xxx.lanl.gov/abs/1511.0348}{{\tt arXiv:1511.0348}}.

\bibitem{Collins:1989gx}
J.~C. Collins, D.~E. Soper, and G.~F. Sterman, {\it {Factorization of Hard
  Processes in QCD}},  {\em Adv.Ser.Direct.High Energy Phys.} {\bf 5} (1988)
  1--91, [\href{http://xxx.lanl.gov/abs/hep-ph/0409313}{{\tt hep-ph/0409313}}].

\bibitem{Collins:1981uk}
J.~C. Collins and D.~E. Soper, {\it {Back-To-Back Jets in QCD}},  {\em
  Nucl.Phys.} {\bf B193} (1981) 381.

\bibitem{Ashman:1987hv}
{\bf European Muon} Collaboration, J.~Ashman {\em et.~al.}, {\it {A Measurement
  of the Spin Asymmetry and Determination of the Structure Function g(1) in
  Deep Inelastic Muon-Proton Scattering}},  {\em Phys. Lett.} {\bf B206} (1988)
  364.

\bibitem{Ashman:1989ig}
{\bf European Muon} Collaboration, J.~Ashman {\em et.~al.}, {\it {An
  Investigation of the Spin Structure of the Proton in Deep Inelastic
  Scattering of Polarized Muons on Polarized Protons}},  {\em Nucl. Phys.} {\bf
  B328} (1989) 1.

\bibitem{Accardi:2012qut}
A.~Accardi, J.~Albacete, M.~Anselmino, N.~Armesto, E.~Aschenauer, {\em
  et.~al.}, {\it {Electron Ion Collider: The Next QCD Frontier - Understanding
  the glue that binds us all}},  \href{http://xxx.lanl.gov/abs/1212.1701}{{\tt
  arXiv:1212.1701}}.

\bibitem{Aschenauer:2013woa}
E.~C. Aschenauer {\em et.~al.}, {\it {The RHIC Spin Program: Achievements and
  Future Opportunities}},  \href{http://xxx.lanl.gov/abs/1304.0079}{{\tt
  arXiv:1304.0079}}.

\bibitem{Aschenauer:2015eha}
E.-C. Aschenauer {\em et.~al.}, {\it {The RHIC SPIN Program: Achievements and
  Future Opportunities}},  \href{http://xxx.lanl.gov/abs/1501.0122}{{\tt
  arXiv:1501.0122}}.

\bibitem{Jaffe:1989jz}
R.~L. Jaffe and A.~Manohar, {\it {The G(1) Problem: Fact and Fantasy on the
  Spin of the Proton}},  {\em Nucl. Phys.} {\bf B337} (1990) 509--546.

\bibitem{Ji:1996ek}
X.-D. Ji, {\it {Gauge-Invariant Decomposition of Nucleon Spin}},  {\em Phys.
  Rev. Lett.} {\bf 78} (1997) 610--613,
  [\href{http://xxx.lanl.gov/abs/hep-ph/9603249}{{\tt hep-ph/9603249}}].

\bibitem{Ji:2012sj}
X.~Ji, X.~Xiong, and F.~Yuan, {\it {Proton Spin Structure from Measurable
  Parton Distributions}},  {\em Phys. Rev. Lett.} {\bf 109} (2012) 152005,
  [\href{http://xxx.lanl.gov/abs/1202.2843}{{\tt arXiv:1202.2843}}].

\bibitem{deFlorian:2014yva}
D.~de~Florian, R.~Sassot, M.~Stratmann, and W.~Vogelsang, {\it {Evidence for
  polarization of gluons in the proton}},  {\em Phys. Rev. Lett.} {\bf 113}
  (2014), no.~1 012001, [\href{http://xxx.lanl.gov/abs/1404.4293}{{\tt
  arXiv:1404.4293}}].

\bibitem{Nocera:2014gqa}
{\bf NNPDF} Collaboration, E.~R. Nocera, R.~D. Ball, S.~Forte, G.~Ridolfi, and
  J.~Rojo, {\it {A first unbiased global determination of polarized PDFs and
  their uncertainties}},  {\em Nucl. Phys.} {\bf B887} (2014) 276--308,
  [\href{http://xxx.lanl.gov/abs/1406.5539}{{\tt arXiv:1406.5539}}].

\bibitem{Aschenauer:2012ve}
E.~C. Aschenauer, R.~Sassot, and M.~Stratmann, {\it {Helicity Parton
  Distributions at a Future Electron-Ion Collider: A Quantitative Appraisal}},
  {\em Phys. Rev.} {\bf D86} (2012) 054020,
  [\href{http://xxx.lanl.gov/abs/1206.6014}{{\tt arXiv:1206.6014}}].

\bibitem{Kirschner:1983di}
R.~Kirschner and L.~Lipatov, {\it {Double Logarithmic Asymptotics and Regge
  Singularities of Quark Amplitudes with Flavor Exchange}},  {\em Nucl.Phys.}
  {\bf B213} (1983) 122--148.

\bibitem{Kirschner:1994rq}
R.~Kirschner, {\it {Reggeon interactions in perturbative QCD}},  {\em Z.Phys.}
  {\bf C65} (1995) 505--510,
  [\href{http://xxx.lanl.gov/abs/hep-th/9407085}{{\tt hep-th/9407085}}].

\bibitem{Kirschner:1994vc}
R.~Kirschner, {\it {Regge asymptotics of scattering with flavor exchange in
  QCD}},  {\em Z.Phys.} {\bf C67} (1995) 459--466,
  [\href{http://xxx.lanl.gov/abs/hep-th/9404158}{{\tt hep-th/9404158}}].

\bibitem{Griffiths:1999dj}
S.~Griffiths and D.~Ross, {\it {Studying the perturbative Reggeon}},  {\em
  Eur.Phys.J.} {\bf C12} (2000) 277--286,
  [\href{http://xxx.lanl.gov/abs/hep-ph/9906550}{{\tt hep-ph/9906550}}].

\bibitem{Kuraev:1977fs}
E.~A. Kuraev, L.~N. Lipatov, and V.~S. Fadin, {\it {The Pomeranchuk
  singlularity in non-Abelian gauge theories}},  {\em Sov. Phys. JETP} {\bf 45}
  (1977) 199--204.

\bibitem{Balitsky:1978ic}
I.~Balitsky and L.~Lipatov, {\it {The Pomeranchuk Singularity in Quantum
  Chromodynamics}},  {\em Sov.J.Nucl.Phys.} {\bf 28} (1978) 822--829.

\bibitem{Bartels:1995iu}
J.~Bartels, B.~Ermolaev, and M.~Ryskin, {\it {Nonsinglet contributions to the
  structure function g1 at small x}},  {\em Z.Phys.} {\bf C70} (1996) 273--280,
  [\href{http://xxx.lanl.gov/abs/hep-ph/9507271}{{\tt hep-ph/9507271}}].

\bibitem{Itakura:2003jp}
K.~Itakura, Y.~V. Kovchegov, L.~McLerran, and D.~Teaney, {\it {Baryon stopping
  and valence quark distribution at small x}},  {\em Nucl. Phys.} {\bf A730}
  (2004) 160--190, [\href{http://xxx.lanl.gov/abs/hep-ph/0305332}{{\tt
  hep-ph/0305332}}].

\bibitem{'tHooft:1974hx}
G.~'t~Hooft, {\it A two-dimensional model for mesons},  {\em Nucl. Phys.} {\bf
  B75} (1974) 461.

\bibitem{Lepage:1980fj}
G.~P. Lepage and S.~J. Brodsky, {\it Exclusive processes in perturbative
  quantum chromodynamics},  {\em Phys. Rev.} {\bf D22} (1980) 2157.

\bibitem{Brodsky:1997de}
S.~J. Brodsky, H.-C. Pauli, and S.~S. Pinsky, {\it {Quantum chromodynamics and
  other field theories on the light cone}},  {\em Phys.Rept.} {\bf 301} (1998)
  299--486, [\href{http://xxx.lanl.gov/abs/hep-ph/9705477}{{\tt
  hep-ph/9705477}}].

\bibitem{Meissner:2007rx}
S.~Meissner, A.~Metz, and K.~Goeke, {\it {Relations between generalized and
  transverse momentum dependent parton distributions}},  {\em Phys. Rev.} {\bf
  D76} (2007) 034002, [\href{http://xxx.lanl.gov/abs/hep-ph/0703176}{{\tt
  hep-ph/0703176}}].

\bibitem{Gribov:1972ri}
V.~N. Gribov and L.~N. Lipatov, {\it {Deep inelastic e p scattering in
  perturbation theory}},  {\em Sov. J. Nucl. Phys.} {\bf 15} (1972) 438--450.

\bibitem{Altarelli:1977zs}
G.~Altarelli and G.~Parisi, {\it {Asymptotic Freedom in Parton Language}},
  {\em Nucl. Phys.} {\bf B126} (1977) 298.

\bibitem{Dokshitzer:1977sg}
Y.~L. Dokshitzer, {\it {Calculation of the Structure Functions for Deep
  Inelastic Scattering and $e^+ e^-$ Annihilation by Perturbation Theory in
  Quantum Chromodynamics}},  {\em Sov. Phys. JETP} {\bf 46} (1977) 641--653.

\bibitem{Chen:1995pa}
Z.~Chen and A.~H. Mueller, {\it {The dipole picture of high-energy scattering,
  the BFKL equation and many gluon compound states}},  {\em Nucl. Phys.} {\bf
  B451} (1995) 579--604.

\bibitem{Dominguez:2011gc}
F.~Dominguez, A.~Mueller, S.~Munier, and B.-W. Xiao, {\it {On the small-x
  evolution of the color quadrupole and the Weizs\'acker-Williams gluon
  distribution}},  {\em Phys.Lett.} {\bf B705} (2011) 106--111,
  [\href{http://xxx.lanl.gov/abs/1108.1752}{{\tt arXiv:1108.1752}}].

\bibitem{Weigert:2000gi}
H.~Weigert, {\it Unitarity at small {B}jorken x},  {\em Nucl. Phys.} {\bf A703}
  (2002) 823--860, [\href{http://xxx.lanl.gov/abs/hep-ph/0004044}{{\tt
  hep-ph/0004044}}].

\bibitem{Rummukainen:2003ns}
K.~Rummukainen and H.~Weigert, {\it Universal features of {JIMWLK} and {BK}
  evolution at small $x$},  {\em Nucl. Phys.} {\bf A739} (2004) 183--226,
  [\href{http://xxx.lanl.gov/abs/hep-ph/0309306}{{\tt hep-ph/0309306}}].

\bibitem{Ermolaev:2003zx} 
  B.~I.~Ermolaev, M.~Greco and S.~I.~Troyan, {\it Running coupling
    effects for the singlet structure function g(1) at small x}, {\em
    Phys.\ Lett.} {\bf B579}, 321 (2004), 
    [\href{http://xxx.lanl.gov/abs/hep-ph/0307128}{{\tt hep-ph/0307128}}].

\bibitem{Gardi:2006rp}
E.~Gardi, J.~Kuokkanen, K.~Rummukainen, and H.~Weigert, {\it Running coupling
  and power corrections in nonlinear evolution at the high-energy limit},  {\em
  Nucl. Phys.} {\bf A784} (2007) 282--340,
  [\href{http://xxx.lanl.gov/abs/hep-ph/0609087}{{\tt hep-ph/0609087}}].

\bibitem{Balitsky:2006wa}
I.~I. Balitsky, {\it {Quark Contribution to the Small-$x$ Evolution of Color
  Dipole}},  {\em Phys. Rev. D} {\bf 75} (2007) 014001,
  [\href{http://xxx.lanl.gov/abs/hep-ph/0609105}{{\tt hep-ph/0609105}}].

\bibitem{Kovchegov:2006vj}
Y.~Kovchegov and H.~Weigert, {\it {Triumvirate of Running Couplings in
  Small-$x$ Evolution}},  {\em Nucl. Phys. {\bf A}} {\bf 784} (2007) 188--226,
  [\href{http://xxx.lanl.gov/abs/hep-ph/0609090}{{\tt hep-ph/0609090}}].

\bibitem{Albacete:2007yr}
J.~L. Albacete and Y.~V. Kovchegov, {\it Solving high energy evolution equation
  including running coupling corrections},  {\em Phys. Rev.} {\bf D75} (2007)
  125021, [\href{http://xxx.lanl.gov/abs/0704.0612}{{\tt 0704.0612}}].

\bibitem{Kovchegov:2008mk}
Y.~V. Kovchegov, J.~Kuokkanen, K.~Rummukainen, and H.~Weigert, {\it
  {Subleading-$N_c$ corrections in non-linear small-$x$ evolution}},  {\em
  Nucl. Phys.} {\bf A823} (2009) 47--82,
  [\href{http://xxx.lanl.gov/abs/0812.3238}{{\tt arXiv:0812.3238}}].

\end{thebibliography}

\providecommand{\href}[2]{#2}\begingroup\raggedright\endgroup

\end{document}